\documentclass{article}
\usepackage{amssymb}
\usepackage{amsmath}
\usepackage{cprotect}
\usepackage{cite}
\usepackage{hyperref}
\usepackage{amsfonts}
\usepackage{comment}
\usepackage{soul}
\usepackage{tikz}
\usepackage{tikz-cd}
\usepackage{subcaption}
\usetikzlibrary{arrows,decorations.markings,decorations.pathreplacing}
\usepackage{bbm}

\usepackage{mathtools,amsthm}
\usepackage[titletoc]{appendix}
\makeatletter
\newtheorem*{rep@theorem}{\rep@title}
\newcommand{\newreptheorem}[2]{%
\newenvironment{rep#1}[1]{%
 \def\rep@title{#2 \ref{##1}}%
 \begin{rep@theorem}}%
 {\end{rep@theorem}}}
\makeatother
\newtheorem{lemma}{Lemma}[section]
\newreptheorem{lemma}{Lemma}

\newreptheorem{conj}{Conjecture}

\theoremstyle{definition}

\newcommand{\OO}{\mathcal{O}}

\input xy
\xyoption{all}
\usepackage{enumitem}

\usepackage{graphicx} 

\def\sqr#1#2{{\vcenter{\vbox{\hrule height.#2pt
            \hbox{\vrule width.#2pt height#1pt \kern#1pt
                  \vrule width.#2pt}\hrule height.#2pt}}}}

\def\sqra#1#2#3{{\vcenter{\vbox{\hrule height.#2pt
            \hbox{\vrule width.#2pt height#1pt \kern5pt 
#3
                  \vrule width.#2pt}\hrule height.#2pt}}}}

\numberwithin{equation}{section}

\numberwithin{table}{section}

\begin{document}

\begin{center}

{\large\bf Noninvertible symmetries in the B model TFT}

Andrei C\u{a}ld\u{a}raru$^1$, Tony Pantev$^2$, Eric Sharpe$^3$, Benjamin Sung$^4$, Xingyang Yu$^3$

        \vspace*{0.1in}
        
        \begin{tabular}{cc}
                {\begin{tabular}{l}
                $^1$ Department of Mathematics\\
                        480 Lincoln Drive\\
                        213 Van Vleck Hall\\
                        Madison, WI 53706 \end{tabular}}
                        &
                {\begin{tabular}{l}
                $^2$ Department of Mathematics\\
                        David Rittenhouse Lab\\
                        209 South 33rd Street\\
                        Philadelphia, PA  19104-6395 \end{tabular}}
                        \\
                        $\,$ & $\,$ \\
                {\begin{tabular}{l}
                $^3$ Department of Physics MC 0435\\
                                850 West Campus Drive\\
                                Virginia Tech\\
                                Blacksburg, VA  24061 \end{tabular}}
                                 &
                {\begin{tabular}{l}
                $^4$ Department of Mathematics\\
                    South Hall, Room 6607\\
                    University of California\\
                    Santa Barbara, CA  93106-3080 \end{tabular}}
        \end{tabular}

        \vspace*{0.2in}

{\tt andreic@math.wisc.edu},
{\tt tpantev@math.upenn.edu},
{\tt ersharpe@vt.edu},
{\tt bsung@ucsb.edu},
{\tt xingyangy@vt.edu}

\end{center}

In this paper we explore noninvertible symmetries in general (not necessarily rational) SCFTs and their topological B-twists for Calabi-Yau manifolds.  We begin with a detailed overview of defects in the topological B model.  For trivial reasons, all defects in the topological B model are topological operators, and define (often noninvertible) symmetries of that topological field theory, but only a subset remain topological in the physical (i.e., untwisted) theory.  For a generic target space Calabi-Yau $X$, we discuss geometric realizations of those defects, as simultaneously A- and B-twistable complex Lagrangian and complex coisotropic branes on $X \times X$, and discuss their fusion products. To be clear, the possible noninvertible symmetries in the B model are more general than can be described with fusion categories.  That said, we do describe realizations of some Tambara-Yamagami categories in the B model for an elliptic curve target, and also argue that elliptic curves can not admit Fibonacci or Haagerup structures.  We also discuss how decomposition is realized in this language.

\begin{flushleft}
March 2025 
\end{flushleft}

\newpage

\tableofcontents

\newpage

\section{Introduction}

Briefly, the point of this paper is to concretely illustrate the role of noninvertible symmetries in nonlinear sigma models for general Calabi-Yau's by 
studying B-branes and the B model topological field theory.  In these theories, B-branes are described in terms of derived categories
(see e.g.~\cite{Sharpe:1999qz,Aspinwall:2001pu,Sharpe:2003dr,Aspinwall:2004jr,Ando:2010nm}),
hence topological line operators
can be realized explicitly via
derived categories.
 The topological line operators that form the basis for (noninvertible) symmetries, abstractly, can themselves be described completely explicitly in the B model as objects in a derived category, and fusion products of line operators, orientation reversal maps, and other operations can also be realized completely explicitly, making the B model a rich playground for understanding the role of noninvertible symmetries in Calabi-Yau compactifications.

Now, typically one defines finite noninvertible symmetries in 2D in terms of fusion categories, but, derived categories are, broadly speaking, richer and more general structures than fusion categories.  We will explicitly illustrate how almost all of the axioms of fusion categorical symmetries can be realized by B model defects, with the important exception of axioms about the category being finitely generated.  This exception stems from the fact that
fusion categories are analogues of finite groups, whereas derived categories for Calabi-Yau's, and for that matter noninvertible symmetries more generally, are not so constrained.  We will see how examples of fusion categories can be realized in derived categories, but, we emphasize, derived categories are not equivalent to fusion categories.

\begin{figure}[h]
    \centering
    \includegraphics[width=10cm]{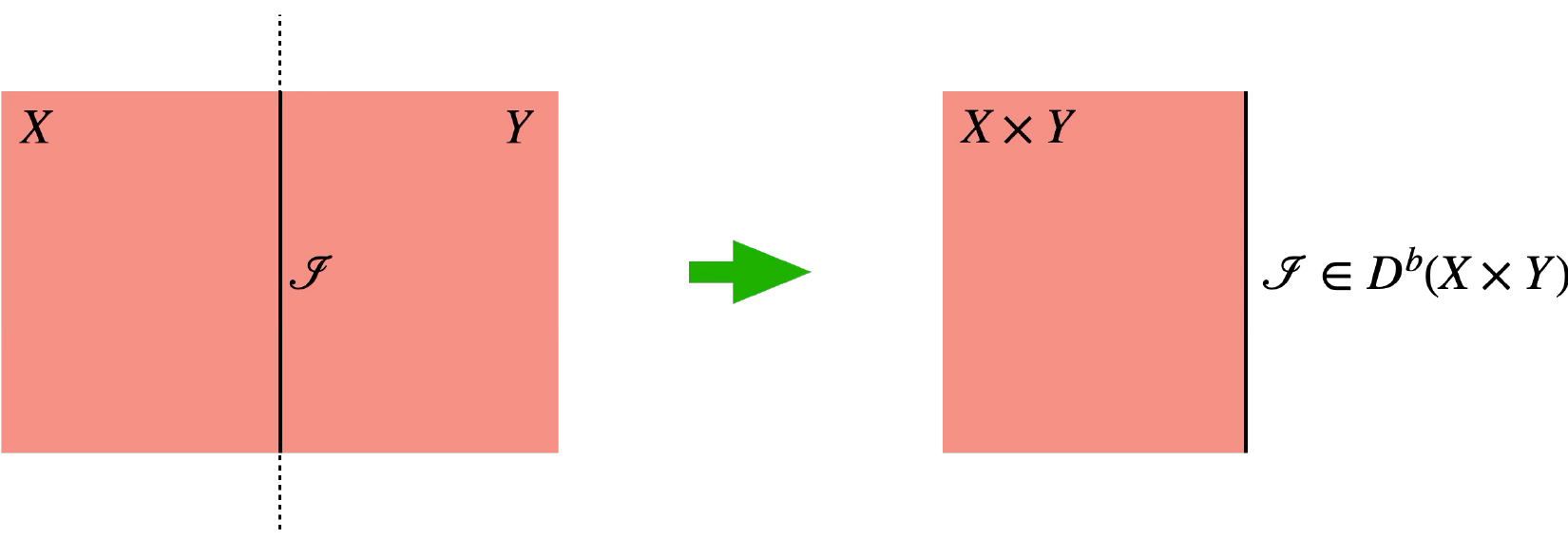}
    \caption{Topological interfaces of B model TFTs can be folded into boundaries of the produced theory, which can naturally be captured by derived categories. The interface is promoted to a defect in the special case $X=Y$.}
    \label{fig:folding}
\end{figure}

The correspondence between topological defects, or more generally topological interfaces, in B model TFTs and derived categories can be understood from the folding trick as follows. Given a topological interface separating two B model TFTs with target spaces $X$ and $Y$, folding the worldsheet along the interface will transform it as a boundary of the tensor product theory with target space $X\times Y$\footnote{More precisely, folding along an interface between two theories $\mathcal{T}_1$ and $\mathcal{T}_2$ will lead to a tensor product theory $\mathcal{T}_1\times \bar{\mathcal{T}}_2$ with left- and right-moving sectors exchanged for the ${\mathcal{T}_2}$ theory. For reading topological interfaces/defects in topological sigma models from derived categories, this exchange matters little. However, there is a subtlety when building the symplectic form on the product space $X\times Y$, which we will discuss in Section \ref{sect:semisimp}.}. This boundary specifies a B-brane wrapping a holomorphic submanifold of the target space $X\times Y$, which, under the B model language, is best described by the derived category of coherent sheaves, denoted by $D^b(X\times Y)$. See Figure \ref{fig:folding} for an illustration. The topological defects for a given B model with target space $X$ fall in the special case with $X=Y$, thus given by $D^d(X\times X)$.

To be clear, the idea that derived categories give a concrete realization of topological line operators in the B model, is already known to experts. Our interest in this paper is to apply these methods to better understand what noninvertible symmetries can arise in Calabi-Yau compactifications.  To that end, as we are not aware of similar presentations in the literature, we will need to explicitly describe the realization of the structure of noninvertible symmetries in this language in order to be able to give precise computations.  The resulting description of noninvertible symmetries will hopefully also be useful for non-experts.

From a 2D TFT perspective, our work complements those constructing TFTs whose symmetries are given by fusion categories, e.g., \cite{Bhardwaj:2017xup, Thorngren:2019iar, Huang:2021zvu} (which generalize the earlier work \cite{Moore:2006dw}). On one hand, those work start with fusion categories $\mathcal{C}$ and their module categories (associated with special symmetric algebra objects) as the defining data to axiomatize $\mathcal{C}$-symmetric 2D TFTs. On the other hand, we take a different perspective where TFTs can be explicitly constructed from topological twists, and then the derived category data underlying them can be extracted to describe noninvertible defects, which are more general than fusion categories. 

From a string theory perspective, our work generally constructs noninvertible symmetries in the worldsheet theory describing the internal geometric sector for superstring compactification, providing an alternative treatment compared to works such as \cite{Cordova:2023qei, Angius:2024evd}. Compared to \cite{Cordova:2023qei}, where noninvertible symmetries are inherited from the RCFT Gepner point via symmetry preserving deformation, we construct noninvertible defects for generic irrational points in the Calabi-Yau moduli space. For compact Calabi-Yau manifolds, the spacetime quantum gravity theory is expected to have related gauged noninvertible symmetries\footnote{In the leading order of the $g_s$ expansion. Noninvertible global symmetries on the worldsheet generally get broken by higher string-loop corrections. See e.g.~\cite{Bachas:2012bj} (and also \cite{Kaidi:2024wio, Heckman:2024obe, Bharadwaj:2024gpj} for more recent discussion.)}. We find evidence for such exotic gauge symmetries by investigating noninvertible gauge transformations of the B-field. For noncompact Calabi-Yau manifolds, it is known that B-type D-branes capture the data of QFTs engineered on D-brane probes\footnote{See e.g.~\cite{Aspinwall:2004jr} for a review.}. Given that our approach in this paper identifies worldsheet defects to the B-branes in spacetime, it is natural to expect our construction implies a more systematic study on noninvertible symmetries in these QFTs\footnote{Similar expectation was pointed out in \cite{Heckman:2022muc}.}, generalizing the results in e.g.~\cite{GarciaEtxebarria:2022vzq, Apruzzi:2022rei, Heckman:2022xgu, Bah:2023ymy, Apruzzi:2023uma, Yu:2023nyn, Franco:2024mxa}.
(See also \cite{Arias-Tamargo:2025xdd} for a recent construction.)

Before outlining the structure of this paper, we would like to point out that interfaces/defects in conformal field theories and topological field theories have been studied in many references, a small partial list of which includes \cite{Carqueville:2023qrk,Frohlich:2006ch,Bachas:2007td,Bachas:2013ora,Petkova:2000ip,Bachas:2004sy,Bachas:2001vj,Quella:2002ct,Kapustin:2010zc,Bachas:2012bj,Brunner:2013ota,Bachas:2013nxa,Brunner:2014lua,Carqueville:2017aoe,Carqueville:2023jhb}, and also in Landau-Ginzburg models and GLSMs,
see for example \cite{Rozansky:2003hz,Khovanov:2004bc,Brunner:2007qu,Brunner:2007ur,Brunner:2008fa,Brunner:2009zt,Brunner:2010xm,Carqueville:2010hu,Brunner:2021cga,Brunner:2023ckg,Brunner:2024quk}.  There is a long history behind the observation that all symmetries of CFTs can be encoded in interfaces/defects, though not every interface/defect is an invertible symmetry.

In section~\ref{sect:bmodel} we discuss the topological B model.  We describe line operators as objects in a derived category, how their fusion products and other operations can be understood via derived categories, and describe pertinent identities that they satisfy.  We also briefly outline analogous results in related topological field theories (such as B-twisted Landau-Ginzburg models, and the A model).
Now, in the B model topological field theory, every defect is topological (meaning, invariant under motions of the defect on the worldsheet).  However, after untwisting to a physical theory, only a subset of those defects remain topological.  

With that in mind, in section~\ref{sect:phys} we turn to topological line operators (of B-model-type) in untwisted nonlinear sigma models.
We review conditions for a defect in a physical theory to be topological, and also discuss semistability and its relation to semisimplicity.

In section~\ref{sect:ellcurve} we turn to examples on elliptic curves.
To simplify computations, we discuss results in K-theory rather than derived categories per se.  In section~\ref{sect:k3} we describe some analogous computations on K3 surfaces, and in section~\ref{sect:cy} we briefly describe two noninvertible symmetry structures arising on general Calabi-Yau's.

Finally, in section~\ref{sect:decomp}, we describe how decomposition
\cite{Hellerman:2006zs,Sharpe:2022ene}
and some condensation defects can be understood in this language.

In appendix~\ref{sect:misc}, we give some miscellaneous technical results on derived categories which are used frequently, and in appendix~\ref{app:ell-int} we describe some computations on elliptic curves that are used in examples in the main text.

Throughout this paper, we will utilize ideas regarding equivalences between `folded' B models and collapsing B model defects onto one another that were previously described in \cite{Ando:2010nm}.

A word on notation.  Throughout this paper, we implicitly assume all operations are derived,
and omit the usual ${\bf R}$, ${\bf L}$ symbols.  For example, in this paper,
\begin{eqnarray}
    {\rm Hom} & \mbox{ means } & {\bf R}{\rm Hom},
    \\
    \otimes & \mbox{ means } & \stackrel{ {\bf L }}{\otimes},
    \\
    p_* & \mbox{ means } & {\bf R}p_*, \\
    p^* & \mbox{ means } & {\bf L}p^*,
\end{eqnarray}
and so forth.  In the same spirit, we shall use ``D-brane'' to refer to any object in a derived category of coherent sheaves, so our ``D-branes'' are really collections of branes, antibranes, and tachyons,
following \cite{Sharpe:1999qz}.

For an introduction to derived categories in mathematics, see for example
\cite{weibel,andreirev}.

We assume throughout this paper that all target spaces are smooth,
either smooth complex manifolds or smooth Deligne-Mumford stacks.

\section{Topological defects in the B model}  \label{sect:bmodel}

In this section, we will describe the structure of topological defects in the B model topological field theory, ultimately as an exercise in the mathematics of derived categories.
We will work through the details explicitly, as we are not aware of literature describing this.  That said, experts on derived categories will probably not be surprised by the constructions described in this section.

\subsection{Review of defects and fusion in the B model}   \label{sect:rev}

Defects have been discussed in a variety of papers, as listed in the introduction.
Briefly, given a nonlinear sigma model with target $X$,
a defect can be understood as an analogue of an open string in the middle of the closed string worldsheet, describing a D-brane on $X \times X$.
More generally, defects can link worldsheets describing different target spaces, in which case
they are sometimes known as interfaces.
If one side is a nonlinear sigma model with target $X$, and the other a nonlinear sigma model with target $Y$, then the defect/interface is a D-brane on $X \times Y$. 

Such defects are constructed using the `folding trick' \cite{Wong:1994np,Bachas:2001vj}.
The idea of the folding trick is to represent an interface between, say, a nonlinear sigma model on $X$ and a nonlinear sigma model on $Y$ by `folding' the worldsheet along the defect, so that the defect is defined on $X \times Y^{\vee}$,
where $Y^{\vee}$ is a space with the peculiar property that a sigma model on $Y^{\vee}$ is the same as a sigma model on $Y$ but with left- and right-movers exchanged. 
In this section we will briefly review how these defects and their fusions are described in the case of the B model.

In the B model, it is most convenient not to work just with D-branes, but rather with
collections of D-branes, antibranes, and tachyons, which can be identified with elements of the derived category of coherent sheaves $D^b(X)$ for the B model with target $X$ \cite{Sharpe:1999qz,Sharpe:2003dr}.
Condensation of tachyons, for example, is interpreted in terms of the cone construction.

In this language, defects/interfaces can be understood in terms of integral
and Fourier-Mukai transforms $\Phi: D^b(X) \rightarrow D^b(Y)$, where $D^b(X)$ denotes the bounded derived category of coherent sheaves on $X$.  Explicitly, given an object $K \in  D^b(X \times Y)$ (known technically as the kernel),
the corresponding integral transform $\Phi(K): D^b(X) \rightarrow D^b(Y)$ is given by
\begin{equation}   \label{eq:defn:int-trans}
    \Phi(K)({\cal E}) \: = \: p_{Y *} \left( K \otimes p_X^* {\cal E} \right), 
\end{equation}
for any ${\cal E} \in D^b(X)$, where $p_{X,Y}$ are projections $p_X: X \times Y \rightarrow X$,
$p_Y: X \times Y \rightarrow Y$, and upper and lower $*$ denote pullback and pushforward respectively.  Throughout this paper, we assume all pushforwards, pullbacks, tensor products, and so forth are derived.  For most of this paper we will also specialize to the case of defects in a single theory, not interfaces between multiple theories, hence we will
usually assume
$Y = X$.

A common example we will frequently encounter has the following form.
Suppose $X = Y$, and $K = \Delta_* V$, for $\Delta: X \rightarrow X \times X$ the diagonal embedding,
and $V \rightarrow X$ any vector bundle on $X$.  Then,
\begin{equation}
    \Phi(K)({\cal E}) \: = \: {\cal E} \otimes V.
\end{equation}

\subsubsection{Integral transforms as defects}

These integral transforms have the following properties which allow us to identify them with defects in the B model:
\begin{enumerate}
    \item Existence of an identity.  Let $\Delta \subset X \times X$ be the diagonal, consisting
    of pairs $(x,x) \in X \times X$ for all $x \in X$.    Then, \cite[example 5.4, section 5.1]{huy}
    \begin{equation}
        \Phi({\cal O}_{\Delta}) \: = \: {\rm Id}_{D^b(X)},
    \end{equation}
    for which reason, we identify the identity defect with ${\cal O}_{\Delta}$.
    (See also \cite{Ando:2010nm} for the same conclusion in a different context related to the B model, and also e.g.~\cite{Bachas:2001vj,Quella:2002ct} for earlier discussions of the diagonal as the identity defect.)
    \item Addition.  Given two sheaves $K_1, K_2 \in D^b(X \times X)$,
    \begin{equation}
        \Phi(K_1 + K_2) \: = \: \Phi(K_1) + \Phi(K_2),
    \end{equation}
    where the sum ($+$) indicates direct sum, see e.g.~\cite[section 1.2]{weibel}. 
    (At the level of complexes, this is a simple degreewise sum, i.e.~$(K_1 + K_2)_n = (K_1)_n \oplus (K_2)_n$.)
    In the notation of fusion products, we will simply write this as $K_1 + K_2$.

    \item Lack of subtraction.  Defects only form a semigroup under $+$, not a group, so no subtraction is expected, and one is not present here without going to K theory.
    Let us walk through this in more detail.
    Let $K \in D^b(X \times X)$.  In K theory, $K + K[1]$ is equivalent to zero, but this is not true of integral transforms:
    \begin{equation}
        \Phi(K + K[1]) \: = \: \Phi(K) + \Phi(K[1]) \: \neq \: 0.
    \end{equation}
    One could implement a subtraction $K - K$ by using the cone construction.
    If $\tilde{K} = {\rm Cone}({\rm Id}: K \rightarrow K)$, then $\Phi(\tilde{K}) = 0$.
    However, the result of the cone construction depends upon the map, and in general there is no choice of map to make for a general difference $K_1 - K_2$.
    \item Fusion products = composition.  Fusion products (more properly, their analogues) can be naturally understood as composition of integral transforms, as is well known in certain quarters, see e.g.~\cite[section 4.3.3]{Bachas:2001vj}, and we review here.  Physically, when two defects collide, one would integrate out degrees of freedom from the `middle.'  Mathematically, given $K_{XY} \in D^b(X \times Y)$ and 
    $K_{YZ} \in D^b(Y \times Z)$, the fusion product (denoted here $*$, to be consistent with mathematics notation for integral transforms) should be
    \begin{equation}
        K_{XY} * K_{YZ} \: = \: q_* \left( p_{XY}^* K_{XY} \otimes p_{YZ}^* K_{YZ} \right)
        \: \in \: D^b(X \times Z).
    \end{equation}
    The pushforward $q_*$ is the mathematical implementation of the `integrating out' above.
    However, this is also how composition of integral transforms is defined \cite[prop. 5.10]{huy}, \cite{mukai1}:
    \begin{equation}
        \Phi(K_{YZ}) \circ \Phi(K_{XY}) \: = \: \Phi(K_3) 
    \end{equation}
    where
    \begin{equation}
        K_3 \: = \: q_* \left( p_{XY}^* K_{XY} \otimes p_{YZ}^* K_{YZ} \right)
        \: \in \: D^b(X \times Z),
    \end{equation}
    and
    \begin{eqnarray}
        p_{XY}: X \times Y \times Z & \rightarrow & X \times Y,
        \\
        p_{YZ}:  X \times Y \times Z & \rightarrow & Y \times Z, 
        \\
        q:  X \times Y \times Z & \rightarrow & X \times Z
    \end{eqnarray}
    are projections.  
    Thus, the mathematical analogue of a fusion product of two defects, coincides with the composition of the corresponding integral transforms.  As a result, we will sometimes write the composition above in the notation of fusion products as
    \begin{equation}
        K_{XY} * K_{YZ} \: = \: K_3,
    \end{equation}
    so that
    \begin{equation}  \label{eq:comp-int-trans}
        \Phi(K_{YZ}) \circ \Phi(K_{XY}) \: = \: \Phi\left(K_{XY} * K_{YZ} \right).
    \end{equation}

    In the special case $X = Y = Z$, $K_{XY} = \Delta_* V_1$, and $K_{YZ} = \Delta_* V_2$, for $V_1$ and $V_2$ vector bundles on $X$, then
    \begin{equation}
        K_3 \: = \: \Delta_* \left( V_1 \otimes V_2 \right),
    \end{equation}
    or more simply,
    \begin{equation}  \label{eq:fusion-diagonals}
        (\Delta_* V_1) * (\Delta_* V_2) = \Delta_* (V_1 \otimes V_2).
    \end{equation}

    \item Non-commutivity.  Given $K_1, K_2 \in D^b(X \times X)$, in general,
    \begin{equation}
        \Phi(K_2) \circ \Phi(K_1) \: \neq \: \Phi(K_1) \circ \Phi(K_2),
    \end{equation}
    or more simply,
    \begin{equation}
        K_1 * K_2 \: \neq \: K_2 * K_1.
    \end{equation}
    For example, suppose $K_1 = \pi_1^* L$ for $L$ a nontrivial line bundle on $X$, and $K_2 = {\cal O}_{X \times X}$.  Then,
    \begin{eqnarray}
        \Phi(K_1)({\cal E}) & = & H^{\bullet}(X, {\cal E} \otimes L) \otimes {\cal O},
        \\
        \Phi(K_2)({\cal E}) & = & H^{\bullet}(X, {\cal E}) \otimes {\cal O}.
    \end{eqnarray}
    As a result,
    \begin{eqnarray}
        \Phi(K_1) \circ \Phi(K_2)({\cal E}) & = & H^{\bullet}(X, {\cal E}) \otimes
        H^{\bullet}(X, L),
        \\
        \Phi(K_2) \circ \Phi(K_1)({\cal E}) & = & H^{\bullet}(X, {\cal E} \otimes L) \otimes
        H^{\bullet}(X, {\cal O}),
    \end{eqnarray}
    and
    \begin{eqnarray}
        K_1 * K_2 & = & H^{\bullet}(X, L) \otimes K_2,
        \\
        K_2 * K_1 & = & H^{\bullet}(X, {\cal O}) \otimes K_1.
    \end{eqnarray}
    In particular, 
    $K_1 * K_2 \neq K_2 * K_1$ and
    \begin{eqnarray}
        \Phi(K_1) \circ \Phi(K_2) & = & \Phi(K_2 * K_1),
        \\
        & \neq & \Phi(K_1 * K_2) \: = \: \Phi(K_2) \circ \Phi(K_1).
    \end{eqnarray}

    Another example may be helpful.  Let $S_{(x,y)}$ denote a skyscraper sheaf\footnote{Skyscraper sheaf is a sheaf supported at a single point.} on $X \times Y$ supported at the point $(x,y)$.
    Let $K_1, K_2 \in D^b(X \times X)$ be given by $K_1 = S_{(x_1,x_2)}$, $K_2 = S_{(x_1',x_2')}$.
    
    Then,
    \begin{itemize}
        \item $\pi_{12} K_1$ is supported at $(x_1,x_2,y)$ for all $y \in X$,
        \item $\pi_{23}^* K_2$ is supported at $(z, x_1', x_2')$ for all $z \in X$,
        \item $\pi_{12}^* K_1 \otimes \pi_{23}^* K_2$ vanishes or $x_2 \neq x_1'$,
        and for $x_2 = x_1'$, is a skyscraper sheaf supported at $(x_1, x_2 = x_1', x_2') \in X \times X \times X$.
    \end{itemize}
    From this we compute that $\pi_{13 *} \left( \pi_{12}^* K_1 \otimes \pi_{23}^* K_2 \right)$ vanishes for $x_2 \neq x_1'$,
    and for $x_2 = x_1'$, 
    \begin{equation}
        \pi_{13 *} \left( \pi_{12}^* K_1 \otimes \pi_{23}^* K_2 \right) \: = \: S_{(x_1, x_2')}.
    \end{equation}

    In the other direction, 
    \begin{itemize}
        \item $\pi_{12}^* K_2$ is supported at $(x_1',x_2',y)$ for all $y \in X$,
        \item $\pi_{23}^* K_1$ is supported at $(z, x_1, x_2)$ for all $z \in X$,
        \item $\pi_{12}^* K_2 \otimes \pi_{23}^* K_1$ vanishes for $x_1 \neq x_2'$, and for $x_1 = x_2'$,
        is a skyscraper sheaf supported at $(x_1', x_1 = x_2', x_2) \in X \times X \times X$.
    \end{itemize}
    From this we compute that $\pi_{13 *}\left( \pi_{12}^* K_2 \otimes \pi_{23}^* K_1 \right)$ vanishes or $x_1 \neq x_2'$,
    and for $x_1 = x_2'$,
    \begin{equation}
        \pi_{13 *}\left( \pi_{12}^* K_2 \otimes \pi_{23}^* K_1 \right) \: = \: S_{(x_1',x_2)}.
    \end{equation}
    In particular,
    \begin{equation}
        K_1 * K_2 \: \neq \: K_2 * K_1.
    \end{equation}
    They are not even quasi-isomorphic; they are simply non-isomorphic elements of $D^b(X \times X)$.

\item Associators.  Given any three kernels $K_1$, $K_2$, $K_3$, there is a canonical isomorphism 
\begin{equation}
    \phi_{1,2,3}: \: (K_1 * K_2) * K_3 \: \stackrel{\cong}{\longrightarrow} \: K_1 * (K_2 * K_3),
\end{equation}
satisfying the pentagon identity
\begin{equation}
    \begin{tikzcd}
         & (K_1 * K_2) * (K_3 * K_4) \ar[dr, "\phi_{1,2,34}"] & 
        \\
        ( (K_1 * K_2) * K_3) * K_4 \ar[ur, "\phi_{12,3,4}"] \ar[d, "\phi_{1,2,3} \otimes {\rm id}_4"]
         & & K_1 * ( K_2 * (K_3 * K_4) ) 
        \\
        (K_1 * (K_2 * K_3) ) * K_4  \ar[rr, "\phi_{1,23,4}"] & &
        K_1 * ( (K_2 * K_3) * K_4) \ar[u, "{\rm id}_1 \otimes \phi_{2,3,4}"]
    \end{tikzcd}
\end{equation}
just as one would expect in a fusion category, which are also referred to as F-symbols in the fusion category and physics literature.

\item Fusion products of D-branes.  So far we have outlined properties of defects and interfaces in string worldsheets.  D-branes themselves can be understood as special cases of interfaces, mapping $D^b(X) \rightarrow
D^b(Y = {\rm point})$ or $D^b({\rm point}) \rightarrow D^b(X)$.  Interpreted as an interface, the D-brane\footnote{
More accurately, brane/antibrane/tachyon collection.
} is an element of 
$D^b(X \times {\rm point}) = D^b(X)$.  The fusion product of two D-branes can then be understood
in the same fashion as composition of interfaces.  If $K_1 \in D^b(X \times {\rm point})$ and
$K_2 \in D^b({\rm point} \times X)$, then
\begin{equation}
    K_1 * K_2 \: = \: q_* \left( \pi_1^* K_1 \otimes \pi_2^* K_2 \right) \: = \:
     \pi_1^* K_1 \otimes \pi_2^* K_2,
\end{equation}
using the fact that $q: X \times {\rm point} \times X \rightarrow X \times X$ is merely the identity map.  

In passing, the reader should note that a fusion product of D-branes is different from a collision of D-branes: The former means fusing the worldsheet defects while the latter means moving D-branes on top of each other in spacetime. For example, given a pair of individual D-branes, the collision will involve an enhancement of a $U(1)^2$ gauge symmetry on the worldvolume to $U(2)$, but the fusion product will not.

\end{enumerate}

Beyond the identity ${\cal O}_{\Delta}$, simple examples are pushforward and pullbacks,
which we explain next.
Let $f: X \rightarrow Y$ be any function,
    and $\Gamma_f \subset X \times Y$ be its graph, meaning the set of pairs
    $\{ (x, f(x)) \in X \times Y \}$.  Let ${\cal O}_f$ denote the structure sheaf of $\Gamma_f$,
    interpreted as an object in $D^b(X \times Y)$.  Then, $f_*: D^b(X) \rightarrow D^b(Y)$ and
    $f^*: D^b(Y) \rightarrow D^b(X)$ are both implemented as integral transforms with kernel ${\cal O}_f$, see \cite[examples 5.4, section 5.1]{huy}.

As a special case, consider ordinary symmetries defined by elements of a group.
 Let $G$ be a group that acts on $X$,
    and let $K_g$ denote the structure sheaf of the graph of the function $g: X \rightarrow X$, then $K_g$ is the defect corresponding to $g$, that implements the action of $g$ on $X$.  The resulting isomorphism of derived categories respects the group law, in the following sense.  If we let $\phi_g: D(X) \rightarrow D(X)$ denote the integral transform associated as above to the graph of $g$, then,  there is an isomorphism
    \begin{equation}
        \psi_{g,h}: \phi_h \circ \phi_g \: \Rightarrow \: \phi_{hg}
    \end{equation}
    which makes the obvious pentagon identity commute.

\subsubsection{Defects as generalized symmetries in B models}
 Now, should all defects be interpreted as symmetries?  For example, consider a function $f: X \rightarrow Y$ that maps all of $X$ to a single point $y \in Y$.  The corresponding integral transform will annihilate the cohomology of $X$.  That is certainly `noninvertible,' but is hardly a `symmetry.'

    More generally, one could ask which kernels in $D^b(X \times X)$ induce isomorphisms $D^b(X) \rightarrow D^b(X)$.  For $X$ Fano or of general type, the only such automorphisms are tensor products with line bundles, and shifts $[n]$, see \cite[theorem 4.6]{bondalorlovsemi}.  
    For $X$ Calabi-Yau, on the other hand, there are typically more.  For example,
    on an elliptic curve, there is an integral transform (with kernel\footnote{
    The Poincar\'e line bundle is more generally interpreted as mapping an abelian variety to its dual, which
    in general need not be itself.  
    } the Poincar\'e line bundle) which maps ${\cal O}_P$
    (the skyscraper sheaf supported at a point $P$) to the line bundle ${\cal O}(P-P_0)$, for
    a fixed point $P_0$, and defines an isomorphism $D^b(X) \rightarrow D^b(X)$, but it is not of the forms described above.

For a different example, a noninvertible line in two dimensions acts on the vacuum by multiplying it by its quantum dimension:
\begin{equation}
    L|\Omega\rangle =\langle L \rangle |\Omega \rangle \neq |\Omega \rangle.
\end{equation}
As a result, the vacuum is not preserved, which gives another sense in which describing these transformations as `symmetries' is, at least, counterintuitive. This is, in fact, one of the main disagreements about whether noninvertible topological defects should be referred to as `symmetries'. Unlike an ordinary group symmetry, a noninvertible operator does not act unitarily on a given Hilbert space $\mathcal{H}$. It should be understood as defining a quantum operation with two steps: 1) unitary action on an enlarged Hilbert space $\mathcal{H}\otimes \mathcal{H}'$ and 2) partial-trace over $\mathcal{H}'$. We refer the reader to \cite{Okada:2024qmk} for more details on this point.

With this subtlety in mind, the perspective we will take is that defects should be interpreted as defining actions, not necessarily analogs of groups themselves.  In particular, groups can have trivial actions without being trivial themselves.  One example is the trivial representation of any group -- the group is not represented faithfully, as all elements act in the same way.  We interpret the defects as defining analogues of group actions, so that the example above of a constant function is an analogue of a nonfaithful representation of a group.  As a result, referring to defects as generating noninvertible `symmetries' is a misnomer from this perspective, but the language is now standard.

Phrased another way, the issue here is the language, not the integral transforms.
The line operators in the B model are integral transforms, defined by kernels.  However, labeling them ``symmetries'' can be misleading, but it is now a widely accepted notion in physics literature.

We close this subsection with the following comments. In passing, we should mention that in principle, mirrors should be computable via homological mirror symmetry \cite{Kontsevich:1995joz}.  If two Calabi-Yau's $X$ and $Y$ are mirror to one another, then homological mirror symmetry maps $D(X)$ to a corresponding Fukaya category of $Y$.  Trivially, given an interface between $X$ and $Y$,
homological mirror symmetry should map $D(X \times Y)$ to the Fukaya category of $X \times Y$,
and in particular should map kernels of integral transforms to the Lagrangian correspondences mentioned
above.

\subsection{Orientation reversal as adjoints}
\label{sect:adj}

We shall sometimes need to reverse the orientation of a given defect $K \in D^b(X \times X)$ for $X$ Calabi-Yau.  The desired operation $\dag$ should have properties including
\begin{equation}
    (K^{\dag})^{\dag} \: \cong \: K,
\end{equation}
for $K \in D^b(X \times X)$,
\begin{equation}
    (K_1 * K_2)^{\dag} \: \cong \: K_2^{\dag} * K_1^{\dag},
\end{equation}
for $K_1, K_2 \in D^b(X \times Y)$,  and
\begin{equation}  \label{eq:dagprop3}
    {\rm Hom}_{X \times X}(A * B, C) \: \cong \: {\rm Hom}_{X \times X}(A, C * B^{\dag})
    \: \cong \: {\rm Hom}_{X \times X}(B , A^{\dag} * C),
\end{equation}
\begin{equation}   \label{eq:dagprop4}
    {\rm Hom}_{X \times X}(A, B * C) \: \cong \: {\rm Hom}_{X \times X}(B^{\dag} * A, C)
    \: \cong \: {\rm Hom}_{X \times X}(A * C^{\dag}, B),
\end{equation}
for $A, B, C \in D^b(X \times X)$.  In this section we will define such an operation $\dag$, utilizing adjoints arising in derived categories,
and show that it satisfies the first property above.  In the next section we will
establish the remaining properties.

First, we recall the left- and right-adjoints to integral transforms. Briefly, following \cite[prop.~5.9]{huy}, \cite{mukai1}, \cite[lemma 1.2]{bondalorlovsemi},
given $K \in D^b(X \times Y)$ defining an integral transform $\Phi(K): D^b(X) \rightarrow D^b(Y)$,
the left and right adjoints, respectively, of the integral transform are the integral transforms of 
\begin{equation}
    K_L \: = \: K^{\vee} \otimes \pi_Y^* \omega_Y[\dim Y], \: \: \:
    K_R \: = \: K^{\vee} \otimes \pi_X^* \omega_X[ \dim X],
\end{equation}
where $\omega_W$ is the canonical bundle of $W$. (Implicitly, the right adjoint functor can be derived from the left adjoint functor by conjugating by Serre functors \cite[theorem 4.6]{andreirev}.) 
More explicitly,
\begin{eqnarray}
    {\rm Hom}_{D^b(X)} \left( \Psi(K_L)({\cal E}), {\cal F} \right) & \cong &
    {\rm Hom}_{D^b(Y)} \left( {\cal E}, \Phi(K)({\cal F}) \right),
    \\
    {\rm Hom}_{D^b(Y)}\left( \Phi(K)({\cal F}), {\cal E} \right) & \cong &
    {\rm Hom}_{D^b(X)}\left( {\cal F}, \Psi(K_R)({\cal E}) \right),
\end{eqnarray}
where for $A \in D^b(X \times Y)$, $\Psi(A)(-) = p_{X *} \left( A \otimes p_Y^* -\right)$ is a functor $D^b(Y) \rightarrow D^b(X)$.

Now, to match physics, we want to work in conventions where the integral transforms are defined with respect to a single direction, so that we do not have to label each adjoint according to whether it should be interpreted as defining a map $D^b(X) \rightarrow D^b(Y)$ or $D^b(Y) \rightarrow D^b(X)$.  Put another way, we want to work in conventions in which we have only $\Phi$ or $\Psi$ in the expressions above, but not both.

A convention of this form is utilized in \cite{Kuznetsov:2009}, which defines adjoints using only $\Phi$, and not $\Psi$.
Following 
\cite[Lemma 3.3]{Kuznetsov:2009}
given $K \in D^b(X \times Y)$ defining an integral transform $\Phi(K): D^b(X) \rightarrow D^b(Y)$, let
\[
\sigma \colon Y \times X \rightarrow X \times Y
\]
denote the transposition morphism. 
Then, for $K \in D^b(X \times Y)$, one can define
$K_L, K_R \in D^b(Y \times X)$ by
\begin{equation}
    K_L \: = \: (\sigma^*K)^{\vee} \otimes \pi_Y^* \omega_Y[\dim Y], \: \: \:
    K_R \: = \: (\sigma^*K)^{\vee} \otimes \pi_X^* \omega_X[ \dim X].
\end{equation}
These have the properties that
\begin{eqnarray}
    {\rm Hom}_{D^b(X)} \left( \Phi(K_L)({\cal E}), {\cal F} \right) & \cong &
    {\rm Hom}_{D^b(Y)} \left( {\cal E}, \Phi(K)({\cal F}) \right),
    \\
    {\rm Hom}_{D^b(Y)}\left( \Phi(K)({\cal F}), {\cal E} \right) & \cong &
    {\rm Hom}_{D^b(X)}\left( {\cal F}, \Phi(K_R)({\cal E}) \right),
\end{eqnarray}
(using the fact that the right-adjoint can be obtained from the left-adjoint by
conjugating by the Serre functor, as in \cite[theorem 4.6]{andreirev}).
In particular, the adjoint functors to $\Phi(K)$ are the integral transforms associated to
$K_L$, $K_R$  \cite[Lemma 3.3]{Kuznetsov:2009}:
\begin{equation}
    \Phi(K)_L \: = \: \Phi(K_L), \: \: \:
    \Phi(K)_R \: = \: \Phi(K_R).
\end{equation}

If $X$ and $Y$ are both Calabi-Yau, then 
\begin{equation}
    K_L \: = \: (\sigma^* K)^{\vee} [\dim Y], \: \: \:
    K_R \: = \: (\sigma^* K)^{\vee}[ \dim X].
\end{equation}
If $X$ and $Y$ are both Calabi-Yau and $X = Y$, then we define\footnote{$K_L$ and $K_R$ are canonically isomorphic, therefore we identify them, in the case $X = Y$ are both Calabi-Yau.}
\begin{equation}
    K^{\dag} \: = \: K_L \: = \: K_R \: = \: (\sigma^* K)^{\vee} [ \dim X ].
\end{equation}
This defines the adjoint or orientation-reversal operation we will use most commonly in this paper in working with defects in the B model.

Now, let us compare the derived dual and the adjoint as possible descriptions of orientation reversal.  One property we require is that the operation square to the identity, and this is true of both.  For derived
duals, it is a standard result that
\begin{equation}
    (A^{\vee})^{\vee} \: = \: {\underline {\rm Hom}}( A^{\vee}, {\cal O}) \: \cong \:
    A.
\end{equation}
For the adjoint operation for a defect in $D^b(X \times X)$ for $X$ Calabi-Yau,
\begin{eqnarray}
    (A^{\dag})^{\dag} & = &
    {\underline {\rm Hom}}(\sigma^* A^{\dag}, {\cal O}) [\dim X],
    \\
    & = & 
    {\underline {\rm Hom}}(\sigma^* (\sigma^* A)^{\vee}[\dim X], {\cal O}) [\dim X],
    \\
    & = &
    {\underline {\rm Hom}}(A^{\vee}, {\cal O}) [-\dim X][\dim X],
    \\
    & = & (A^{\vee})^{\vee} = A.
\end{eqnarray}
(Note that since $\sigma$ is an isomorphism, $\sigma^*$ does not have any higher derived
functors, and so $(\sigma^* A)^{\vee} = \sigma^* (A^{\vee})$.)

Now, we also want to require that the identity defect be invariant
under this operation, and this determines the adjoint operation $\dag$
as the operation of choice.
So, consider the identity defect ${\cal O}_{\Delta} \in D^b(X \times X)$.
We require that reversing the orientation of the identity defect should return the original defect.
From \cite[example 14]{Ando:2010nm}, \cite[example 5.19]{huy}, for any space $X$,
\begin{eqnarray}
    \left( {\cal O}_{\Delta} \right)^{\vee} \: = \: \left( \Delta_* {\cal O}_{X} \right)^{\vee} 
    & \cong & \Delta_* \omega_X^{-1}[ - \dim X],
    \\
    & \cong & \Delta_* {\cal O}_X \otimes \pi_1^* \omega_X^{-1}[- \dim X]
    \: \cong \: \Delta_* {\cal O}_X \otimes \pi_2^* \omega_X^{-1}[- \dim X],
\end{eqnarray}
where $\omega_X$ is the canonical bundle of $X$. 
Now, for any vector bundle $V \rightarrow X$, 
\begin{equation}  \label{eq:diagonalvspi1}
    \Delta_* V \: = \: \Delta_* \left( \Delta^* \pi_1^* V \right) \: = \:
    \Delta_* {\cal O}_X \otimes \pi_1^* V,
\end{equation}
using the fact that $\pi_1 \circ \Delta = 1$ and the projection formula,
hence
\begin{equation}   \label{eq:diagonal:derdual}
    \left( {\cal O}_{\Delta} \right)^{\vee} \: = \: \Delta_* {\cal O}_X \otimes \pi_1^*  \omega_X^{-1}[ - \dim X ].
\end{equation}
In particular, even for $X$ a Calabi-Yau, we have
\begin{equation}  
    ({\cal O}_{\Delta})^{\vee} \: = \: {\cal O}_{\Delta}[-\dim X] \: \neq \:
    {\cal O}_{\Delta},
\end{equation}
and so the identity defect is not invariant under the derived dual.
If we consider the adjoint instead, we find
\begin{equation}
    \left( {\cal O}_{\Delta} \right)^{\dag} \: = \: \left( \sigma^* {\cal O}_{\Delta} \right)^{\vee} \otimes\pi_1^* \omega_X [+ \dim X]
    \: = \: \left(  {\cal O}_{\Delta} \right)^{\vee} \otimes\pi_1^* \omega_X [+ \dim X]
    \: = \: {\cal O}_{\Delta},
\end{equation}
and so under the adjoint operation, the identity defect is invariant (regardless of whether $X$ is Calabi-Yau).
This is one reason that we identify orientation-reversal with the adjoint $\dag$,
and not with the derived dual.

For later use, let us mention some useful computational results.

Suppose $Y$ is a smooth (or local complete intersection) subvariety of $X$, then, by applying Koszul resolutions locally, one can argue that
\begin{equation}
    {\cal O}_Y^{\vee} \: = \: \wedge^{\rm top} N_{Y/X} [ - {\rm codim}\, Y/X]
\end{equation}
where $N_{Y/X}$ is the normal bundle to $Y$ in $X$, and codim $Y/X$ is the codimension of $Y$ in $X$.  A formal argument, which establishes the result above in greater generality, is as follows.

Let $i: Y \rightarrow X$ be a closed embedding. As it is a proper morphism (i.e., fibers are compact) it admits a dualizing complex $\omega$ in $D^b(Y)$.  (In the special case that $Y$ is a local complete intersection then $\omega = \wedge^{\rm top}(N_{Y/X})[-c]$ for $c = {\rm codim}\, Y/X$.) 
From Grothendieck duality, there is a functor $i^! : D^b(X) \rightarrow D^b(Y)$, which is related to the usual pull-back $i^*$, and has the properties:
\begin{itemize}
\item $i^!(F) = i^*(F) \otimes \omega$,
\item $i^!$ is right adjoint to $i_*$, meaning
\begin{equation}
    {\rm Hom}_{D^b(Y)} (F, i^! G)  =  {\rm Hom}_{D^b(X)} (i_*F, G),
    \: \: \:
    i_* {\underline{\rm Hom}}_{D^b(Y)} (F, i^! G)  =  {\underline{\rm Hom}}_{D^b(X)} (i_* F, G).
\end{equation}
\end{itemize}
In the present case, take $F = {\cal O}_Y$, $G = {\cal O}_X$,
so ${\cal O}_Y^{\vee} = {\underline{\rm Hom}}_{D^b(X)}( i_* F, G)$ (the right-hand-side of the
above).  On the left-hand side, $i^! {\cal O}_X = i^* {\cal O}_X \otimes \omega = \omega$.
In the special case that $Y$ is a local complete intersection, the previous description applies.

\subsection{Junction and defect operators}

So far we have outlined formally how interfaces and defects naturally correspond to kernels of integral transforms, on products $X \times Y$.  Now, a junction between defects and interfaces can naturally carry operators, known as junction operators.  At any junction, the junction operators only depend upon the cyclic ordering of the intersecting interfaces / defects.

In the B model, such spaces of junction local operators are naturally associated with Ext groups, cohomologies of Hom's between kernels, as we now explain.

First, given kernels $K_1$, $K_2$ for two integral transforms $\Phi(K_{1,2}): D^b(X) \rightarrow D^b(Y)$, natural transformations 
    $\Phi(K_1) \Rightarrow \Phi(K_2)$ correspond\footnote{
A technical aside.  In order for natural transformations to be in one-to-one correspondence with Hom's, we must be working in dg categories, and not merely derived categories \cite[theorem 8.15]{toen1}, a distinction we largely suppress in this paper.   
    } to morphisms Hom$(K_1,K_2)$, see for example \cite[section 2.2]{andreimukai2}, \cite[section 3.1]{bridgeland99}.
For example, given three defects $K_1$, $K_2$, $K_3$, the junction operators at the triple intersection correspond to elements of Hom$(K_1 * K_2, K_3)$ (see e.g.~\cite{Chang:2018iay,Perez-Lona:2023djo,Perez-Lona:2024sds}), which immediately corresponds to natural transformations
    $\Phi(K_1) \circ \Phi(K_2) \Rightarrow \Phi(K_3)$.

There are several points to clarify.  First, we give a more explicit description of the junction operators associated to any intersection, and clarify how it only depends upon the cyclic ordering.
Later, we will discuss the grading carried by Hom spaces in this context.

\subsubsection{Basic identities}

Consider a junction of several oriented defects, cyclically ordered as $K_1, \cdots K_m$ incoming defects followed by $L_1, \cdots, L_m$ outgoing defects, as illustrated below:
\begin{center}
    \begin{tikzpicture}
        \filldraw[color=black] (1,1) circle [radius=0.08];
        \draw[thick,->] (1,1) -- (0.5,0.5);  \draw[thick] (0.5,0.5) -- (0,0); 
        \draw (0.1,0.5) node {$L_m$};
        \draw[thick,->] (0,2) -- (0.5,1.5);  \draw[thick] (0.5,1.5) -- (1,1);
        \draw (0.1,1.5) node {$K_1$};
        \draw[thick,->] (1,1) -- (1,0.5);  \draw[thick] (1,0.5) -- (1,0);
        \draw (1.3,0.2) node {$L_2$};
        \draw[thick,->] (1,2) -- (1,1.5);  \draw[thick] (1,1.5) -- (1,1);
        \draw (0.7,2) node {$K_2$};
        \draw[thick,->] (1,1) -- (1.5,0.5);  \draw[thick] (1.5,0.5) -- (2,0);
        \draw (2,1.6) node {$K_n$};
        \draw[thick,->] (2,2) -- (1.5,1.5);  \draw[thick] (1.5,1.5) -- (1,1);
        \draw (2,0.4) node {$L_1$};
        \draw (2,1.1) node {$\vdots$};
        \draw (0.6,0.2) node {$\cdots$};
    \end{tikzpicture}
\end{center}
At such a junction, the junction operators are counted by
\begin{equation}
    {\rm Hom}(K_1 * \cdots * K_n, L_1 * \cdots * L_m)
\end{equation}
where the $*$ denotes the composition of kernels (fusion products of defects) discussed previously in subsection~\ref{sect:rev}. 
We can flip the orientation of any given defect $K$ by replacing it with its adjoint $K^{\dag}$.

Later in section~\ref{sect:junction-ops} we will argue that the Hom's above are cyclically symmetric (so that the choice of starting point is irrelevant), and that one can move any given defect from one side to the other if one simultaneously takes an adjoint ($\dag$).
Before doing that, however, we must establish
 some preliminaries, some basic identities
satisfied by the Hom spaces, that we will use in section~\ref{sect:junction-ops} to establish the cyclic property and so forth.
The purpose of this section is to establish those preliminary results. 

The following statements are true:  
\begin{itemize}
    \item For $A \in D^b(X \times Y)$, $B \in D^b(Y \times Z)$, $C \in D^b(X \times Z)$,
    \begin{equation}  \label{eq:cycle1}
        {\rm Hom}_{X \times Z}\left( A * B, C \right) \: \cong \: 
        {\rm Hom}_{X \times Y}\left( A, C * B_R \right).
    \end{equation}
    \item For $A \in D^b(X \times Y)$, $B \in D^b(Y \times Z)$, $C \in D^b(X \times Z)$,
    \begin{equation}  \label{eq:cycle2}
        {\rm Hom}_{X \times Z}\left( A * B, C \right) \: \cong \:
        {\rm Hom}_{Y \times Z}\left( B, A_L * C \right).
    \end{equation}
    \item For $A \in D^b(X \times Z)$, $B \in D^b(X \times Y)$, $C \in D^b(Y \times Z)$,
    \begin{equation}  \label{eq:cycle3}
        {\rm Hom}_{X \times Z}\left( A, B * C \right) \: \cong \:
        {\rm Hom}_{Y \times Z}\left( B_R * A, C \right).
    \end{equation}

    \item For $A \in D^b(X \times Z)$, $B \in D^b(X \times Y)$, $C \in D^b(Y \times Z)$,
    \begin{equation}   \label{eq:cycle4}
        {\rm Hom}_{X \times Z}\left( A, B * C \right) \: \cong \:
        {\rm Hom}_{X \times Y}\left(A * C_L, B \right).
    \end{equation}

    \item For $A \in D^b(X \times Y)$, $B \in D^b(Y \times Z)$,
    \begin{equation} \label{eq:cycle5}
        \left( A * B \right)_{L} \: \cong \: B_{L} * A_{L},
        \: \: \:
        \left( A * B \right)_{R} \: \cong \: B_{R} * A_{R}.
    \end{equation}

    \item In the special case that $X$ is a Calabi-Yau, for $A, B \in D^b(X \times X)$,
    \begin{eqnarray}  \label{eq:symtrace1}
        {\rm Hom}_{X \times X} \left( {\cal O}_{\Delta}, A * B \right) & = &
        {\rm Hom}_{X \times X} \left( {\cal O}_{\Delta}, B * A \right),
        \\
        {\rm Hom}_{X \times X}\left( A * B, {\cal O}_{\Delta} \right) & = &
        {\rm Hom}_{X \times X}\left( B * A, {\cal O}_{\Delta} \right).
        \label{eq:symtrace2}
    \end{eqnarray}
    This is slightly counterintuitive because, as discussed earlier, $A * B \neq B * A$ in general.
    However, the reader might feel more comfortable comparing $A$ and $B$ to square matrices, and the Hom to a trace:
    although matrix products do not commute, nevertheless ${\rm Tr}\, A B = {\rm Tr}\, B A$.
\end{itemize}

In passing, analogous statements can be found in \cite[prop.~2.4]{Horja:2001cp},
and analogous statements involving derived tensor products instead of
fusion products are given in e.g.~\cite[section 3.3]{huy}.

In the special case that $X = Y = Z$ is a Calabi-Yau, the identities above
trivially reduce to the remaining dagger identities~(\ref{eq:dagprop3}),
(\ref{eq:dagprop4}).

In the remainder of this subsection, we will derive these identities.
In the next subsection, subsection~\ref{sect:junction-ops},
we will apply these to show our to build a space of junction operators that only depends upon the cyclic ordering of the intersecting defects, meaning for example
that we will 
establish the cyclic property
\begin{equation}
    {\rm Hom}_{X \times X}\left( K_1 * \cdots * K_n, {\cal O}_{\Delta} \right)
    \: \cong \:
    {\rm Hom}_{X \times X}\left( K_2 * \cdots * K_n * K_1, {\cal O}_{\Delta} \right)
\end{equation}
and related identities.

First, we consider equations~(\ref{eq:cycle1}), (\ref{eq:cycle2}).
We begin by writing, for both cases,
\begin{eqnarray}
    {\rm Hom}_{X \times Z}( A * B, C ) & = &
    \Gamma_{X \times Z}\left( (A * B)^{\vee} \otimes C \right),
    \\
    & = &
    \Gamma_{X \times Z}\left( \left( \pi_{XZ *} \left( \pi_{XY}^* A \otimes\pi_{YZ}^* B \right) \right)^{\vee} \otimes C \right),
\end{eqnarray}
We compute
\begin{eqnarray}
    \left( \pi_{X Z *} \left( \pi_{XY}^* A \otimes \pi_{YZ}^* B \right)
    \right)^{\vee}
    & = &
    \pi_{XZ *}\left( \left( \pi_{XY}^* A \otimes \pi_{YZ}^* B \right)^{\vee} \otimes \pi_Y^* \omega_Y [ \dim Y] \right),
    \\
    & = &
    \pi_{XZ *}\left( \pi_{XY}^* A^{\vee} \otimes \pi_{YZ}^* B^{\vee} \otimes \pi_Y^* \omega_Y [\dim Y] \right).
\end{eqnarray}
hence
\begin{eqnarray}
    {\rm Hom}_{X \times Z}( A * B, C ) & = &
    \Gamma_{X \times Z}\left( \pi_{XZ *} \left( \pi_{XY}^* A^{\vee} \otimes\pi_{YZ}^* B^{\vee} \otimes \pi_Y^* \omega_Y[\dim Y] \right)  \otimes C \right),
    \\
    & = &
    \Gamma_{X \times Y \times Z} \left(
     \pi_{XY}^* A^{\vee} \otimes \pi_{YZ}^* B^{\vee} \otimes \pi_Y^* \omega_Y [\dim Y]  \otimes \pi_{XZ}^* C \right).
\end{eqnarray}
In the last step we used the fact that $\Gamma$ is the pushforward to a point, and pushing forward to $X \times Z$ and then to a point is the same
as pushing forward immediately to a point. Next, we convert this to an expression on $X \times Z \times Y$, for which we need to map $B$ from 
$D^b(Y \times Z)$ to $D^b(Z \times Y)$.  To that end, we utilize 
two closely-related projection maps which are
\begin{equation}
    \pi_{YZ}: \: X \times Z \times Y  \: \longrightarrow \: Y \times Z,
    \: \: \:
    \pi_{ZY}: \: X \times Z \times Y \: \longrightarrow \: Z \times Y,
\end{equation}
and the transposition map
\begin{equation}
    \sigma_{YZ}: \: Z \times Y \: \longrightarrow \: Y \times Z,
\end{equation}
which obey $\sigma_{YZ} \circ \pi_{ZY} = \pi_{YZ}$.
Continuing,
\begin{eqnarray}
    {\rm Hom}_{X \times Z}( A * B, C ) & = &
    \Gamma_{X \times Z \times Y} \left(
     \pi_{XY}^* A^{\vee} \otimes \pi_{ZY}^* (\sigma_{YZ}^* B)^{\vee} \otimes \pi_Y^* \omega_Y [\dim Y]  \otimes \pi_{XZ}^* C \right),
    \\
    & = &
    \Gamma_{X \times Y}\left( \pi_{XY *} \left(
    \pi_{XY}^* A^{\vee} \otimes \pi_{ZY}^* (\sigma_{YZ}^* B)^{\vee} \otimes \pi_Y^* \omega_Y[\dim Y] \otimes \pi_{XZ}^* C \right) \right),
    \\
    & = &
    \Gamma_{X \times Y}\left( A^{\vee} \otimes \pi_{XY *} \left(  \pi_{ZY}^* (\sigma_{YZ}^* B)^{\vee} \otimes \pi_Y^* \omega_Y[\dim Y] \otimes \pi_{XZ}^* C \right) \right),
    \\
    & = &
    \Gamma_{X \times Y}\left( A^{\vee} \otimes \left( C * B_R \right) \right),
    \\
    & = &
    {\rm Hom}_{X \times Y}\left(A, C * B_R \right),
\end{eqnarray}
which is (\ref{eq:cycle1}).  (Note we used the projection formula in the
simplification above.)

To recover identity~(\ref{eq:cycle2}), we simplify very similarly:
\begin{eqnarray}
        {\rm Hom}_{X \times Z}( A * B, C ) & = &
    \Gamma_{X \times Y \times Z} \left(
     \pi_{XY}^* A^{\vee} \otimes \pi_{YZ}^* B^{\vee} \otimes \pi_Y^* \omega_Y [\dim Y]  \otimes \pi_{XZ}^* C \right),
    \\
    & = &
    \Gamma_{Y \times X \times Z} \left(
    \pi_{YX}^* (\sigma_{XY}^* A)^{\vee} \otimes \pi_{YZ}^* B^{\vee} \otimes \pi_Y^* \omega_Y [\dim Y]  \otimes \pi_{XZ}^* C \right),
    \\
    & = &
    \Gamma_{Y \times Z}\left( \pi_{YZ *} \left(
    \pi_{YX}^* (\sigma_{XY}^* A)^{\vee} \otimes \pi_{YZ}^* B^{\vee} \otimes \pi_Y^* \omega_Y[\dim Y] \otimes \pi_{XZ}^* C \right) \right),
    \\
    & = &
    \Gamma_{Y \times Z}\left( B^{\vee} \otimes \pi_{YZ *}\left(
    \pi_{YX}^* (\sigma_{XY}^* A)^{\vee} \otimes \pi_Y \omega_Y[\dim Y] \otimes \pi_{XZ}^* C \right) \right),
    \\
    & = &
    \Gamma_{Y \times Z}\left( B^{\vee} \otimes \left( A_L * C \right) \right), 
    \\
    & = &
    {\rm Hom}_{Y \times Z}\left( B, A_L * C \right).
\end{eqnarray}

The identities (\ref{eq:cycle3}) and (\ref{eq:cycle4}) can be recovered
very similarly.  For completeness, we give the highlights below.
\begin{eqnarray}
    {\rm Hom}_{X \times Z}\left(A, B * C \right)
    & = &
    \Gamma_{X \times Z}\left( A^{\vee} \otimes \pi_{XZ *}\left( \pi_{XY}^* B \otimes \pi_{YZ}^* C \right) \right),
    \\
    & = &
    \Gamma_{X \times Y \times Z}\left( \pi_{XZ}^* A^{\vee} \otimes \pi_{XY}^* B \otimes \pi_{YZ}^* C \right),
    \\
    & = &
    \Gamma_{Y \times X \times Z}\left( \pi_{XZ}^* A^{\vee} \otimes \pi_{YX}^* (\sigma_{XY}^* B) \otimes \pi_{YZ}^* C \right).
\end{eqnarray}

To get (\ref{eq:cycle3}),  we write 
\begin{eqnarray}
    {\rm Hom}_{X \times Z}\left(A, B * C \right)
    & = &
    \Gamma_{Y \times Z}\left( \pi_{YZ *}\left( \pi_{XZ}^* A^{\vee} \otimes \pi_{YX}^* (\sigma_{XY}^* B \right) \otimes C \right),
    \\
    & = &
    \Gamma_{Y \times Z}\left( \left( \pi_{YZ *}\left( \pi_{XZ}^* A^{\vee} \otimes \pi_{YX}^* (\sigma_{XY}^* B) \right) \right)^{\vee \vee} \otimes C \right),
\end{eqnarray}
and compute
\begin{eqnarray}
    \left( \pi_{YZ *}\left( \pi_{XZ}^* A^{\vee} \otimes \pi_{YX}^* (\sigma_{XY}^* B) \right) \right)^{\vee}
    & = &
    \pi_{YZ *}\left( \pi_{XZ}^* A^{\vee \vee} \otimes \pi_{YX}^* (\sigma_{XY}^* B)^{\vee} \otimes \pi_X^* \omega_X[\dim X] \right),
    \\
    & = & B_R * A,
\end{eqnarray}
hence
\begin{eqnarray}
    {\rm Hom}_{X \times Z}\left(A, B * C \right)
    & = &
    \Gamma_{Y \times Z}\left( \left( B_R * A \right)^{\vee} \otimes C \right)
     \: = \:
     {\rm Hom}_{Y \times Z}\left( B_R * A, C \right).
\end{eqnarray}

To get (\ref{eq:cycle4}), we write
\begin{eqnarray}
    {\rm Hom}_{X \times Z}\left(A, B * C \right)
    & = &
    \Gamma_{X \times Y \times Z}\left( \pi_{XZ}^* A^{\vee} \otimes \pi_{XY}^* B \otimes \pi_{YZ}^* C \right),
    \\
    & = &
    \Gamma_{X \times Z \times Y}\left(  \pi_{XZ}^* A^{\vee} \otimes \pi_{XY}^* B \otimes \pi_{ZY}^* (\sigma_{YZ}^* C) \right),
    \\
    & = &
    \Gamma_{X \times Y}\left( \pi_{XY *}\left( \pi_{XZ}^* A^{\vee} \otimes \pi_{ZY}^* (\sigma_{YZ}^* C) \right) \otimes B \right),
    \\
    & = &
    \Gamma_{X \times Y}\left( \left( \pi_{XY *}\left( \pi_{XZ}^* A^{\vee} \otimes \pi_{ZY}^* (\sigma_{YZ}^* C) \right) \right)^{\vee \vee} \otimes B \right),
\end{eqnarray}
and compute
\begin{eqnarray}
    \left( \pi_{XY *}\left( \pi_{XZ}^* A^{\vee} \otimes \pi_{ZY}^* (\sigma_{YZ}^* C) \right) \right)^{\vee}
    & = &
    \pi_{XY *}\left( \pi_{XZ}^* A^{\vee \vee} \otimes \pi_{ZY}^* (\sigma_{YZ}^* C)^{\vee} \otimes \pi_Z^* \omega_Z[\dim Z] \right),
    \\
    & = &
    A* C_L,
\end{eqnarray}
hence
\begin{eqnarray}
    {\rm Hom}_{X \times Z}\left(A, B * C \right)
    & = &
    \Gamma_{X \times Y}\left( \left( A * C_L \right)^{\vee} \otimes B \right) \: = \:
    {\rm Hom}_{X \times Y}\left( A * C_L, B \right).
\end{eqnarray}

We can also see these identities graphically.
To do so, we will need to use some canonically-defined maps.  To clarify conventions, we describe them below.  First, from adjunction, for any $A \in D^b(X \times Y)$ and ${\cal E} \in D^b(X)$,
\begin{equation}
    {\rm Hom}_{X}\left( (\Phi(A)_L \circ \Phi(A))({\cal E}), {\cal E}\right) \: \cong \:
    {\rm Hom}_Y\left( \Phi(A)({\cal E}), \Phi(A)({\cal E}) \right),
\end{equation}
for the integral transform $\Phi$ defined in~(\ref{eq:defn:int-trans}),
hence the identity on $\Phi(A)({\cal E})$ is related to a map $(\Phi(A)_L \circ \Phi(A))({\cal E}) \rightarrow {\cal E}$.  Since $\Phi(A)_L \circ \Phi(A) = \Phi(A * A_L)$
from~(\ref{eq:comp-int-trans}), and all maps are natural, we have a natural map
\begin{equation} \label{eq:natl1}
    A * A_L \: \Longrightarrow \: {\rm Id}.
\end{equation}
(Again, the idea is standard, but the details of the expression are convention-dependent.)
Similarly,
\begin{equation}
    {\rm Hom}\left( \Phi(A_L)({\cal F}), \Phi(A_L)({\cal F}) \right) \: \cong \:
    {\rm Hom}\left( {\cal F}, (\Phi(A) \circ \Phi(A_L))({\cal F}) \right)
\end{equation}
for ${\cal F} \in D^b(Y)$
implies a natural transformation
\begin{equation}  \label{eq:natl3}
    {\rm Id} \: \Longrightarrow \: A_L * A.
\end{equation}
Similarly for right adjoints,
\begin{equation}
    {\rm Hom}_Y\left( \Phi(A)({\cal E}), \Phi(A)({\cal E}\right) \: \cong \:
    {\rm Hom}_X\left( {\cal E}, (\Phi(A)_R \circ \Phi(A))({\cal E}\right),
\end{equation}
implies the existence of a natural map
\begin{equation}  \label{eq:natl2}
    {\rm Id} \: \Longrightarrow  \: A * A_R,
\end{equation}
and
\begin{equation}
    {\rm Hom}\left( ( \Phi(A) \circ \Phi(A_R)) {\cal F}), {\cal F}\right) \: \cong \:
    {\rm Hom}\left( \Phi(A_R)({\cal F}), \Phi(A_R)({\cal F}) \right)
\end{equation}
implies the existence of a natural map
\begin{equation}  \label{eq:natl4}
    A_R * A \: \Longrightarrow \: {\rm Id}.
\end{equation}

Given the maps above, we can now describe the identities of this section graphically.
Let $A \in D^b(X \times Y)$, $B \in D^b(Y \times Z)$, and $C \in D^b(X \times Z)$, then we can represent
\begin{equation}
    {\rm Hom}_{X \times Z}\left( A * B, C \right)
\end{equation}
as follows:
\begin{center}
    \begin{tikzpicture}
        \draw (0.5,1.5) -- (0.5,2);  \draw (0.7,2) node {$A$};
        \draw (1.5,1.5) -- (1.5,2);  \draw (1.7,2) node {$B$};
        \draw (0.5,1.5) arc [start angle = -180, end angle = 0, radius = 0.5];
        \draw (1,0) -- (1,1);
        \draw (1.3,0) node {$C$};
        \draw (0.2,1.4) node {$X$};
        \draw (1,1.4) node {$Y$};
        \draw (1.8,1.4) node {$Z$};
    \end{tikzpicture}
\end{center}
As drawn\footnote{
These diagrams should be read left-to-right and bottom-to-top, which unfortunately is the exact opposite of the convention in \cite{aw}.
}, this can be interpreted as merely a schematic picture of the Hom above, but later, we will also interpret
the Hom above as the space of junction operators at the intersection of three interfaces $A$, $B$, $C$,
with suitable orientations, linking the three spaces $X$, $Y$, $Z$. Now, using~(\ref{eq:natl2}), namely,
${\rm Id} \Rightarrow B * B_R$, 
so we can rewrite the diagram above as
\begin{center}
    \begin{tikzpicture}
        \draw (0.5,1.5) -- (0.5,3);  \draw (0.7,2) node {$A$};
        \draw (1.5,1.5) -- (1.5,2);  \draw (1.7,2) node {$B$};
        \draw (0.5,1.5) arc [start angle = -180, end angle = 0, radius = 0.5];
        \draw (1,0) -- (1,1);
        \draw (1.3,0) node {$C$};
        \draw (0.2,1.4) node {$X$};
        \draw (1,1.4) node {$Y$};
        \draw (2,1.4) node {$Z$};
        \draw (1.5,2) arc [start angle = 180, end angle = 1, radius = 0.5];
        \draw (2.5,2) -- (2.5,0);
        \draw (2.8,0) node {$B_R$};
        \draw[dashed] (2,2.5) -- (2,3);
        \draw (2.3,3) node {Id};
        \draw (2.8,1.4) node {$Y$};
    \end{tikzpicture}
\end{center}
and as these diagrams are meant to be equivalent, we have equation~(\ref{eq:cycle1}), namely
\begin{equation}
    {\rm Hom}_{X\times Z}\left(A * B, C\right) \: \cong \: {\rm Hom}_{X \times Y}\left(A, C * B_R\right).
\end{equation}
Similarly, turning the interface $A$ around and using the existence of the natural transformation~(\ref{eq:natl3}), namely
${\rm Id} \Rightarrow A_L * A$, we have that the first diagram is equivalent to
\begin{center}
    \begin{tikzpicture}
        \draw (2,1.5) -- (2,2);  \draw (2.2,2) node {$A$};
        \draw (3,1.5) -- (3,3);  \draw (3.2,2) node {$B$};
        \draw (2,1.5) arc [start angle = -180, end angle = 0, radius = 0.5];
        \draw (2.5,0) -- (2.5,1);
        \draw (2.8,0) node {$C$};
        \draw (1.5,1.4) node {$X$};
        \draw (2.5,1.4) node {$Y$};
        \draw (3.5,1.4) node {$Z$};
        \draw (1,2) arc [start angle = 180, end angle = 0, radius = 0.5];
        \draw (1,2) -- (1,0);
        \draw[dashed] (1.5,2.5) -- (1.5,3);   \draw (1.8,3) node {Id};
        \draw (0.5,1.4) node {$Y$};
        \draw (0.7,0) node {$A_L$};
    \end{tikzpicture}
\end{center}
and as these diagrams are meant to be equivalent, we have equation~(\ref{eq:cycle2}), namely
\begin{equation}
    {\rm Hom}_{X \times Z}\left( A * B, C \right) \: \cong \:
    {\rm Hom}_{Y \times Z}\left( B, A_L * C \right).
\end{equation}

We can also represent~(\ref{eq:cycle3}) and (\ref{eq:cycle4}) graphically.
For $A \in D^b(X \times Z)$, $B \in D^b(X \times Y)$, $C \in D^b(Y \times Z)$, we can represent
\begin{equation}
    {\rm Hom}_{X \times Z}\left( A, B*C \right)
\end{equation}
graphically by the diagram
\begin{center}
    \begin{tikzpicture}
        \draw (0.5,0) -- (0.5,0.5);  \draw (0.7,0) node {$B$};
        \draw (1.5,0) -- (1.5,0.5);  \draw (1.7,0) node {$C$};
        \draw (0.5,0.5) arc [start angle = 180, end angle = 0, radius = 0.5];
        \draw (1,1) -- (1,2);   \draw (1.2,2) node {$A$};
        \draw (0.2,0.7) node {$X$};
        \draw (1,0.7) node {$Y$};
        \draw (1.8,0.7) node {$Z$};
    \end{tikzpicture}
\end{center}
Now, there is the natural transformation~(\ref{eq:natl4}), namely, $B_R * B \Rightarrow {\rm Id}$, so by turning $B$ around, we can rewrite the
diagram above as
\begin{center}
    \begin{tikzpicture}
        \draw (2,1) -- (2,1.5);  \draw (2.2,1) node {$B$};
        \draw (3,0) -- (3,1.5);  \draw (3.2,1) node {$C$};
        \draw (2,1.5) arc [start angle = 180, end angle = 0, radius = 0.5];
        \draw (2.5,2) -- (2.5,3);   \draw (2.7,3) node {$A$};
        \draw (1.5,1.7) node {$X$};
        \draw (2.5,1.7) node {$Y$};
        \draw (3.3,1.7) node {$Z$};
        \draw (1,1) arc [start angle = -180, end angle = 0, radius = 0.5];
        \draw (1,1) -- (1,3);   \draw (1.3,3) node {$B_R$};
        \draw[dashed] (1.5,0.5) -- (1.5,0);  \draw (1.8,0) node {Id};
        \draw (0.5,1.7) node {$Y$};
    \end{tikzpicture}
\end{center}
and as these diagrams are meant to be equivalent, we have equation~(\ref{eq:cycle3}), namely
\begin{equation}
    {\rm Hom}_{X \times Z}\left( A, B * C \right) \: \cong \:
    {\rm Hom}_{Y \times Z}\left( B_R * A, C \right).
\end{equation}

Similarly, using the natural transformation~(\ref{eq:natl1}), namely, $C * C_L \Rightarrow {\rm Id}$, and turning around the $C$ line,
we have the equivalent diagram
\begin{center}
    \begin{tikzpicture}
        \draw (0.5,0) -- (0.5,1.5);  \draw (0.7,1) node {$B$};
        \draw (1.5,1) -- (1.5,1.5);  \draw (1.7,1) node {$C$};
        \draw (0.5,1.5) arc [start angle = 180, end angle = 0, radius = 0.5];
        \draw (1,2) -- (1,3);   \draw (1.2,3) node {$A$};
        \draw (0.2,1.7) node {$X$};
        \draw (1,1.7) node {$Y$};
        \draw (2,1.7) node {$Z$};
        \draw (1.5,1) arc [start angle = -180, end angle = 0, radius = 0.5];
        \draw (2.5,1) -- (2.5,3);  \draw (2.8,3) node {$C_L$};
        \draw[dashed] (2,0) -- (2,0.5);  \draw (2.3,0) node {Id};
        \draw (3,1.7) node {$Y$};
    \end{tikzpicture}
\end{center}
and as these diagrams are meant to be equivalent, we have equation~(\ref{eq:cycle4}), namely
\begin{equation}
    {\rm Hom}_{X \times Z}\left( A, B * C \right) \: \cong \:
    {\rm Hom}_{X \times Y}\left( A * C_L, B \right).
\end{equation}

Next, we turn to identity~(\ref{eq:cycle5}).  
This is a straightforward consequence of adjunction, and the composition~(\ref{eq:comp-int-trans}).  For example,
\begin{equation}
    {\rm Hom}\left( \Phi( A*B )_L({\cal E}), {\cal F} \right) \: \cong \:
    {\rm Hom}\left( {\cal E}, \Phi(A*B)({\cal F}) \right),
\end{equation}
but $\Phi(A*B) = \Phi(B) \circ \Phi(A)$, 
and
\begin{equation}
    {\rm Hom}\left(( \Phi(A)_L \circ \Phi(B)_L )({\cal E}, {\cal F}\right)
    \: \cong \:
    {\rm Hom}\left( \Phi(B)_L({\cal E}), \Phi(A)({\cal F}) \right)
    \: \cong \:
    {\rm Hom}\left( {\cal E}, ( \Phi(B) \circ \Phi(A) )({\cal F}) \right),
\end{equation}
hence $\Phi(A*B)_L \cong \Phi(A)_L \circ \Phi(B)_L \cong \Phi(A_L) \circ \Phi(B_L) \cong \Phi(B_L * A_L)$, and using the isomorphism between integral transforms and kernels establishes half of~(\ref{eq:cycle5}).  The remainder is established similarly.

In particular, for $X = Y$ a Calabi-Yau, we have immediately from the expressions
above that
\begin{equation}
    (K_1 * K_2)^{\dag} \: \cong \: K_2^{\dag} * K_1^{\dag}
\end{equation}
for $K_1, K_2 \in D^b(X \times X)$, as desired.

Finally, we turn to identities~(\ref{eq:symtrace1}), (\ref{eq:symtrace2}).  Specialize to $X$ a Calabi-Yau, and let $A, B \in D^b(X \times X)$, so that $A^{\dag}$, $B^{\dag}$ are defined.
Then,
\begin{equation}
    {\rm Hom}_{X \times X}\left( A^{\dag}, B \right) \: = \: 
    {\rm Hom}_{X \times X} \left( {\cal O}_{\Delta} * A^{\dag}, B \right) \: = \:
    {\rm Hom}_{X \times X} \left( {\cal O}_{\Delta}, B * A \right),
\end{equation}
using identity~(\ref{eq:cycle1}) and the fact that $(A^{\dag})^{\dag} = A$ on a Calabi-Yau.
Similarly,
\begin{equation}
    {\rm Hom}_{X \times X}\left( A^{\dag}, B \right) \: = \: 
    {\rm Hom}_{X \times X} \left( A^{\dag} * {\cal O}_{\Delta}, B \right) \: = \:
    {\rm Hom}_{X \times X} \left( {\cal O}_{\Delta}, A * B \right),
\end{equation}
using identity~(\ref{eq:cycle2}).  Thus,
\begin{equation}
    {\rm Hom}_{X \times X} \left( {\cal O}_{\Delta}, A * B \right) \: = \:
    {\rm Hom}_{X \times X} \left( {\cal O}_{\Delta}, B * A \right).
\end{equation}
Similarly,
\begin{equation}
    {\rm Hom}_{X \times X} \left( A, B^{\dag} \right) \: = \:
    {\rm Hom}_{X \times X} \left( A, B^{\dag} * {\cal O}_{\Delta} \right) \: = \:
    {\rm Hom}_{X \times X} \left( B * A, {\cal O}_{\Delta} \right)
\end{equation}
using identity~(\ref{eq:cycle3}), and
\begin{equation}
    {\rm Hom}_{X \times X} \left( A, B^{\dag} \right) \: = \:
    {\rm Hom}_{X \times X}\left( A, {\cal O}_{\Delta} * B^{\dag} \right) \: = \:
    {\rm Hom}_{X \times X} \left( A * B, {\cal O}_{\Delta}) \right),
\end{equation}
using identity~(\ref{eq:cycle4}), hence
\begin{equation}
    {\rm Hom}_{X \times X}\left( A * B, {\cal O}_{\Delta} \right) \: = \:
    {\rm Hom}_{X \times X}\left( B * A, {\cal O}_{\Delta} \right).
\end{equation}

\subsubsection{Junction operators and cyclic identities}  
\label{sect:junction-ops}

Given the identities of the previous section, we can now define spaces of operators appearing at junctions of defects, known as junction operators, and check basic properties.  For example, we will apply the identities derived in the previous section to argue a cyclic property.
For simplicity, we assume that all of the defects are on a single Calabi-Yau $X$,
meaning that they are all objects in $D^b(X \times X)$.

Briefly, given a set of defects intersecting at a single point, each with an orientation (ingoing / outgoing with respect to the point), to define the space of junction operators, we pick one defect, call it $K_1$, and then, proceeding clockwise\footnote{
We could also proceed counterclockwise; either direction can be used, so long as one specifies one's conventions.
} around the junction, we form a product
of the defects and their adjoints.  The space of junction operators is then
\begin{equation}
    {\rm Hom}_{X \times X}( {\cal O}_{\Delta}, P ),
\end{equation}
where $P$ is the product.

For example, consider the junction illustrated below:
\begin{center}
\begin{tikzpicture}
    \draw[thick,->] (1,1) -- (1.5,1.5);  \draw[thick] (1.5,1.5) -- (2,2);
    \draw[thick,->] (1,1) -- (0.5,0.5);  \draw[thick] (0.5,0.5) -- (0,0);
    \draw[thick,->] (1,1) -- (0.5,1.5);  \draw[thick] (0.5,1.5) -- (0,2);
    \draw[thick,->] (1,1) -- (1.5,0.5);  \draw[thick] (1.5,0.5) -- (2,0);
    \filldraw[color=black] (1,1) circle [radius=0.08];
    \draw (2,1.6) node {$K_1$};
    \draw (2,0.4) node {$K_2$};
    \draw (0,0.4) node {$K_3$};
    \draw (0,1.6) node {$K_4$};
\end{tikzpicture}
\end{center}
This junction describes four outgoing defects.  To assign a space of junction operators, we must specify an order in which to traverse the defects -- clockwise or counterclockwise.
If we proceed clockwise from the top left defect, we get
\begin{equation}
    {\rm Hom}\left( {\cal O}_{\Delta}, K_1 * K_2 * K_3 * K_4 \right).
\end{equation}
We will see shortly that this is the same counting as for defect opertors at the end of the defect
$K_1 * K_2 * K_3 * K_4$, which graphically corresponds to rotating all four defects on top of one another:
\begin{center}
    \begin{tikzpicture}
        \draw[thick,->] (0,0.25) -- (0.5,0.25);  \draw[thick] (0.5,0.25) -- (1,0.25);
        \draw[thick,->] (0,0.2) -- (0.5,0.2);  \draw[thick] (0.5,0.2) -- (1,0.2);
        \draw[thick,->] (0,0.15) -- (0.5,0.15);  \draw[thick] (0.5,0.15) -- (1,0.15);
        \draw[thick,->] (0,0.1) -- (0.5,0.1);  \draw[thick] (0.5,0.1) -- (1,0.1);
        \filldraw[color=black] (0,0.17) circle [radius=0.1];
    \end{tikzpicture}
\end{center}
(See \cite{Ando:2010nm} for related constructions involving defects in the B model.)

More generally, if there are $n$ defects, $K_1, \cdots, K_n$, in that order, all outgoing, then the space of junction operators is
\begin{equation}
    {\rm Hom}_{X \times X}\left( {\cal O}_{\Delta}, K_n * K_{n-1} * \cdots * K_1 \right).
\end{equation}
If some of the defects are outgoing, then we take their adjoints. 

We have described this in terms of outgoing defects, but the results are equivalent to
a set of incoming defects.  Successively applying equation~(\ref{eq:cycle3}), the
Hom group above matches
\begin{eqnarray}
    {\rm Hom}_{X \times X}\left( {\cal O}_{\Delta}, K_n * K_{n-1} * \cdots * K_1 \right)
    & \cong &
    {\rm Hom}_{X \times X}\left( K_n^{\dag}, K_{n-1} * \cdots * K_1 \right),
    \\
    & \cong & 
    {\rm Hom}_{X \times X}\left( K_{n-1}^{\dag} * K_n^{\dag}, K_{n-2} * \cdots * K_1 \right),
    \\
    & \cong & \cdots \: \cong \:
    {\rm Hom}_{X \times X}\left( K_1^{\dag} * K_2^{\dag} * \cdots * K_{n-1}^{\dag} * K_n^{\dag},
    {\cal O}_{\Delta} \right).
\end{eqnarray}
This presentation of the Hom group can be interpreted as describing the same set of defects as all incoming, with the opposite orientation.

Furthermore, we can also demonstrate that the result is cyclic, independent of the choice of
starting position.  For example, using identity~(\ref{eq:cycle3}) then identity~(\ref{eq:cycle1}),
we have
\begin{eqnarray}
    {\rm Hom}_{X \times X}\left( {\cal O}_{\Delta}, K_n * K_{n-1} * \cdots * K_1 \right)
    & \cong &
    {\rm Hom}_{X \times X}\left( K_n^{\dag}, K_{n-1} * \cdots K_1 \right),
    \\
    & \cong &
    {\rm Hom}_{X \times X}\left( {\cal O}_{\Delta}, K_{n-1} * \cdots * K_1 * K_n \right).   \label{eq:cyclic1}
\end{eqnarray}
Similarly, using identity~(\ref{eq:cycle2}) followed by identity~(\ref{eq:cycle4}), we have
\begin{eqnarray}
    {\rm Hom}_{X \times X}\left( K_1 * \cdots * K_n, {\cal O}_{\Delta} \right)
    & \cong &
    {\rm Hom}_{X \times X}\left( K_2 * \cdots * K_n, K_1^{\dag} \right),
    \\
    & \cong & 
    {\rm Hom}_{X \times X}\left( K_2 * \cdots * K_n * K_1, {\cal O}_{\Delta} \right).   \label{eq:cyclic2}
\end{eqnarray}

Junction operators will appear more systematically in the axioms we review
in subsection~\ref{sect:axioms}.

\subsubsection{Defect local operators}  \label{sect:defect-ops}

We can now use junction operators to derive defect local operators, which are defined to be operators appearing at the end of a defect\footnote{Defect local operators are also referred to as twisted sector operators, or operators in the defect Hilbert space, especially in CFT literature.}.  For example, given a two-point junction of defects $A$, $B$, 
\begin{center}
\begin{tikzpicture}
    \draw[thick,->] (1,2) -- (1,1.5);   \draw[thick] (1,1.5) -- (1,1);
    \filldraw[color=black] (1,1) circle [radius=0.08];
    \draw[thick,->] (1,1) -- (1,0.5);  \draw[thick] (1,0.5) -- (1,0);
    \draw (0.5,1) node {$X$};   \draw (1.5,1) node {$Y$};
    \draw (1.3,2) node {$A$};   \draw (1.3,0) node {$B$};
\end{tikzpicture}
\end{center}
counted by Hom$(A,B)$,
we can use the identities above to rotate $A$ onto $B$,  giving the equivalent diagrams
\begin{center}
    \begin{tikzpicture}
        \draw[thick,->] (1,2) -- (1,1.5);   \draw[thick] (1,1.5) -- (1,1);
        \filldraw[color=black] (1,1) circle [radius=0.08];
        \draw[thick,->] (1,1) -- (1,0.5);  \draw[thick] (1,0.5) -- (1,0);
        \draw (0.5,1) node {$X$};   \draw (1.5,1) node {$Y$};
        \draw (1.3,2) node {$A$};   \draw (1.3,0) node {$B$};
        \draw (2.5,1) node {$=$};
        \draw[thick,->] (4,2) -- (4,1.5);   \draw[thick] (4,1.5) -- (4,1);
        \filldraw[color=black] (4,1) circle [radius=0.08];
        \draw[thick,->] (4,1) -- (4,0.5);  \draw[thick] (4,0.5) -- (4,0);
        \draw (3.5,1) node {$X$};   \draw (4.5,1) node {$Y$};
        \draw (3.7,2) node {$A$};   \draw (3.7,0) node {$B$};
        \draw[thick] (4,2) arc [start angle = 180, end angle = 0, radius = 0.5];
        \draw[thick,->] (5,2) -- (5,1);  \draw[thick] (5,1) -- (5,0);
        \draw (5.3,0) node {$A_R$};
        \draw (5.5,1) node {$X$};
        \draw (6.5,1) node {$=$};
        \draw[thick,dashed] (8,2) -- (8,1);  \draw (8.3,2) node {Id};
        \filldraw[color=black] (8,1) circle [radius=0.08];
        \draw[thick,->] (8,1) -- (8,0.5);  \draw[thick] (8,0.5) -- (8,0);
        \draw (8.6,0) node {$B * A_R$};
        \draw (7.5,1) node {$X$};  \draw (8.5,1) node {$X$};
    \end{tikzpicture}
\end{center}
\begin{center}
    \begin{tikzpicture}
        \draw[thick,->] (1,2) -- (1,1.5);   \draw[thick] (1,1.5) -- (1,1);
        \filldraw[color=black] (1,1) circle [radius=0.08];
        \draw[thick,->] (1,1) -- (1,0.5);  \draw[thick] (1,0.5) -- (1,0);
        \draw (0.5,1) node {$X$};   \draw (1.5,1) node {$Y$};
        \draw (1.3,2) node {$A$};   \draw (1.3,0) node {$B$};
        \draw (2.5,1) node {$=$};
        \draw[thick,->] (5,2) -- (5,1.5);   \draw[thick] (5,1.5) -- (5,1);
        \filldraw[color=black] (5,1) circle [radius=0.08];
        \draw[thick,->] (5,1) -- (5,0.5);  \draw[thick] (5,0.5) -- (5,0);
        \draw (4.5,1) node {$X$};   \draw (3.5,1) node {$Y$};
        \draw (5.3,2) node {$A$};   \draw (5.3,0) node {$B$};       
        \draw[thick] (4,2) arc [start angle = 180, end angle = 0, radius = 0.5];
        \draw[thick,->] (4,2) -- (4,1);  \draw[thick] (4,1) -- (4,0);
        \draw (3.7,0) node {$A_L$};
        \draw (5.5,1) node {$Y$};
        \draw (6.5,1) node {$=$};
        \draw[thick,dashed] (8,2) -- (8,1);  \draw (8.3,2) node {Id};
        \filldraw[color=black] (8,1) circle [radius=0.08];
        \draw[thick,->] (8,1) -- (8,0.5);  \draw[thick] (8,0.5) -- (8,0);
        \draw (8.6,0) node {$A_L * B$};
        \draw (7.5,1) node {$Y$};  \draw (8.5,1) node {$Y$};
    \end{tikzpicture}
\end{center}
or algebraically
\begin{eqnarray}
    {\rm Hom}_{X \times Y}(A,B) & = & {\rm Hom}_{X \times Y}( {\cal O}_{\Delta} * A, B) \: = \: {\rm Hom}_{X \times X}\left({\cal O}_{\Delta}, B * A_R \right),
    \\
    & = & {\rm Hom}_{X \times Y}( A * {\cal O}_{\Delta}, B ) \: = \: {\rm Hom}_{Y \times Y} \left( {\cal O}_{\Delta}, A_L * B \right).
\end{eqnarray}
(See \cite{Ando:2010nm} for related constructions involving defects in the B model.)

With this in mind, we now see that the defect operators at the end of a defect $A$ are given by
\begin{equation}
    {\rm Hom}( {\cal O}_{\Delta}, A).
\end{equation}

We will enumerate and verify properties of defect operators in subsection~\ref{sect:axioms}.

\subsubsection{Grading}  \label{sect:grading}

So far, we have argued that Hom's between kernels are the natural B model realization of vector spaces of junction operators, which include defect operators as a special case.
However, these Hom spaces also carry a natural grading, which reflects the underlying $N=2$ algebra
structure of the B model, and more specifically, the presence of a $U(1)_R$ symmetry.
Phrased another way, these Hom's have a grading and a differential, so we can take cohomology at any fixed degree.  This cohomology is precisely global Ext groups, of that degree.

Frequently in this paper, we will suppress the grading, and use the notation Hom to indicate Ext groups of every degree (or more precisely, chain complexes of vector spaces whose cohomology gives Ext group elements in each degree), but it will play an important role sometimes.

\subsubsection{Examples}

In this paper we will frequently encounter defects of the form $\Delta_* V$,
for $V \rightarrow X$ a vector bundle, and $\Delta: X \rightarrow X \times X$ the
diagonal.    Ext groups between defects of this form can be computed from
a spectral sequence \cite{Katz:2002gh}
\begin{equation}
    E_2^{p,q} \: = \: H^p\left( X, {\rm Hom}(V_1, V_2) \otimes \wedge^q N_{\Delta/X \times X} \right)
    \: \Longrightarrow \: {\rm Ext}^{p+q}_{X \times X}\left( \Delta_* V_1,
    \Delta_* V_2 \right).
\end{equation}
For the diagonal,
\begin{equation}
    N_{\Delta/X \times X} \: = \: TX,
\end{equation}
so this spectral sequence can be written as
\begin{equation}  \label{eq:diagvect:ss}
    E_2^{p,q} \: = \: H^p\left( X, {\rm Hom}(V_1, V_2) \otimes \wedge^q TX \right)
    \: \Longrightarrow \: {\rm Ext}^{p+q}_{X \times X}\left( \Delta_* V_1,
    \Delta_* V_2 \right).
\end{equation}

For example, if $V_1 = V_2 = {\cal O}_X$, the trivial line bundle, then we have
\begin{equation}
    H^p(X, \wedge^q TX) \: \Longrightarrow \: {\rm Ext}^{p+q}_{X \times X} \left( {\cal O}_{\Delta}, {\cal O}_{\Delta} \right).
\end{equation}
In this case, the spectral sequence trivializes\footnote{
As discussed in \cite{Katz:2002gh}, the differentials involve the Atiyah class of the
bundle on the support.  In this case, that bundle is trivial, so the Atiyah class vanishes,
as do the differentials.
}, and so 
\begin{equation}
    {\rm Ext}^n_{X \times X}\left( {\cal O}_{\Delta}, {\cal O}_{\Delta} \right)
    \: = \: \sum_{p+q=n} H^p(X, \wedge^q TX ),
\end{equation}
a standard expression for the Hochschild cohomology and local operators of the closed string B model, as expected.

If $X$ is an orbifold $[X/G]$ for finite $G$, nearly the same story applies
except that one takes $G$-invariants, as discussed in
\cite{Katz:2002jh}.

\subsection{Comparison to axioms for defects and junction operators}
\label{sect:axioms}

In this section, we will compare the properties of defects and Hom groups to the properties of junction and defect operators listed in \cite[section 2.1]{Chang:2018iay}.  To be clear, as mentioned previously, the noninvertible symmetries of the B model are not the same as a fusion category, not even as its trivial continuous generalization\footnote{By saying `trivial continuous generalization', we mean the direct product of a continuous group and a continuous group.}.  Nevertheless, we shall see that defects and Hom groups do satisfy very similar identities.  

In each case below, we begin with an axiom or property listed in \cite[section 2.1]{Chang:2018iay}, and then compare to properties of defects in the B model as we have described them here.

\begin{enumerate}
    \item Isotopy invariance:  Briefly, this states that on a flat surface, all physical observables are invariant under continuous deformations of topological defect lines that are ambient isotopies of the graph embedding.  This is a consequence of topological defects in general, and holds automatically for all defects in the B model.
    \item Defect local operators:  Defect local operators are point-like operators that can exist at the ends of a topological defect line.  We use ${\cal H}_L$ to denote the space of defect operators that can arise at the end of a topological defect line $L$.  
    These spaces are required to obey the following:
    \begin{itemize}
        \item The spaces ${\cal H}_{K_1 * \cdots * K_n}$ are invariant under cyclic permutations, meaning
        \begin{equation}
            {\cal H}_{K_1 * \cdots * K_n} \: \cong \:
            {\cal H}_{K_2 * \cdots * K_n * K_1},
        \end{equation}
        \item For the trivial topological defect line $I$, ${\cal H}_I$ is required to be the same as the space of bulk local operators. 
    \end{itemize}

    We discussed defect operators in the B model in subsection~\ref{sect:defect-ops},
    where we argued that 
    ${\cal H}_L = {\rm Hom}({\cal O}_{\Delta}, L)$ (in the convention that $L$ is outgoing from the endpoint).  
    We demonstrated the cyclic property above in section~\ref{sect:junction-ops},
    equation~(\ref{eq:cyclic1}). 
    Next, when $L = {\cal O}_{\Delta}$, to identify ${\cal H}_{ {\cal O}_{\Delta}}$ with bulk local operators, we will use the fact \cite[section 3.4]{Ando:2010nm} that closed string B model states can be identified with elements
of either Hochschild cohomology
\begin{equation}
    {\rm Ext}^{\bullet}_{X \times X}\left( {\cal O}_{\Delta}, {\cal O}_{\Delta} \right)
\end{equation}
or Hochschild homology
\begin{equation}
    {\rm Ext}^{\bullet}_{X \times X}\left( {\cal O}_{\Delta}^{\vee}, {\cal O}_{\Delta} \right)
\end{equation}
(which on a Calabi-Yau match up to a grading shift). In particular, the Hochschild-Kostant-Rosenberg isomorphism \cite[theorem 6.3]{andreirev} says
\begin{equation}
    {\rm Ext}^{\bullet}_{X \times X}\left( {\cal O}_{\Delta}^{\vee}, {\cal O}_{\Delta} \right)
    \: = \: \bigoplus_{p+q=\bullet} H^p\left(X, \Omega_X^q\right),
\end{equation}
and from the fact that
\begin{equation}
    \Omega_X^q \: = \: K_X \otimes \wedge^{n-q} TX,
\end{equation}
for $n$ the dimension of $X$, on a Calabi-Yau we have that
\begin{equation}
    {\rm Ext}^{\bullet}_{X \times X}\left( {\cal O}_{\Delta}^{\vee}, {\cal O}_{\Delta} \right)
    \: = \: \bigoplus_{p+q=\bullet} H^p\left(X, \wedge^{n-q} TX \right).
\end{equation}
These Ext group elements are precisely the cohomology of the Hom's that we identify with the defect local operators, as discussed in section~\ref{sect:defect-ops}, and so we see that 
the space of defect local operators associated with the identity ${\cal H}_{\Delta} = {\rm Hom}( {\cal O}_{\Delta}, {\cal O}_{\Delta})$ coincides with $H^{\bullet}(X, \wedge^{\bullet} TX)$, the space of closed-string B model bulk operators, as expected from \cite[section 2.1]{Chang:2018iay}.

    \item Junction vectors:  Junction operators are operators inserted at junctions of defects.  Given a junction defined by a set of incoming lines $K_1, \cdots, K_n$, if we let $V_{K_1 * K_2 * \cdots K_n}$ denote the vector space of junction operators associated to these lines, then the vector spaces are required to be cyclically symmetric in the sense that
    \begin{equation}
        V_{K_1 * \cdots * K_n} \: \cong \: V_{K_2 * \cdots * K_n * K_1}.
    \end{equation}

    We described the space of junction operators in the B model in section~\ref{sect:junction-ops}.  As described there, for incoming lines,
    \begin{equation}
        V_{K_1 * \cdots * K_n} \: = \: {\rm Hom}\left( K_1 * \cdots * K_n, {\cal O}_{\Delta} \right),
    \end{equation}
    and the desired cyclic symmetry was established in equation~(\ref{eq:cyclic2}).

    \item Correlation functions: There exists a notion of correlation functions of defects.  In the B model, correlation functions of defects certainly exist, and indeed, we will compute examples of correlation functions of defects in the B model in section~\ref{sect:corr-fns}.  

    In general, there is also an ``isotopy anomaly,'' in which under a deformation on a curved surface, a topological defect line may pick up a curvature-dependent phase.
    (See for excample \cite[section 3]{Chang:2020aww} for a discussion of the isotopy anomaly in two-dimensional theories, and its relation to the gravitational anomaly and contact terms.)

    Here, because we are working in the B model, this has a comparatively simple understanding:  because it is a twisted theory, worldsheet curvature only enters the B model via zero-mode counting, hence we do not expect to see an isotopy anomaly (save perhaps in subtle contact terms).  A physical, rigidly supersymmetric theory in two dimensions is different, as their formulation on curved spaces often requires adding curvature terms to the action and supersymmetry transformations.  However, those same curvature terms also obstruct the existence of topological twists.
    See for example \cite{Adams:2011vw,Jia:2013foa} for work on curvature terms in rigidly supersymmetric theories, and \cite[section 4]{Jia:2013foa} for a discussion of topological twists in such theories.

    \item Direct sum:  Given two topological defect lines $L_1$, $L_2$, there is a direct sum $L_1 + L_2$, and spaces of junction operators and correlation functions are linear with respect to this addition.  Indeed, in derived categories, there is an additive structure:  given any two kernels $K_1, K_2 \in D^b(X \times Y)$,
    there is a sum $K_1 + K_2 \in D^b(X \times Y)$.  Furthermore, Hom's are linear with respect to addition.
    For example,
    \begin{equation}
        {\rm Hom}(L_1 + L_2, L_3) \: = \: {\rm Hom}(L_1, L_3) + {\rm Hom}(L_2, L_3),
    \end{equation}
    \begin{equation}
        (L_1 + L_2) * L_3 \: = \: L_1 * L_3 + L_2 * L_3.
    \end{equation}

    This axiom in \cite[section 2.1]{Chang:2018iay} also mentions two other conditions on topological defect lines:
    \begin{itemize}
        \item Every topological defect line is semisimple.
        \item There are only finitely many types of topological defect lines.
    \end{itemize}
    In general terms, we do not expect these conditions to hold in the more general case of derived categories.

    \item Conjugation:  There exists a conjugation map $\dag$, which is required to obey
    \begin{equation}  \label{eq:conj:axiom}
        {\cal H}_L \cong {\cal H}_{L^{\dag}}^*
    \end{equation}
    for any defect $L \in D^b(X \times X)$. 
    Now, as explained in section~\ref{sect:adj}, conjugation is given in derived categories by the adjoint $\dag$.  We can see that the adjoint $\dag$ in derived categories
    obeys~(\ref{eq:conj:axiom}) as a consequence of Serre duality:
    \begin{eqnarray}
        {\cal H}_L & = & {\rm Hom}\left( {\cal O}_{\Delta}, L \right),
        \\
        & \cong & 
        {\rm Hom}\left( L, {\cal O}_{\Delta}\otimes \omega_X[ \dim X] \right)^*,
        \\
        & \cong &
        {\rm Hom}\left( {\cal O}_{\Delta}, L^{\dag}[\dim X] \right)^*.
    \end{eqnarray}
    In the expressions above, $\omega_X$ is the canonical bundle of $X$, which is trivial for $X$ Calabi-Yau.  Now, this expression has an extra factor of $[\dim X]$, but that
    only shifts the $U(1)_R$ grading, which is not a part of the axioms of \cite[section 2]{Chang:2018iay}.  As a result, if we omit the grading, then ${\cal H}_L \cong {\cal H}_{L^{\dag}}^*$ as desired.

    \item Locality:  This is in essence the requirement that OPEs exist.
    There are separate axioms for defect local operators and junction operators,
    which we describe below.
    \begin{enumerate}
        \item {\it Defect local operators.}  A series of defects $L_1, \cdots, L_n$, ending in defect operators $\psi_1, \cdots, \psi_n$, respectively, is equivalent to a junction operator $\tilde{\psi}$ joining each defect.  This is illustrated below for the case of two defects:
        \begin{center}
            \begin{tikzpicture}
                \draw[thick] (1,1.5) -- (3,1.5);
                \draw[thick] (1,0.5) -- (3,0.5);
                \filldraw[color=black] (1,1.5) circle [radius=0.08];
                \filldraw[color=black] (1,0.5) circle [radius=0.08];
                \draw (2,1.8) node {$L_1$}; \draw (2,0.2) node {$L_2$};
                \draw (0.6,1.5) node {$\psi_1$}; \draw (0.6,0.5) node {$\psi_2$};
                \draw (4,1) node {$=$};
                \draw[thick] (5.5,1) arc (120:90:5);  \draw (6.7,1.8) node {$L_1$};
                \draw[thick] (5.5,1) arc (240:270:5);  \draw (6.7,0.2) node {$L_2$};
                \filldraw[color=black] (5.5,1) circle [radius=0.08];
                \draw (5.1,1) node {$\tilde{\psi}$};
            \end{tikzpicture}
        \end{center}

        We describe this in derived categories below, for the case of two defects.
        Given $\psi_i \in {\rm Hom}(L_i, {\cal O}_{\Delta})$, locality for defect operators is the statement that there is
    a natural pairing
    \begin{equation} \label{eq:loc:hompair}
        {\rm Hom}\left(L_1, {\cal O}_{\Delta}\right) \otimes {\rm Hom}\left( L_2, {\cal O}_{\Delta}\right) 
        \: \longrightarrow \:
        {\rm Hom}\left(L_1 * L_2, {\cal O}_{\Delta}\right).
    \end{equation}

    This map can be derived from the 
    following map, which exists for any
three objects $A, B, C \in D^b(X \times Y)$:
\begin{equation}  \label{eq:basicmap} 
    {\rm Hom}_{X \times Y}(A, B) \otimes
    {\rm Hom}_{X \times Y}(B, C) \: \longrightarrow \:
    {\rm Hom}_{X \times Y}(A, C).
\end{equation}
    Then, 
    for any two objects $L_1, L_2 \in D^b(X \times X)$, for $X$ a Calabi-Yau, there is a map
    \begin{equation}
        {\rm Hom}_{X \times X}\left(L_1, {\cal O}_{\Delta} \right)
        \otimes
        {\rm Hom}_{X \times X}\left( {\cal O}_{\Delta}, L_2^{\dag} \right)
        \: \longrightarrow \:
        {\rm Hom}_{X \times X}\left( L_1, L_2^{\dag} \right).
    \end{equation}
    However, from identities~(\ref{eq:cycle1})-(\ref{eq:cycle4}),
    \begin{eqnarray}
        {\rm Hom}_{X\times X}\left( {\cal O}_{\Delta}, L_2^{\dag} \right)
        & = & {\rm Hom}_{X \times X}\left( L_2, {\cal O}_{\Delta} \right),
        \\
        {\rm Hom}_{X \times X}\left( L_1, L_2^{\dag} \right) & = &
        {\rm Hom}_{X \times X}\left( L_1 * L_2, {\cal O}_{\Delta} \right),
    \end{eqnarray}
    hence we have the map~(\ref{eq:loc:hompair}) above.  The general case can be derived by iterating this construction.

    \item {\it Junction operators.}  Given a pair of junction operators $\psi_1$, $\psi_2$ at either end of a line $L$, there exists an equivalent picture in which the line has been collapsed and replaced with a single junction operator $\tilde{\psi}$.  We illustrate this below in the special case of three lines on either side, with the collapsing intermediate line being $L_4$:
    \begin{center}
        \begin{tikzpicture}
            \draw[thick,->] (1,1) -- (2,1);  \draw[thick] (2,1) -- (3,1);
            \draw (2,0.7) node {$L_4$};
            \filldraw[color=black] (1,1) circle [radius=0.08];  
            \draw (1.2,1.3) node {$\psi_1$};
            \filldraw[color=black] (3,1) circle [radius=0.08];
            \draw (2.8,0.7) node {$\psi_2$};
            \draw[thick,->] (1,1) -- (0.5,0.5);  \draw[thick] (0.5,0.5) -- (0,0);  
            \draw (0.1,1.5) node {$L_7$};
            \draw[thick,->] (1,1) -- (0.5,1.5);  \draw[thick] (0.5,1.5) -- (0,2);
            \draw (0.1,0.5) node {$L_6$};
            \draw[thick,->] (1,1) -- (1,0.5);  \draw[thick] (1,0.5) -- (1,0);
            \draw (1.3,0.2) node {$L_5$};
            \draw[thick,->] (3,1) -- (3,1.5);  \draw[thick] (3,1.5) -- (3,2);
            \draw (2.7,2) node {$L_1$};
            \draw[thick,->] (3,1) -- (3.5,0.5);  \draw[thick] (3.5,0.5) -- (4,0);
            \draw (4,1.6) node {$L_2$};
            \draw[thick,->] (3,1) -- (3.5,1.5);  \draw[thick] (3.5,1.5) -- (4,2);
            \draw (4,0.4) node {$L_3$};
            \draw (5,1) node {$=$};
            \filldraw[color=black] (7,1) circle [radius=0.08];
            \draw[thick,->] (7,1) -- (6.5,0.5);  \draw[thick] (6.5,0.5) -- (6,0);  
            \draw (6.1,1.5) node {$L_7$};
            \draw[thick,->] (7,1) -- (6.5,1.5);  \draw[thick] (6.5,1.5) -- (6,2);
            \draw (6.1,0.5) node {$L_6$};
            \draw[thick,->] (7,1) -- (7,0.5);  \draw[thick] (7,0.5) -- (7,0);
            \draw (7.3,0.2) node {$L_5$};
            \draw[thick,->] (7,1) -- (7,1.5);  \draw[thick] (7,1.5) -- (7,2);
            \draw (6.7,2) node {$L_1$};
            \draw[thick,->] (7,1) -- (7.5,0.5);  \draw[thick] (7.5,0.5) -- (8,0);
            \draw (8,1.6) node {$L_2$};
            \draw[thick,->] (7,1) -- (7.5,1.5);  \draw[thick] (7.5,1.5) -- (8,2);
            \draw (8,0.4) node {$L_3$};
            \draw (7.4,1) node {$\tilde{\psi}$};
        \end{tikzpicture}
    \end{center}

        In derived categories, this is a consequence of the product~(\ref{eq:basicmap}).
        In terms of the diagram above, locality for junction operators is the statement
        that there should exist a corresponding map of junction operators
    \begin{equation}
        {\rm Hom}\left( {\cal O}_{\Delta}, L_4 * L_5 * L_6 * L_7 \right) 
        \otimes
        {\rm Hom}\left( L_4, L_1 * L_2 * L_3 \right)
        \: \longrightarrow \:
        {\rm Hom}\left( {\cal O}_{\Delta}, L_1 * L_2 * L_3 * L_5 * L_6 * L_7 \right).
        \nonumber 
    \end{equation}
    Using the identities~(\ref{eq:cycle1})-(\ref{eq:cycle4}),
    we can rewrite the first factor as
    \begin{equation}
        {\rm Hom}\left( {\cal O}_{\Delta}, L_4 * L_5 * L_6 * L_7 \right)
        \: = \:
        {\rm Hom}\left( L_7^{\dag} * L_6^{\dag} * L_5^{\dag} , L_4\right),
    \end{equation}
    and then applying the product~(\ref{eq:basicmap}) we have a map
    \begin{eqnarray}
        {\rm Hom}\left( L_7^{\dag} * L_6^{\dag} * L_5^{\dag}, L_4 \right) 
        \otimes
        {\rm Hom}\left( L_4, L_1 * L_2 * L_3 \right)
        & \longrightarrow &
        {\rm Hom}\left(  L_7^{\dag} * L_6^{\dag} * L_5^{\dag},  L_1 * L_2 * L_3 \right),
        \nonumber \\
        & & \: = \:
        {\rm Hom} \left({\cal O}_{\Delta}, L_1 * L_2 * L_3 * L_5 * L_6 * L_7 \right)
        \nonumber
    \end{eqnarray}
    which is clearly the space of junction operators associated with the `collapsed' diagram.  The general case can be derived identically.

    \end{enumerate}

    \item Partial fusion:  Topological defect lines obey an analogue of an OPE of the following form.  Locally, a correlation function with a pair of parallel topological defect lines is equivalent to a correlation function in which the lines have been fused to $L_1 * L_2$ with a set of junction vectors inserted at the endpoints, as in the figure below:
    \begin{center}
        \begin{tikzpicture}
            \draw[thick,->] (0.4,1.05) -- (0.2,1.54); \draw[thick] (0.2,1.54) -- (0,2);
            \draw[thick,->] (0.4,0.9) -- (0.2,0.44); \draw[thick] (0.2,0.44) -- (0,0);
            \draw[thick,->] (1.6,1.05) -- (1,1.05);  \draw[thick] (1,1.05) -- (0.4,1.05);
            \draw[thick,->] (1.6,0.9) -- (1,0.9); \draw[thick] (1,0.9) -- (0.4,0.9);
            \draw[thick,->] (2,2) -- (1.8,1.54); \draw[thick] (1.8,1.54) -- (1.6,1.05);
            \draw[thick,->] (2,0) -- (1.8,0.44); \draw[thick] (1.8,0.44) -- (1.6,0.9);
            \draw (0.3,2) node {$L_1$}; \draw (1,1.4) node {$L_1$}; \draw (1.7,2) node {$L_1$};
            \draw (0.3,0) node {$L_2$}; \draw (1,0.6) node {$L_2$}; \draw(1.7,0) node {$L_2$};
            \draw (3,1) node {$=$};  \draw (4,1) node {$\sum_i$};
            \draw[thick,->] (4.9,1) -- (4.7,1.5);  \draw[thick] (4.7,1.5) -- (4.5,2);
            \draw[thick,->] (4.9,1) -- (4.7,0.5);  \draw[thick] (4.7,0.5) -- (4.5,0);
            \draw[thick,->] (6.1,1) -- (5.5,1);  \draw[thick] (5.5,1) -- (4.9,1);
            \draw[thick,->] (6.5,2) -- (6.3,1.5); \draw[thick] (6.3,1.5) -- (6.1,1);
            \draw[thick,->] (6.5,0) -- (6.3,0.5); \draw[thick] (6.3,0.5) -- (6.1,1);
            \draw (4.8,2) node {$L_1$};   \draw (6.2,2) node {$L_1$};
            \draw (4.8,0) node {$L_2$};  \draw (6.2,0) node {$L_2$};
            \draw (5.5,1.3) node {$L_1 * L_2$};
            \filldraw[color=black] (4.9,1) circle [radius=0.08];
            \filldraw[color=black] (6.1,1) circle [radius=0.08];
            \draw (4.6,1) node {$v_i$};  \draw (6.4,1) node {$v_i$};
        \end{tikzpicture}
    \end{center}
    As a special case, a pair of topological defect lines  $L_1$, $L_2$ wrapping a circle on a cylinder (hence, without endpoints) can be fused 
    to $L_1 * L_2$, also wrapping the cylinder.

    The latter case, parallel TDL's wrapping a cylinder, should be clear in a topological field theory, as there are no endpoints to take into account.  Understanding the role of the endpoints is slightly more subtle.  In principle, we can understand this by thinking of the arrangement
    \begin{center}
    \begin{tikzpicture}
        \draw[thick,->] (0.4,1) -- (0.2,1.5);  \draw[thick] (0.2,1.5) -- (0,2);
            \draw[thick,->] (0.4,1) -- (0.2,0.5);  \draw[thick] (0.2,0.5) -- (0,0);
            \draw[thick,->] (2.8,1) -- (1.6,1);  \draw[thick] (1.6,1) -- (0.4,1);
            \draw[thick,->] (3.2,2) -- (3,1.5); \draw[thick] (3,1.5) -- (2.8,1);
            \draw[thick,->] (3.2,0) -- (3,0.5); \draw[thick] (3,0.5) -- (2.8,1);
            \draw[dashed] (1,0) -- (1,2);
    \end{tikzpicture}
    \end{center}
    where we think of the dashed line as the location of a spacelike slice.  Since the space of junction operators at that point is
    \begin{equation}
        {\rm Hom}(L_1 * L_2, L_1 * L_2),
    \end{equation}
    which includes the identity operator, we can insert the identity operator.  Writing the identity in the form of a sum over a complete set of states:
    \begin{equation}
        1 \: = \: \sum_i | v_i \rangle \, \langle v_i |
    \end{equation}
    leads to the expression at top.

    The necessity of having junction operators at the endpoints can also be understood mathematically, in terms of making functors encountered by various paths consistent.  To that end, consider a composition of integral transforms corresponding to defects encountered along a path.  The dashed line in the figure above can be understood as associated with the functor $\Phi(L_1 * L_2)$, for example, and the dashed line below
    \begin{center}
    \begin{tikzpicture}
        \draw[thick,->] (0.4,1) -- (0.2,1.5);  \draw[thick] (0.2,1.5) -- (0,2);
            \draw[thick,->] (0.4,1) -- (0.2,0.5);  \draw[thick] (0.2,0.5) -- (0,0);
            \draw[thick,->] (2.8,1) -- (1.6,1);  \draw[thick] (1.6,1) -- (0.4,1);
            \draw[thick,->] (3.2,2) -- (3,1.5); \draw[thick] (3,1.5) -- (2.8,1);
            \draw[thick,->] (3.2,0) -- (3,0.5); \draw[thick] (3,0.5) -- (2.8,1);
            \draw[dashed] (0.1,0) -- (0.1,2);
    \end{tikzpicture}
    \end{center}
    can be understood in terms of the composition $\Phi(L_2) \circ \Phi(L_1)$.  Now, the dashed line is arbitrary, and can be deformed; however, although the functors $\Phi(L_1 * L_2)$ and $\Phi(L_2) \circ \Phi(L_1)$ are isomorphic, they are not necessarily the same.  To map one to the other, one needs a natural transformation,
    which here can be identified with an element of
    \begin{equation}
        {\rm Hom}(L_1 * L_2 * L_2^{\dag}, L_1) \: \cong \: {\rm Hom}(L_1 * L_2, L_1 * L_2).
    \end{equation}
    This is, of course, the space of junction operators.

    \item Modular covariance:  Briefly, this axiom says that correlation functions of defect operators attached to networks of defects on a $T^2$ should be invariant under modular transformations.  The defect operators themselves will be invariant -- but the network of one-dimensional defects on the $T^2$ may change.  In principle, this should follow from modular properties of the B model topological field theory, in essence the fact that the B model is well-defined on $T^2$.  We leave any further details for future work.

\end{enumerate}

Altogether, we see that defects in the B model, with the junction and defect operators defined earlier, satisfy most of the axioms of \cite[section 2.1]{Chang:2018iay},
the exceptions involving finiteness and semisimplicity.

\subsection{Actions of defects on bulk states}   \label{sect:bulk}

So far we have discussed how defects act on the D-branes (the derived category $D(X)$).
However, defects also act on bulk (closed string) states, and it is this sense in which defects can be used to define analogues of noninvertible symmetry actions.

Arbitrary integral transforms act on Hochschild homology, though only those that induce an equivalence $D(X) \stackrel{\sim}{\rightarrow} D(X)$ act on Hochschild cohomology.

The action of integral transforms on cohomology is described in \cite{andreimukai2},
\cite[section 5.2]{huy}.
Briefly, the action of a defect defined by kernel $K \in D^b(X \times Y)$ on cohomology is given by the map
\begin{equation}
    \varphi(K): \: H^{\bullet}(X,{\mathbb Q}) \: \longrightarrow \: H^{\bullet}(Y,{\mathbb Q})
\end{equation}
given by\footnote{
Depending upon circumstances, the $\hat{A}$ genus may be more natural than the Todd genus used above.  However, on a Calabi-Yau, the Todd and $\hat{A}$ genera match.
In principle, another option is to replace the square root of the Todd genus with the complex Gamma class, see for example \cite[section 2]{Halverson:2013qca}, \cite{iritani1,iritani2,Iritani:2023pzf,Iritani:2023ngp,Katzarkov:2008hs} and references therein.
As the precise action on closed string states will not play an essential role in this paper, we will use the square root of the Todd genus for simplicity, and leave a more careful examination for future work.
}
\begin{equation}
    \varphi(K)(\mu) \: = \: \pi_{Y *}\left( {\rm ch}(K)\wedge \sqrt{ {\rm td}(T(X \times Y))} \wedge \pi_X^* \left( \mu   \right) \right),
\end{equation}
for $\mu \in H^{\bullet}(X,{\mathbb Q})$.
Formally, this can be derived from the requirement that  $\varphi(K)$ makes
the following diagram commute:
\begin{equation}   \label{eq:int-transform-hochschild}
    \xymatrix{
    D(X) \ar[r]^{\Phi(K)} \ar[d]^{v_X} & D(Y) \ar[d]^{v_Y} 
    \\
    H^{\bullet}(X,{\mathbb Q}) \ar[r]^{\varphi(K)} & H^{\bullet}(Y,{\mathbb Q}),
    }
\end{equation}
where on any space $W$, $v_W(a) = {\rm ch}(a) \sqrt{{\rm td}(TW)}$ for $a \in D(W)$.
We can see this as follows.
From commutivity of the diagram above, for ${\cal F} \in D(X)$, we have
\begin{eqnarray}
    \varphi(K)\left( v_X( {\cal F} ) \right)  & = & {\rm ch}( \Phi(K)({\cal F}) )
    \sqrt{ {\rm td}(TY) },
    \\
    & = & {\rm ch}\left( \pi_{Y *} \left( K \otimes \pi_X^* {\cal F} \right) \right)
    \sqrt{ {\rm td}(TY) },
    \\
    & = & \pi_{Y *} \left( {\rm ch} \left( \pi_X^* {\cal F} \otimes K \right) \pi_X^* {\rm td}(TX) \right)
    \sqrt{ {\rm td}(TY) },
\end{eqnarray}
using Grothendieck-Riemann-Roch, hence
\begin{eqnarray}
     \varphi(K)\left( v_X( {\cal F}) \right) 
    & = &
    \pi_{Y *} \left( \pi_X^*( v_X({\cal F}) ) {\rm ch}(K) \sqrt{ {\rm td}(T(X \times Y)) } \right),
    \\
    & = & \pi_{Y *} \left( \pi_X^* \left( v_X({\cal F}) \right) v_{X \times Y}(K) \right),
\end{eqnarray}
and replacing $v_X({\cal F})$ with $\mu$ gives the $\varphi(K)$ above.

Another variation of this diagram and map is described in \cite[section 2.2]{bridge97}; these are all equivalent to Grothendieck-Riemann-Roch, as described in \cite[section 5.2, cor. 5.29]{huy}.  We work with the version above as it was used in \cite{andreimukai2} to argue that, with the $\varphi$ above, the Cardy condition in the B model is equivalent to Hirzebruch-Riemann-Roch.

In general, the map $\varphi(K)$ respects neither the usual cohomological grading $\bullet$ nor Hodge decompositions.  However, it does respect the grading by columns of the Hodge decomposition:
\begin{equation}
    \varphi(K): \: \bigoplus_{q-p=i} H^{p,q}(X) \: \longrightarrow \:
    \bigoplus_{q-p=i} H^{p,q}(Y).
\end{equation}

Now, let us apply this to the special case of a defect defined by the kernel $\Delta_* V \in D^b(X \times X)$.
For any $\mu \in H^{\bullet}(X,{\mathbb Q})$,
\begin{eqnarray}   \label{eq:diag-act-a}
    \mu & \mapsto & \varphi(\Delta_* V)(\mu) \: = \:
    \pi_{2 *} \left( {\rm ch}(\Delta_* V) \sqrt{{\rm td}(T(X \times X))} \pi_1^* \mu \right).
\end{eqnarray}

To compute this, we will need ${\rm ch}(\Delta_* V)$.
Using Grothendieck-Riemann-Roch for $\Delta$, we have
\begin{equation}
    {\rm ch}(\Delta_* V) \: = \: \Delta_* \left( {\rm ch}(V) {\rm td}(TX) 
    \Delta^*\left( {\rm td}(T(X \times X))^{-1} \right) \right).
\end{equation}
Since the Todd class is multiplicative,
\begin{equation}
    \Delta^* {\rm td}( T(X \times X)) \: = \: \left( {\rm td}(TX) \right)^2,
\end{equation}
hence
\begin{equation}
     {\rm ch}(\Delta_* V) \: = \: \Delta_* \left( {\rm ch}(V) {\rm td}(TX)^{-1} \right).
\end{equation}

Now, we plug this back into equation~(\ref{eq:diag-act-a}), to find
\begin{eqnarray}
    \mu & \mapsto & \pi_{2 *}\left( \Delta_*\left( {\rm ch}(V) {\rm td}(TX)^{-1} \right)  \sqrt{{\rm td}(T(X \times X))} \pi_1^* \mu \right),
    \\
    & & \: = \: \pi_{2 *}\left( \Delta_*\left( {\rm ch}(V) \, {\rm td}(TX)^{-1} {\rm td}(TX) \right)  \pi_1^* \mu \right),
    \\
    & & \: = \:  \pi_{2 *}\left( \Delta_*\left( {\rm ch}(V) \right)  \pi_1^* \mu \right),
    \\
    & & \: = \: \pi_{2 *}\left( \Delta_*\left( {\rm ch}(V) \left( \pi_1^* \mu \right)|_{\Delta} \right) \right),
    \\
    & & \:= \: \pi_{2 *}\left( \Delta_* \left( {\rm ch}(V) \mu \right) \right),
    \\
    & & \: = \: {\rm ch}(V) \cdot \mu,
\end{eqnarray}
where we have used the projection formula, the observation
\begin{eqnarray}
    \left( \pi_1^* \mu \right) \Delta_* \alpha & = &
    \Delta_*\left( \left( \pi_1^* \mu \right)|_{\Delta} \alpha \right),
    \\
    & = & \Delta_*\left( \mu \alpha \right),
\end{eqnarray}
and, in the last step, the fact that 
$\pi_{2 *} \Delta_* = (\Delta \circ \pi_2)_* = {\rm Id}$.

Thus, the integral transform defined by $K = \Delta_* V$ induces a map
\begin{equation}
    \mu \mapsto \mu \cdot {\rm ch}(V)
\end{equation}
on $\mu \in H^{\bullet}(X)$.

As one special case, if $K = {\cal O}_{\Delta}$ (the identity element), then the action on
cohomology is the identity: $\mu \mapsto \mu$, as expected.

As another special case,
if $\mu = {\rm ch}({\cal F})$ for some object ${\cal F} \in D(X)$, then we see that the integral transform associated to $\Delta_* V$ maps
\begin{equation}
    {\rm ch}({\cal F}) \: \mapsto \: {\rm ch}({\cal F}) {\rm ch}(V) \: = \:
    {\rm ch}( {\cal F} \otimes V),
\end{equation}
as expected since on $D^b(X)$, the integral transform acts as ${\cal F} \mapsto {\cal F} \otimes V$.  This serves as a consistency check.

So far we have seen that for $X$ a smooth projective variety, the defect $\Delta_* V$ induces a map
on cohomology of the form ${\rm ch}({\cal F}) \mapsto {\rm ch}({\cal F}) {\rm ch}(V)$.  Similar arguments also apply when $X$ is a smooth Deligne-Mumford stack; one simply replaces
ch$(V)$ with ch$^{\rm rep}(V)$.

\subsection{Nonfiniteness and semisimplicity}
\label{sect:semisimple0}

Two of the defining properties of a fusion category are finiteness and semisimplicity. Finiteness means there are finitely many simple objects. Semisimplicity means every object can be written as a direct sum of simple objects. According to our previous discussion, it is clear that defects in B model captured by derived categories are not finite.

Now, characterizing `simple' objects in a derived category is not entirely simple.
For example, using Grothendieck's equivalence, in any short exact sequence
\begin{equation}
    0 \: \longrightarrow \: A \: \longrightarrow \: B \: \longrightarrow \: C \: \longrightarrow 0,
\end{equation}
one has an equivalence $B \sim A + C$.  For example, if $X$ is an elliptic curve, then
\begin{equation}
    0 \: \longrightarrow \: {\cal O}(-\Delta) \: \longrightarrow \: {\cal O} \:
    \longrightarrow \: {\cal O}_{\Delta} \: \longrightarrow \: 0,
\end{equation}
and so the trivial line bundle ${\cal O} \sim {\cal O}(-\Delta) + {\cal O}_{\Delta}$.

Later in section~\ref{sect:semisimp} we will propose that simple objects be identified with
objects that are stable with respect to a fixed Bridgeland stability condition.
As making sense of this requires picking e.g.~a K\"ahler metric, working with stability conditions in principle takes us outside the realm of topological field theories, and so we defer
this discussion to a later section.

\subsection{Correlation functions of lines and quantum dimensions}
\label{sect:corr-fns}

Correlation functions of lines can be computed in the B model 
(or in principle in any topological field theory) using the taffy
methods\footnote{
See also e.g.~\cite{Ganor:1994rm,Bars:1994xi,Bars:1994qm,Sonnenschein:2020jbe,Dutta:2023uxe} and references therein for discussions of folds in other contexts.
}  of \cite{Ando:2010nm}, which observed (and checked in numerous examples) that since the topological field theory is independent of the metric, we can `flatten' the worldsheet.  In terms of spacelike sections, we collapse circles into intervals, where we imagine the interval as hosting both halves of a circle.  If the circle was propagating on $X$, then the interval is propagating on $X \times X$.  At an edge, now seen as a fold, where the two halves of the original circle meet, the boundary is the diagonal\footnote{
One subtlety we will gloss over is the fact that, depending upon the orientation of the open string, the boundary diagonal could be described by either ${\cal O}_{\Delta}$ or its
derived dual ${\cal O}_{\Delta}^{\vee}$, as discussed in \cite{Ando:2010nm}.
The difference is just a grading shift, which will play no role in our analysis, ultimately cancelling out of computations, so we will suppress that detail in this section.
}, ${\cal O}_{\Delta}$.  We have tried to illustrate this process in the diagram below.

\begin{center}
\begin{tikzpicture}
    \draw (1.5,0) circle (0.75);
    \draw (1.5,0.5) node {$X$};  \draw (1.5,-0.5) node {$X$};
    \draw (3,0) node {$=$};
    \draw (5,0) ellipse (1 and 0.3);
    \draw (5,0.5) node {$X$};  \draw (5,-0.5) node {$X$};
    \draw (7,0) node {$=$};
    \draw[ultra thick] (8.5,0) -- (10.5,0);
    \draw[fill = black] (8.5,0) circle (0.05);
    \draw[fill = black] (10.5,0) circle (0.05);
    \draw (9.5,-0.3) node {$X \times X$};
    \draw (8.1,0) node {${\cal O}_{\Delta}$};
    \draw (10.9,0) node {${\cal O}_{\Delta}$};
\end{tikzpicture}
\end{center}

For one example, a sphere (hosting a nonlinear sigma model with target $X$)
hosting a line operator $K$ along the equator (with $K \in D^b(X \times X)$)
can be flattened to an open string disk diagram, hosting a sigma model with target $X \times X$, with $K$ defining the boundary of the disk diagram. This is in fact an alternative way, compared to the folding trick in Figure \ref{fig:folding}, to see why defects are described by $D^b(X\times X)$.

\begin{center}
    \begin{tikzpicture}
        \draw (2,0) circle (1);
        \draw[ultra thick] (1.05,0) arc (230:310:1.5);
        \draw (2,-1.3) node {$X$};
        \draw (2,0) node {$K$};
        \draw (4,0) node {$=$};
        \draw[ultra thick] (6,0) circle (1);
        \draw (6,-1.3) node {$X \times X$};
        \draw (6,0.6) node {$K$};
    \end{tikzpicture}
\end{center}

In principle, the disk diagram above (an open string diagram on $X \times X$) can be computed using the methods of \cite{Aspinwall:2004bs}.

Next, we turn to one-loop diagrams, as illustrated below:
\begin{center}
    \begin{tikzpicture}
        \draw (2,0) ellipse (2 and 0.8);
        \draw[ultra thick] (2,0) ellipse (1.5 and 0.5);
        \draw (2.3,0) arc (60:120:0.6);
        \draw (1.4,0.1) arc (240:300:1.2);
    \end{tikzpicture}
\end{center}
As before, we flatten the worldsheet torus hosting a sigma model with target $X$
onto an annulus diagram hosting a sigma model with target $X \times X$.
The torus partition function for a sigma model with target $X$ and a defect $K$ inserted along one loop is the same as that of an annulus diagram for a sigma model with target $X \times X$ with one boundary on the identity ${\cal O}_{\Delta}$ and $K$ at the other boundary.  Similarly, from \cite{Ando:2010nm}, the torus partition function for a sigma model with target $X$ and no defect is the same as that of an annulus diagram for a sigma model with target $X \times X$ and 
both boundaries on the diagonal ${\cal O}_{\Delta}$.  We illustrate the first of these two flattenings below.
\begin{center}
    \begin{tikzpicture}
        \draw[ultra thick] (2,0) ellipse (2 and 0.8);
        \draw (2.3,0) arc (60:120:0.6);
        \draw (1.4,0.1) arc (240:300:1.2);  
        \draw (3.8,-0.7) node {$K$};
        \draw (2,-0.5) node {$X$};
        \draw (5,0) node {$=$};
        \draw[ultra thick] (7.5,0) circle (1.5);
        \draw (7.5,0) circle (0.5);
        \draw (7.5,0.7) node {${\cal O}_{\Delta}$};
        \draw (9.3,0) node {$K$};
        \draw (7.5,-1) node {$X \times X$};
    \end{tikzpicture}
\end{center}
On the left is a torus (for a sigma model with target $X$) with a defect $K$ wrapped around the outer edge; on the right is an annulus (for a sigma model with target $X \times X$) with one boundary given by $K$, and the other by ${\cal O}_{\Delta}$.

In the B model, the partition function for an annulus with target a space $Y$ with one boundary on ${\cal E}$ and the other on ${\cal F}$ is (see e.g.~\cite[equ'n (158)]{Aspinwall:2004jr},
\cite[section 9.2]{Hori:2013ika})
\begin{equation}
\sum_i (-)^i \dim {\rm Ext}_{X \times X}^i\left( {\cal E}, {\cal F} \right).
\end{equation}
Thus, for the annulus of interest above with a defect $K$ wrapping the otter boundary, the annulus diagram is given by
\begin{equation}
\sum_i (-)^i \dim {\rm Ext}_{X \times X}^i\left( {\cal O}_{\Delta}, K \right),
\end{equation}
which can also be regarded as the twisted torus partition function twisted by $K$.
If we then normalize it by the partition function for just the torus
(which is computed by an annulus diagram on $X \times X$ with boundaries
${\cal O}_{\Delta}$, ${\cal O}_{\Delta}$), then we get an expression for the one-loop vev of the line associated to $K$:
\begin{equation} \label{eq:qdim:Ext}
    \langle K \rangle_{g=1} \: = \:  \frac{
    \sum_i (-)^i \dim {\rm Ext}^i_{X \times X}\left( {\cal O}_{\Delta}, K \right)
    }{
    \sum_j (-)^j \dim {\rm Ext}^j_{X \times X} \left( {\cal O}_{\Delta}, {\cal O}_{\Delta} \right)
    },
\end{equation}

We remark that the above computation should be distinguished from the quantum dimension of the line operator associated to $K$.
Briefly, the quantum dimension for a line defect is mathematically described in, e.g., \cite[section 4]{ce}, as the ``Frobenius-Perron dimension,'' where it is computed as the eigenvalues of a matrix encoding the action of any given line on elements of a basis of simple lines. Physically speaking, the quantum dimension is given by the cylinder vacuum expectation value of the line defect 
\begin{equation}
    \langle K \rangle_{\text{cyl}}=\langle \Omega | K |\Omega \rangle,
\end{equation}
where the vacuum state $\Omega$ in our case is given by the element $\mathcal{O}_\Delta$. Diagrammatically, we can visualize this computation as a cylinder with a line operator inserted and acting on the vacuum state.  
\begin{center}
    \begin{tikzpicture}
        \draw (0.1,0) ellipse (0.1 and 0.3);
        \draw (0.1,0.3) -- (2.1,0.3);   \draw (0.1,-0.3) -- (2.1,-0.3);  
        \draw (2.1,-0.3) arc (-30:30:0.6);
        \draw[ultra thick] (1.1,-0.3) arc (-30:30:0.6);
    \end{tikzpicture}
\end{center}
If we were to restrict to a basis of simple lines, this would result in the matrix whose eigenvalues define the quantum dimension.

If one were to close the cylinder above to a loop, so as to take a trace,
the result would be the twisted partition function denoted by the following the diagram
\begin{center}
    \begin{tikzpicture}
        \draw (2,0) ellipse (2 and 0.8);
        \draw[ultra thick] (2,-0.07) arc (30:-30:0.75);
        \draw (2.3,0) arc (60:120:0.6);
        \draw (1.4,0.1) arc (240:300:1.2);
    \end{tikzpicture}
\end{center}

Also, again because we are working in a topological field theory, we can interpret the defect as wrapping either noncontractible curve, as deforming the metric changes the picture.
We illustrate the idea below:
\begin{center}
    \begin{tikzpicture}
        \draw (0,0) -- (0,1);  \draw (0,0) -- (1,0);  
        \draw (1,0) -- (1,1);  \draw (0,1) -- (1,1);
        \draw[ultra thick] (0.5,0) -- (0.5,1);
        \draw (2,0.5) node {$=$};
        \draw (3,-1) -- (3,2);  \draw (3,-1) -- (4,-1);
        \draw (3,2) -- (4,2);  \draw (4,-1) -- (4,2);
        \draw[ultra thick] (3.5,-1) -- (3.5,2);
        \draw (5,0.5) node {$=$};
        \draw (6,0) -- (6,1); \draw (6,0) -- (10,0);
        \draw (10,0) -- (10,1);  \draw (6,1) -- (10,1);
        \draw[ultra thick] (8,0) -- (8,1);
    \end{tikzpicture}
\end{center}
The middle diagram describes a defect wrapped along the outer edge of a torus;
the right-most diagram describes a defect wrapped from inside to outside.
As a result, in the topological field theory, we expect
\begin{center}
    \begin{tikzpicture}
        \draw (2,0) ellipse (2 and 0.8);
        \draw[ultra thick] (2,-0.07) arc (30:-30:0.75);
        \draw (2.3,0) arc (60:120:0.6);
        \draw (1.4,0.1) arc (240:300:1.2);
        \draw (5,0) node {$=$};
        \draw (8,0) ellipse (2 and 0.8);
        \draw[ultra thick] (8,0) ellipse (1.5 and 0.5);
        \draw (8.3,0) arc (60:120:0.6);
        \draw (7.4,0.1) arc (240:300:1.2);
    \end{tikzpicture}
\end{center}
and so the annulus diagram we have computed, giving $\langle K \rangle_{g=1}$, could be interpreted as the trace of a matrix, which has a submatrix whose eigenvalues give the quantum dimension.

Such a result is consistent with expectations:  the alternating sums of the dimension of Ext groups between ${\cal O}_{\Delta}$ and $K$ is expected, on general grounds, to behave as some sort of trace of $K$.  Furthermore, it is a straightforward consequence of the description in terms of Ext groups that
\begin{equation}
    \langle K + L \rangle \: = \: \langle K \rangle \: + \: \langle L \rangle,
\end{equation}
also as one would expect intuitively from a trace.

Let us compute the one-loop $\langle K \rangle_{g=1}$ for the case $K = \Delta_* V$, for $V \rightarrow X$ a vector bundle of rank $r$.  Now,
\begin{eqnarray}
    \sum_i (-)^i {\rm Ext}_{X \times X}^i \left( {\cal O}_{\Delta}, \Delta_* V \right)
    & = &
    \sum_{p,q} (-)^{p+q} \dim H^p\left( X, V \otimes \wedge^q TX \right),
    \\
    & = &
    \sum_q (-)^q \chi\left( X, V \otimes \wedge^q TX \right),
\end{eqnarray}
where we have used the spectral sequence~(\ref{eq:diagvect:ss}) and, for any ${\cal E} \in D^b(X)$,
\begin{equation}
    \chi(X, {\cal E}) \: = \: \sum_i (-)^i \dim H^i(X, {\cal E}).
\end{equation}

Consider the case that $\dim X = 1$, and assume $X$ is a general Riemann surface.
We use the fact that
\begin{equation}
    {\rm td}(TX) \: = \: 1 \: + \: \frac{1}{2} c_1(TX)
\end{equation}
to compute
\begin{eqnarray}
    \chi(V) & = & \int_X \left( \frac{r}{2} c_1(TX) + c_1(V) \right),
    \\
    \chi(V \otimes TX) & = & \int_X \left( \frac{3}{2} r c_1(TX) + c_1(V) \right),
\end{eqnarray}
where $r$ is the rank of $V$.  From this we find the index is given by
\begin{eqnarray}
    \chi(V) - \chi(V \otimes TX) & = & -r \int_X c_1(TX) \: = \: -r \,\chi(X),
\end{eqnarray}
hence from~(\ref{eq:qdim:Ext}), we have that $\langle K \rangle_{g=1}$ is
\begin{equation}
    \frac{
      \sum_i (-)^i \dim {\rm Ext}^i_{X \times X}\left( {\cal O}_{\Delta}, \Delta_* V \right)
    }{
    \sum_j (-)^j \dim {\rm Ext}^j_{X \times X} \left( {\cal O}_{\Delta}, {\cal O}_{\Delta} \right)  
    }
    \: = \:
    \frac{
    \sum_q (-)^q \chi(X, V \otimes \wedge^q TX)
    }{
    \sum_p (-)^p \chi(X, \wedge^p TX)
    }
    \: = \:
    \frac{
    -r \chi(X)
    }{
    - \chi(X)
    }
    \: = \:
    r,
\end{equation}
the rank of $V$.
In the special case that $X$ is an elliptic curve, the numerator and denominator separately vanish, so the computation is only defined in a limiting sense.

Next, consider the case that $\dim X = 2$.
We use the fact that
\begin{equation}
    {\rm td}(TX) \: = \: 1 \: + \: \frac{1}{12} c_2(TX)
\end{equation}
to compute
\begin{eqnarray}
    \chi(V) & = & \int_X \left( \frac{r}{12} c_2(TX) + \frac{1}{2} c_1(V)^2 - c_2(V) \right),
    \\
    \chi(V \otimes TX) & = & \int_X \left( - \frac{5}{6} r c_2(TX) + c_1(V)^2 - 2 c_2(V) \right)
\end{eqnarray}
where $r$ is the rank of $V$,
to find that the index is
\begin{eqnarray}
    \chi(V) - \chi(V \otimes TX) + \chi(V \otimes \wedge^2 TX)
    & = & 2 \chi(V) - \chi(V \otimes TX),
    \\
    & = & r \int_X c_2(TX) \: = \: r \, \chi(X),
\end{eqnarray}
where $\chi(X)$ denotes the Euler characteristic of $X$.
From~(\ref{eq:qdim:Ext}), we have that $\langle \Delta_* V \rangle$ is
\begin{equation}
    \frac{
      \sum_i (-)^i \dim {\rm Ext}^i_{X \times X}\left( {\cal O}_{\Delta}, \Delta_* V \right)
    }{
    \sum_j (-)^j \dim {\rm Ext}^j_{X \times X} \left( {\cal O}_{\Delta}, {\cal O}_{\Delta} \right)  
    }
    \: = \:
    \frac{
    \sum_q (-)^q \chi(X, V \otimes \wedge^q TX)
    }{
    \sum_p (-)^p \chi(X, \wedge^p TX)
    }
    \: = \:
    \frac{
    r \chi(X)
    }{
    \chi(X)
    }
    \: = \:
    r,
\end{equation}
the rank of $V$.

We can establish the result more generally in higher dimensions as follows.
First, from Serre duality, for any vector bundle ${\cal E}$ on a Calabi-Yau $X$ of dimension $n$, we can write
\begin{equation}
    \chi(X, {\cal E}) \: = \: \sum_i (-)^i \dim H^i(X, {\cal E}) \: = \:
    \sum_i (-)^i \dim H^{n-i}(X, {\cal E}^*) \: = \: (-)^n \chi(X, {\cal E}^*)
\end{equation}
hence 
\begin{eqnarray}
    \sum_i (-)^i \dim {\rm Ext}^i_{X \times X}\left( {\cal O}_{\Delta}, \Delta_* V \right)
    & = &
    \sum_i (-)^i \chi(X, V \otimes \wedge^i TX),
    \\
    & = &
    (-)^n \sum_i (-)^i \chi(X, V^* \otimes \wedge^i T^*X).
\end{eqnarray}
Next,
from \cite[theorem 10.1.1]{hirzebruch}, for any vector bundle ${\cal E}$,
\begin{equation}
    {\rm td}({\cal E}) \wedge \sum_i (-)^i {\rm ch}\left( \wedge^i {\cal E}^* \right)
    \: = \: c_{\rm top}({\cal E}).
\end{equation}
As a result,
\begin{eqnarray}
    \sum_i (-)^i \chi(X, V^* \otimes \wedge^i T^*X) & = &
    \int_X {\rm td}(TX) \wedge {\rm ch}(V^*) \wedge \sum_i (-)^i {\rm ch}( \wedge^i T^*X),
    \\
    & = & \int_X {\rm ch}(V^*) \wedge c_{\rm top}(TX),
    \\
    & = & r \int_X c_{\rm top}(TX) \: = \: r \,\chi(X),
\end{eqnarray}
where $r$ is the rank of $V$.
Putting this together, we find from~(\ref{eq:qdim:Ext}) that $\langle \Delta_* V \rangle$, for $X$ any\footnote{
Any Calabi-Yau of dimension bigger than one, and in a limiting sense for the special case of elliptic curves.
} Calabi-Yau, is
\begin{eqnarray}
        \frac{
    \sum_i (-)^i \dim {\rm Ext}^i_{X \times X}\left( {\cal O}_{\Delta}, \Delta_* V \right)
    }{
    \sum_j (-)^j \dim {\rm Ext}^j_{X \times X} \left( {\cal O}_{\Delta}, {\cal O}_{\Delta} \right)
    }
    & = &
    \frac{
    (-)^n r \chi(X)
    }{
    (-)^n \chi(X),
    } \: = \: r,
\end{eqnarray}
the rank of $V$.

As a consistency check, consider the vev of 
the identity ${\cal O}_{\Delta}$.  One expects that $\langle {\cal O}_{\Delta} \rangle = 1$, which is easy to verify from the expression~(\ref{eq:qdim:Ext}).  

In fact, we see that the one-loop correlation function $\langle K\rangle_{g=1}$ shares a similar property as the quantum dimension of the line defect associated with $K$: For any invertible line $K = \Delta_* V$  defined by a rank-one bundle $V \rightarrow X$,
$\langle K \rangle_{g=1} = 1$; However, for higher-rank $V$ defining noninvertible $K$, $\langle K \rangle_{g=1} \neq 1$.

\subsection{Subtleties in truncating derived categories to fusion categories}

In many cases, the B-model TFT with target space $X$ admits simple finite group symmetries (such as geometric isometries of $X$) or even finite symmetries described by some more general fusion categories. One expects these simple global symmetries to be implemented also by line defects in the derived category $D^b(X\times X)$. However, there exists a subtlety in describing lines associated to objects in fusion categories, which we explore
as follows.

For simplicity, let us consider $G$ be a finite group acting on a space $X$.
One way to describe $G$ in the language of fusion categories is in terms of the
$G$-graded category of vector spaces Vec$(G)$, as was utilized in e.g.~\cite{Chang:2018iay}.  (See also \cite{Perez-Lona:2023djo,Perez-Lona:2024sds} and references therein.)  If in a slight abuse of notation we let $g$ denote the line in Vec($G$) corresponding to $g \in G$, then junction operators obey
\begin{equation}\label{eq: home space for group elements}
        {\rm Hom}_{ {\rm Vec}(G)}(g,h) \: = \: {\mathbb C} \, \delta_{g,h},
\end{equation}
corresponding to a simple two-point vertex.

Now, such line operators can be realized in derived categories.  For $g \in G$, let $\Gamma_g \subset X \times X$ denote the graph of $g$ as a map $X \rightarrow X$.  The associated integral transform $\Phi(\Gamma_g)$ implements the action of $g$ on $D^b(X)$, so $\Gamma_g \in D^b(X \times X)$ is the natural object to identify with elements $g \in G$.
That said, the Hom spaces are slightly different from those of Vec($G$).  In the special case\footnote{
We can see this as follows.  If $G$ acts freely, then $\Gamma_g$ and $\Gamma_h$ are disjoint if $g \neq h$.  In the case $g = h$, we use the fact that there is an isomorphism $\Gamma_g \cong \Delta$ to write
\begin{equation}
    {\rm Hom}_{D^b(X \times X)}(\Gamma_g,\Gamma_g) \: = \: {\rm Hom}(\Delta,\Delta),
\end{equation}
which is equivalent to the closed-string B model states, as reviewed in section~\ref{sect:axioms}.
} that $G$ acts freely,
\begin{equation}\label{eq: hom space for derived categories}
    {\rm Hom}_{D^b(X \times X)}(\Gamma_g, \Gamma_h) \: = \: V \delta_{g,h},
\end{equation}
for 
\begin{equation}
    V \: = \: \oplus_i HH^i(X),
\end{equation}
the totality of the Hochschild cohomology (the closed-string B model states).  If $G$ does not act freely on $X$, then there can be contributions to Hom$(\Gamma_g,\Gamma_h)$ if $g \neq h$.
(For example, if $G$ does not act freely, then the intersection $\Gamma_g \cap \Delta \cong X^g$, the fixed-point locus of $g \in G$, which leads to nonzero Hom's.)  As a quick consistency check, one can show\footnote{
This can be seen in at least two ways.  For one, since $\Gamma_g * \Gamma_{g^{-1}} = {\cal O}_{\Delta} = \Gamma_{g^{-1}} * \Gamma_g$, we see $\Gamma_g$ is an equivalence and $\Gamma_{g^{-1}}$ is its inverse, which must coincide with the adjoint.  Another argument is as follows.  For $\iota: Y \hookrightarrow X$ a smooth subvariety, ${\cal E} \rightarrow Y$ a vector bundle,
\begin{equation}
    \underline{\rm Hom}_X\left( \iota_* {\cal E}, {\cal O}_X \right) \: = \:
    \iota_* \underline{\rm Hom}_Y\left({\cal E}, \iota^* {\cal O}_X \right) \: = \:
    \iota_* \underline{\rm Hom}_Y\left( {\cal E}, \omega_{Y/X}[dim Y/X] \right).
\end{equation}
Now, $\Gamma_g = \iota_* {\cal O}_X$ for $\iota: X \hookrightarrow X \times X$ as
$x \mapsto (x, g \cdot x)$.  Then, we have
\begin{equation}
    \left( \iota_* {\cal O}_X \right)^{\vee} \: = \:
    \underline{\rm Hom}_{X \times X}\left( \iota_* {\cal O}_X, {\cal O}_{X \times X} \right)
    \: = \: \iota_* \underline{\rm Hom}_X\left( {\cal O}_X, {\cal O}_X[-\dim X]\right)
    \: = \: \iota_* \left( {\cal O}_X[- \dim X] \right) \: = \: \Gamma_g[- \dim X].
\end{equation}
To get $\Gamma_g^{\dag}$, one also takes the transposition (which changes $g$ to $g^{-1}$)
and shift the result by $[\dim X]$, hence $\Gamma_g^{\dag} = \Gamma_{g^{-1}}$.
} that $\Gamma_g^{\dag} = \Gamma_{g^{-1}}$, hence 
\begin{equation}
    {\rm Hom}\left(\Gamma_g, \Gamma_g \right) \: = \: {\rm Hom}\left(\Gamma_g * \Gamma_g^{\dag}, {\cal O}_{\Delta} \right) \: = \: {\rm Hom}\left( {\cal O}_{\Delta}, {\cal O}_{\Delta} \right),
\end{equation}
the space of closed-string states, consistent with the axioms for defect operators listed
in section~\ref{sect:axioms}.

Straightforwardly, one can consider a  fusion category $\mathcal{C}$ with noninvertible simple objects with a similar Hom space formula as (\ref{eq: home space for group elements})
\begin{equation}
    \text{Hom}_\mathcal{C}(L_i, L_j)=\mathbb{C}\delta_{i.j},
\end{equation}
and enrich it into the Hom space for the derived category (\ref{eq: hom space for derived categories}).

This difference in Hom space is as expected:  by comparison, the fusion categorical structure in e.g.~\cite{Chang:2018iay} is a truncation of the structure in derived categories, so that in the setting of derived categories, one sees higher categorical contributions, reflecting the details of the actions, data which is not present in the fusion category $\mathcal{C}$. Put another way, the Ext group elements, the graded elements of the Hom in the derived category, implicitly encode back reaction, which is not encoded in the structures of e.g.~\cite{Chang:2018iay}.

\subsection{Strongly symmetric boundary B-branes}

Following \cite[section 2.3]{Choi:2023xjw}, a boundary state $| B \rangle$
is said to be
strongly symmetric with respect to a fusion category symmetry if, for each line $L$ in the fusion category,
\begin{equation}
    L | B \rangle \: = \: \langle L \rangle | B \rangle,
\end{equation}
where $\langle L \rangle$ denotes the quantum dimension of $L$.

This is closely analogous to the notion of an eigensheaf in derived categories. Therefore, it is natural to utilize it as a notion of strongly symmetry boundary B-brane states with respect to the B-model defects.

Given some kernel $K$ defining an integral transform (a line operator),
an element $B \in D^b(X)$ is said to be an eigensheaf of the integral transform $\Phi(K)$ if there exists some vector bundle ${\cal E} \rightarrow X$ such that
\begin{equation}
    \Phi(K)(B) \: = \: B \otimes {\cal E}.
\end{equation}
The vector bundle ${\cal E}$ is referred to as the eigenvalue.

An example is as follows.  Let $X$ be a space with a $G$ action, then for all $g \in G$, the graph $\Gamma_g \in D^b(X \times X)$ has the property that for any $G$-equivariant sheaf $S \in D^b(X)$,
\begin{equation}
    \Phi(\Gamma_g)(S) \: = \: S,
\end{equation}
corresponding to the fact that $\Gamma_g$ has quantum dimension 1.
In this case, $S$ is an eigensheaf of the integral transform $\Phi(\Gamma_g)$, with eigenvalue ${\cal O}$.

\subsection{Analogous constructions in other topological field theories}
\label{sect:othertft}

In this section we will briefly review defects in other topological field theories,
beyond just B-twisted nonlinear sigma models with trivial $B$ fields.

\begin{itemize}

\item {\bf Defects in theories twisted by topologically nontrivial $B$ fields.}
\label{sect:Bfield}

As is well known (see e.g.~\cite[section 13.4]{huy}, \cite{andreithesis,Caldararu:2003kt}), in the presence of topologically nontrivial $B$ fields, 
one works with `twisted' bundles and sheaves.  This twisting is defined by an $\alpha \in H^2(X, {\cal O}_X^*)$, which we will assume is torsion\footnote{If $\alpha$ is not torsion, if $\alpha$ has infinite order, then the only allowed bundles have infinite ranks.  For simplicity, we assume in this section that $\alpha$ is torsion.}.  Briefly, if one represents $\alpha$ by a Cech cocycle, that cocycle defines the twisting on triple overlaps, in the sense that e.g.~transition functions only close up to that cocycle.  We let $D(X,\alpha)$ denote the resulting derived category of 
coherent sheaves twisted by $\alpha$.

Briefly, let us denote the derived category of $\alpha$-twisted sheaves on $X$ by
$D(X,\alpha)$.  This still exists, though the reader should note that for nontrivial $\alpha$, there are no line bundles -- the only $\alpha$-twisted vector bundles all have rank divisible by the order of $\alpha$.

The tensor product of an object of $D(X,\alpha)$ and an element of $D(X,\beta)$
is an element of $D(X, \alpha \beta)$.  As a result, for example,
one can tensor elements of $D(X,\alpha)$ by elements of $D(X)$ to get elements of $D(X,\alpha)$,
so $D(X,\alpha)$ is akin to a module over $D(X)$.  However,
one cannot tensor $D(X,\alpha)$ with itself to get
elements of $D(X,\alpha)$ (unless $\alpha^2 = 0$), unlike $D(X)$.

Integral transforms $D(X,\alpha) \rightarrow D(Y,\beta)$ can be defined very
similarly to the case of no twisting \cite[prop. 13.21]{huy}, with a kernel that is an element of 
$D^b(X \times Y, \pi_1^* \alpha^{-1} \otimes \pi_2^* \beta)$.

For more information on derived categories of twisted sheaves, see for example
\cite{andreithesis}.

\item {\bf Defects in B-twisted nonlinear sigma models on gerbes.}

Next, we consider string propagation on gerbes \cite{Pantev:2005wj}.
Briefly, a gerbe is a special kind of stack, and can be represented physically as a gauge
theory with a trivially-acting subgroup of the gauge group.

As has been disucssed extensively elsewhere (see e.g.~\cite{Hellerman:2006zs,Sharpe:2022ene}),
strings on gerbes are equivalent to strings on disjoint unions of spaces together with $B$ fields,
which is the prototypical example of `decomposition.'  (This resolves one of several apparent inconsistencies in the physics of string propagation on gerbes and stacks, as was discussed
in e.g.~\cite{Pantev:2005wj,Hellerman:2006zs}.)  At the level of sheaf theory, if we let $K$ denote the subgroup of the gauge group that acts trivially on the underlying space, then the (equivariant) sheaves are characterized, in part, by representations of $K$, corresponding to the restriction of the equivariant structure to $K$.  This then implies that the derived category of the gerbe is equivalent to a disjoint union (an orthogonal decomposition) of derived categories,
as discussed in e.g.~\cite{Hellerman:2006zs}.  

A precise statement of decomposition in this language, in simple cases, and kernels of integral transforms describing e.g.~projections, are given in section~\ref{sect:decomp}.

\item {\bf Defects in B-twisted Landau-Ginzburg models}
\label{sect:rev:lg-defects}

Defects also exist and have similar properties in Landau-Ginzburg theories.
For this article, we define a Landau-Ginzburg model to be a nonlinear sigma model with target a space or stack $X$,
with a superpotential $W: X \rightarrow {\mathbb C}$.  (Sometimes the term is reserved for the special case that $X$ is a vector space,
and other cases are described as hybrid Landau-Ginzburg models, but for simplicity, we will simply use the term Landau-Ginzburg model to refer to all such cases.)  See for example \cite{Herbst:2004ax,Guffin:2008kt} for massless spectrum computations in such theories for general $X$.

One can define defects in Landau-Ginzburg models much as in nonlinear sigma models.
Suppose one has a (B-twisted) nonlinear sigma model with target a Calabi-Yau $X$ and superpotential $W_X$,
and another (B-twisted) nonlinear sigma model with target a Calabi-Yau $Y$ and superpotential $W_Y$,
then a defect can be constructed which is given by a (in general sheafy \cite{Addington:2012zv}) matrix factorization of
\begin{equation}
    W \: = \: \pi_X^* W_X \: - \: \pi_Y^* W_Y
\end{equation}
on $X \times Y$, where $\pi_X: X \times Y \rightarrow X$ and $\pi_Y: X \times Y \rightarrow Y$ are projections,
see for example \cite{Rozansky:2003hz,Khovanov:2004bc,Brunner:2007qu,Brunner:2007ur}. 
The identity defect is ${\cal O}_{\Delta}$,  $\Delta \subset X \times X$ is the diagonal. (The restriction $W|_{\Delta} = 0$, so no matrix factorization is required.)
Integral transforms in these theories are discussed in mathematics in
e.g.~\cite{pv1,pv2}, and in physics in e.g.~\cite{Behr:2010ug,Behr:2014bta,Behr:2020gqw}.

\item {\bf A model analogues.}

In most of this paper we focus on the B model.
However, analogues also exist in the A model.
For example, in A-twisted nonlinear sigma models without a superpotential,
analogues of integral transforms are Lagrangian correspondences,
as described in e.g.~\cite{ww1,ww2,ww3}.  
A-branes\footnote{
The discussion in \cite{Hori:2000ck} predates the existence of matrix factorizations as a tool to solve the Warner problem in the B model.  However, so far as we are aware, there is no known analogue of matrix factorizations consistent with boundary supersymmetry in A twists, so the result of \cite{Hori:2000ck} is still applicable.
} in Landau-Ginzburg models are discussed in \cite{Hori:2000ck} (and A-twisting of closed-string
Landau-Ginzburg models is discussed in e.g.~\cite{Guffin:2008kt,Fan:2007ba,Fan:2007vi}). Invertible defects in the A model on gerbes and their spacetime interpretation as global structures of the Chern-Simons theory on A-branes are investigated in \cite{Pantev:2024kva}.
Briefly, an A-brane is a middle-dimensional Lagrangian submanifold (or coisotropic more generally), whose image under the superpotential
$W: X \rightarrow {\mathbb C}$ is a straight line.  Mathematically, these are understood via Fukaya-Seidel categories, see e.g.~\cite{fs1,Gaiotto:2015zna,Gaiotto:2015aoa}.
We leave a detailed discussion for other work.

\end{itemize}

\section{Topological defects in the untwisted theory} 
\label{sect:phys}

In general, a defect in B-model TFT is not guaranteed to come from a topological defect in the untwisted physical $\mathcal{N}=(2,2)$ theory. In this section, we discuss topological defects in the untwisted theory. We will focus on the case where defects preserve worldsheet supersymmetry, show that a defect is topological when it is simultaneously A- and B-twistable. We further identify geometric interpretations of these topological defects as complex Lagrangian and complex coisotropic branes.

\subsection{Topological versus conformal defects}

Consider an interface between two CFTs. Denote the stress-energy tensors for these two theories as $T_1,\bar{T}_1$ and $T_2, \bar{T}_2$, respectively. A conformal interface $I$ is then defined as 
\begin{equation}
    I(T_1-\bar{T}_1)=(T_2-\bar{T}_2)I.
\end{equation}
A topological interface is a special conformal interface satisfying a stronger condition
\begin{equation}
    IT_1=T_2I, ~I\bar{T}_1=\bar{T}_2I.
\end{equation}
Recall that when the two theories separated by the interface is the same theory $\mathcal{T}$, the interface operator is promoted to a defect operator $I\rightarrow D$ of the theory $\mathcal{T}$. Denote the stress-energy tensor of the theory $\mathcal{T}$ as $T$; the conformal defect $D_c$ and topological defect $D_t$ can then be written as 
\begin{equation}
\begin{split}
    \text{conformal defect $D_c$:}~&D_c(T-\bar{T})=(T-\bar{T})D_c,\\
   \text{topological defect $D_t$:}~&D_tT=TD_t, ~D_t\bar{T}=\bar{T}D_t.
\end{split}
\end{equation}

Within a topological field theory from a supersymmetric theory via topological twists, all compatible supersymmetric interfaces are independent
of their position, up to $Q$-exact contributions, and so all such interfaces are topological.
For example, within the topological B model for a nonlinear sigma model with target $X$,
all elements of $D^b(X \times X)$ define topological interfaces.  However, only a subset
of those are also topological in the untwisted physical theory, and we discuss conditions
for them to remain topological next.

\subsection{SUSY topological defects are simultaneously A- and B-twistable} 

In the context of $\mathcal{N}=(2,2)$ SCFT, a supersymmetric topological interface is an interface that is compatible with both A- and B-type topological twists, as has been discussed in e.g.~\cite{Petkova:2000ip,Bachas:2004sy,Brunner:2007qu,Kapustin:2010zc}
(see also \cite{Bachas:2001vj,Quella:2002ct}). As we will spend a lot of time exploring geometric examples of this result, for completeness we review the reasoning. Recall the $\mathcal{N}=(2,2)$ superconformal algebra (SCA) has the following generators
\begin{equation}
    T, G^+, G^-, J
\end{equation}
as well as their anti-holomorphic partners. The spins of these operators are 2, 3/2, and 1, respectively. Also, $G^\pm$ carries $U(1)$ charge $\pm 1$ under $J$. For the holomorphic sector, one can write down two topological twists given by the following distinguished isomorphism
\begin{equation}\label{eq: top twist of SCA}
\begin{split}
    &T\rightarrow T+\frac{1}{2}\partial J,~ J\rightarrow J,~ G^{\pm}\rightarrow G^{\pm},\\
    \text{or}~&T\rightarrow T-\frac{1}{2}\partial J,~ J\rightarrow -J,~ G^{\pm}\rightarrow G^{\pm}.
\end{split}
\end{equation}
The twisted $G^{+}$ and $G^{-}$ now have integer spins being bosonic. For example, under the twist in the first line of (\ref{eq: top twist of SCA}), $G^{+}$ and $G^{-}$ have spin 2 and 1, respectively. The twisted SCA now enjoys the correspondence to the bosonic CFT algebra:
\begin{equation}
    (T, G^-, J, G^+) \leftrightarrow (T, Q, J_\text{ghost}, b),
\end{equation}
where $Q$ and $b$ are, respectively, the BRST current and the antighost associated with the diffeomorphism on the bosonic string worldsheet. One can straightforwardly write down the similar correspondence for the second twist in (\ref{eq: top twist of SCA})
\begin{equation}
    (T,G^+, J, G^-) = (T,Q,J_{\text{ghost}}, b).
\end{equation}

The A-twist and B-twist are defined by how the two topological twists in (\ref{eq: top twist of SCA}) are performed relatively on the holomorphic and anti-holomorphic sectors of the $\mathcal{N}=(2,2)$ SCA. We can use the resulting BRST current $Q$ to denote the A-twist and B-twist as\footnote{One can also consider twists given by $\bar{A}: (G^-, \bar{G}^-)$ and $\bar{B}: (G^-, \bar{G}^+)$. However, they are trivially related to the A- and B-twists, respectively, by taking an overall complex conjugation.}
\begin{equation}
\begin{split}
     \text{A-twist:}~&(Q, \bar{Q})=(G^+, \bar{G}^+),\\
    \text{B-twist:}~&(Q, \bar{Q})=(G^+, \bar{G}^-).
\end{split}
\end{equation}

Now, let us consider conformal defects of $\mathcal{N}=(2,2)$ SCFT. By their compatibility with A-twist and B-twist, we can define A-type defect as 
\begin{equation}
\begin{split}
    &\hat{D}_c(T-\bar{T})=(T-\bar{T})\hat{D},\\
    &\hat{D}_c(J-\bar{J})=(J-\bar{J})\hat{D},\\
    &\hat{D}_c(G^{\pm}+\bar{G}^{\pm})=(G^{\pm}+\bar{G}^{\pm})\hat{D}_c,
\end{split}
\end{equation}
and similarly B-type defects as
\begin{equation}
\begin{split}
    &\hat{D}_c(T-\bar{T})=(T-\bar{T})\hat{D},\\
    &\hat{D}_c(J+\bar{J})=(J+\bar{J})\hat{D},\\
    &\hat{D}_c(G^{\pm}+\bar{G}^{\mp})=(G^{\pm}+\bar{G}^{\mp})\hat{D}_c.
\end{split}
\end{equation}
It is possible for a defect to be both A- and B-type. The resulting condition reads
\begin{equation}
\begin{split}
    \hat{D}_tT=T\hat{D}_t, ~\hat{D}_t\bar{T}=\bar{T}\hat{D}_t,\\
    \hat{D}_tJ=aJ\hat{D}_t, ~\hat{D}_t\bar{J}=\bar{a}\bar{J}\hat{D}_t,\\
    \hat{D}_tG^{\pm}=G^{\pm a}\hat{D}_t, ~\hat{D}_t\bar{G}^{\pm}=\bar{G}^{\pm a}\hat{D}_t.
\end{split}
\end{equation}
The first line of the above condition tells us this is indeed a topological defect.

So far we have considered conditions for topological defects to be compatible with worldsheet supersymmetry. See also \cite{Kapustin:2010zc}.
One can also consider topological defects which are not supersymmetric -- topological defects which commute
with the stress tensor, but need not behave well with respect to supersymmetry, see 
for example \cite{Fuchs:2007tx}.
In this paper we will focus on supersymmetric topological defects, which are simultaneously A- and B-twistable.

\subsection{Geometric criteria and examples}

As we just reviewed, there has been an extensive discussion of topological defects
in the literature, and the fact that simultaneously A- and B-twistable branes furnish examples of topological defects, where the A twisting is described with respect to
the symplectic form\footnote{
In principle, in the bulk, such a symplectic form would describe kinetic terms with indefinite metrics. Here, however, we are only using this symplectic form to describe the A-twist along the boundary, and no such corresponding ill-defined kinetic terms appear in the theory.
} $\pi_1^* \omega - \pi_2^* \omega$ on $X \times X$,
see for example \cite{Petkova:2000ip,Bachas:2004sy,Brunner:2007qu,Bachas:2001vj,Quella:2002ct,Kapustin:2010zc}.
In this discussion we will review geometric implications and realizations of simultaneous A- and B-twistability, which have not been so extensively discussed, and which we will
utilizer later in this paper.

We will review two types of examples of simultaneously A- and B-twistable defects:
complex Lagrangian submanifolds of $X \times X$, and, separately, more general complex coisotropic
submanifolds of $X \times X$.  For simplicity, we will focus on individual sheaves, but similar considerations should apply to complexes of branes, antibranes, ant tachyons.
We leave a precise formulation as objects in $D^b(X \times X)$ which can also be
interpreted as elements of a derived Fukaya category for future work
(see also e.g.~\cite{joyce}).

\subsubsection{Complex Lagrangian examples}

For a fixed Calabi-Yau $X$, in this section we will focus
on examples of defects that are complex and Lagrangian\footnote{
As noted in e.g.~\cite[section 2.1]{Rozansky:2003hz}, the Lagrangian condition is with respect to the difference of the pullbacks of the symplectic (here, K\"ahler) forms:
\begin{equation}    \label{eq:cpx-lg-condition}
\omega \: = \: \pi_1^* \omega_1  \: - \: \pi_2^* \omega_2.
\end{equation}
We will see later that this nicely meshes with the relevant mathematics.
} in $X \times X$, with bundles whose curvature $F$ obey 
\begin{equation}   \label{eq:FvsBs}
F + \pi_1^* B_1 - \pi_2^* B_2
\end{equation}
on their support\footnote{
In special cases, this can be more complicated.  For example, for Lagrangian branes inside Calabi-Yau three-folds,
it is known that open string disk instanton corrections to the Chern-Simons action modify the critical locus,
to be flat away from a union of $S^1$'s on the Lagrangian, where the curvature can have delta-function support.
} $S$.  Note that in general, this means $F$ can be nonzero so long as it can be absorbed into the difference of the pullbacks of the $B$ fields.

In the remainder of this section, we will discuss specific concrete examples of simultaneously 
A- and B-twistable defects realized as complex Lagrangians in $X \times X$.

We can construct examples of complex Lagrangian submanifolds using the fact
\cite[section 3.4, prop. 3.8]{dasilva} that a diffeomorphism $f: X_1 \rightarrow X_2$ is a symplectomorphism\footnote{
Meaning, preserves the symplectic form.
}
if and only if the graph of $f$ is a Lagrangian submanifold of $X_1 \times X_2$ with symplectic form $\pi_1^* \omega_1 - \pi_2^* \omega_2$.  In particular, if $f: X \rightarrow X$ is any automorphism of $X$ that preserves the K\"ahler form, then the graph of $f$ is complex Lagrangian
in $X \times X$, and so defines a B model defect that remains topological after untwisting.

An important (albeit trivial) example is the diagonal $\Delta: X \times X$.  We have already discussed in section~\ref{sect:rev}
how this acts as the identity among $B$ model defects.  It is also the graph of the identity map
$X \rightarrow X$, and as such it is a Lagrangian submanifold of $X$ with respect to the symplectic form given by the difference of the pullbacks of the K\"ahler forms described
above.  As a result, the identity defect is both A and B twistable, and so, unsurprisingly,
remains topological as a defect even after untwisting to the physical theory.

Another important example arises as a defect describing (part of) T-duality on elliptic curves.
The defect is $\Delta_* {\cal O}(1)$, for ${\cal O}(1)$ a nontrivial line bundle whose
$c_1$ defines a shift of the $B$ field.  Here, the bundle is not flat, but this complex brane is still A-twistable (in addition to B-twistable), by virtue of equation~(\ref{eq:FvsBs}): the defects interpolates between $B = 0$ and $B$ nontrivial,
using~(\ref{eq:diagonalvspi1}), here
\begin{equation}
    \Delta_* {\cal O}(1) \: = \:  \Delta_* \left( \Delta^* \pi_2^* {\cal O}(1) \right)
    \: = \: \Delta_* \left( \pi_2^* {\cal O}(1) |_{\Delta} \right).
\end{equation}
The difference between the pullbacks of the two $B$ fields on either factor cancels against the bundle curvature, so that equation~(\ref{eq:FvsBs}) is satisfied, and the resulting defect is simultaneously A- and B-twistable.

We can also consider deformations of the identity operator.
Now, the normal bundle of the diagonal $\Delta \subset X \times X$ is
identified\footnote{
We would like to thank S.~Katz for a useful discussion of this matter.
} with the tangent bundle of $X$, and if $X$ does not admit a continuous family of
automorphisms containing the identity, then the tangent bundle has no sections, so there are no deformations,
not even any infinitesimal sections.  However, if $X$ is an elliptic curve, or any variety admitting a continuous family of automorphisms containing the identity, and $f$ is any such automorphism deforming the identity, then the graph of $f$ deforms the diagonal.

In addition to deforming the underlying variety, we can also consider nontrivial flat connections on $\Delta$, as a Lagrangian with a flat connection is A-twistable, hence will define a defect
that is topological in the untwisted theory.  
For example, if $X$ is an elliptic curve, then the space of flat connections on 
$\Delta$ is the Jacobian of that curve, another elliptic curve.

Another important familiy of special cases will be defects in nonlinear sigma models with
target an elliptic curve, realizing maps $E \rightarrow E$.
Let $E$ be an elliptic curve, and $f: E \rightarrow E$ a map, an
endomorphism (not necessarily an isomorphism).  Consider the graph of $f$ on $E \times E$ with symplectic form $\omega = \pi_1^* \omega_0 - \pi_2 \omega_0$,
where $\omega_0$ is a symplectic form on $E$.  If $\omega_0$ is invariant\footnote{
E.S.~would like to thank R.~Donagi for an explanation of this point.
} under $f$: $f^* \omega_0 = \omega_0$, then the restriction of $\omega$ to the the graph of $f$ will vanish, and so the graph could be a Lagrangian submanifold of $E \times E$.  

To be specific, take $\omega_0$ to be translation-invariant, meaning
$\omega_0 \propto dz \wedge d \overline{z}$,  Now, for
any map $f:E \rightarrow E$, $f^* dz = ({\rm deg}\, f) dz$ (as $dz$ denotes an element of $H^1$, so
pullback just multiplies by the degree).  As a result,
\begin{equation}
    f^* (dz \wedge d \overline{z}) \: = \: ({\rm deg}\, f)^2 \, dz \wedge d \overline{z},
\end{equation}
and as deg $f \geq 0$, we see that precisely when deg $f = 1$, the map $f$ will preserve $\omega_0$,
and so only then can the graph of $f$ be Lagrangian.  An endomorphism of degree 1 is an automorphism,
and its graph is middle-dimensional,
hence we see that precisely in the case that $f: E \rightarrow E$ is an isomorphism will its graph be a complex Lagrangian submanifold, and hence define a topological defect in the physical nonlinear sigma model.
(In particular, there are no examples of non-isomorphism maps $f$ defining topological defects.)

Before proceeding, we should mention that
a useful reference on complex Lagrangian submanifolds is
\cite{Hitchin:1999hmi}.

\subsubsection{Complex coisotropic branes}

Moving on, in addition to Lagrangian submanifolds, coisotropic submanifolds with non-flat
bundles can also define A-twistable branes \cite{Kapustin:2001ij,Aldi:2005hz,Herbst:2010js,Chan:2018osu,Qin:2020nng} and defects \cite{Kapustin:2010zc}, generalizing branes and defects on Lagrangian submanifolds with flat bundles.

For later use, let us define A-twistable coisotropic defects more precisely, following e.g.~\cite{Kapustin:2010zc}.
Consider defects in a nonlinear sigma model with target a Calabi-Yau $n$-fold $X$, and let $\omega$ be the symplectic form
\begin{equation}\label{eq: symplectic form for folding}
    \omega \: = \: \pi_1^* \omega_1 \: - \: \pi_2^* \omega_2
\end{equation}
on $M = X \times X$, where $\pi_{1,2}$ are the projections to either factor, and $\omega_{1,2}$ are the (same) symplectic
form on either factor.
In general, an (A-twistable) coisotropic defect is a brane supported on a submanifold $S \subset X \times X$ of dimension
$2n+2k$ with a $U(1)$ bundle of curvature $F$, with possible $B$ field, where 
\begin{itemize}
    \item $S$ is a coisotropic submanifold of $X \times X$, meaning that
    $(T S)^{\perp} \subset T S$, for
    \begin{equation}
        ( T_p S )^{\perp} \: = \: \{ v \in T_p S \, | \, \omega(v,-) = 0 \} \: = \: {\rm ker}\, \omega|_S,
    \end{equation}
    \item $F + B|_{S} = 0$ on $(TS)^{\perp}$, where $F + B|_{S}$ is viewed as a map
    $T S \rightarrow T^* S$,
    \item $\omega^{-1} (F + B|_{S})$ squares to the identity.  Explicitly,
    \begin{equation}
        \left( \omega^{-1} (F + B|_{S}) \right)^2 \: = \: +1,
    \end{equation}
    or equivalently,
    \begin{equation}  \label{eq:coiso-cond3}
        \omega - \left(F + B|_S\right) \omega^{-1} \left(F + B|_S\right) \: = \: 0
    \end{equation}
    on $T S$.   
\end{itemize}
The last condition implies that $k$ is even, as the number of eigenvalues of $\omega^{-1}(F + B|_S)$ is evenly split
between $+1$ and $-1$ \cite[appendix A, proof of prop.~6]{Kapustin:2010zc}.  
(It is also sometimes referred to as an almost paracomplex structure,
see for example \cite{kk1}.)

Coisotropic branes are defined very similarly to coisotropic defects \cite{Kapustin:2001ij,Kapustin:2010zc}.  In both cases,
the brane is supported on a coisotropic submanifold.  For ordinary branes, the symplectic form is the K\"ahler form,
whereas here the symplectic form is the difference of pullbacks, which is not K\"ahler.  Also, for ordinary
coisotropic branes, $( \omega^{-1} F)^2 = -1$, and so defines an almost complex structure on $TS / (TS)^{\perp}$, whereas here, $(\omega^{-1} F)^2 = +1$ instead.
The differences are ultimately due to the folding trick and are explained in
\cite{Kapustin:2010zc}.

A-branes on Lagrangian submanifolds are special cases of the coisotropic branes above.
A Lagrangian submanifold is the same as a submanifold which is both isotropic and coisotropic, meaning
for example that $(TS)^{\perp} \cong TS$, and in
particular is a special case of a coisotropic submanifold.  If for example $B = 0$ and $F$ is flat,
the second condition is satisfied, and since $TS \cong (TS)^{\perp}$, the third condition is trivial,
so we see that an A-brane wrapped on a Lagrangian submanifold with flat bundle is a special case of the coisotropic
branes described above.

Next, as a different example, consider the case that $n=1$ ($X$ an elliptic curve) and $S = M = X \times X$, so that $k=1$.
In this case, $S = M$ and $(TS)^{\perp} = 0$.  Assume also for simplicity that $B = 0$.
The last condition for a coisotropic defect can be written in this case as
\begin{equation}
\left( \omega^{-1} F \right)^2 = + I.
\end{equation}
This implies that the eigenvalues of $(\omega^{-1} F)^2$ are $+1$, hence the eigenvalues of
$\omega^{-1} F$ are $\pm 1$. 
Following essentially the same argument as \cite[appendix A, proof of prop.~6]{Kapustin:2010zc}, this implies
\begin{equation}  \label{eq:coiso}
    F \wedge \omega \: = \: 0, \: \: \: F \wedge F \: = \: - \omega \wedge \omega.
\end{equation}
(Very similar reasoning was used in \cite{Kapustin:2001ij} to analyze the corresponding case of coisotropic branes,
which reached in that case a closely related conclusion, differing from~(\ref{eq:coiso}) by a sign in the second equation.)

The necessity of coisotropic A-branes on $E \times E$ for $E$ an elliptic curve, 
as implied by consistency of mirror symmetry, is 
described
in \cite[section 3]{Kapustin:2001ij}, and detailed constructions of coisotropic branes on $E \times E$ are given in \cite{Aldi:2005hz}.  

The constructions of coisotropic branes on $E \times E$ discussed in \cite{Kapustin:2001ij,Aldi:2005hz} assume, however, that the symplectic form is the K\"ahler form, proportional to the sum of the pullbacks of K\"ahler forms on either factor of the elliptic curve $E$.  For our purposes, to describe defects, we take the symplectic form to be the difference of the pullbacks, so we will require different examples than in \cite{Kapustin:2001ij,Aldi:2005hz}.

To construct an example of a brane on $E \times E$ which is coisotropic with respect to the relevant symplectic structure, we use the
intersection numbers given in appendix~\ref{app:ell-int}.  Take $S = E \times E$ itself, and
write the cohomology class of a symplectic form $\omega$ as\footnote{
Alternatively, the cohomology class of a more general symplectic form could be written
\begin{equation}
    \omega \: = \: x_1 (E \times p) + x_2 (p \times E) + x_3 \Delta.
\end{equation}
The K\"ahler cone can be computed from the condition that $\omega \cdot C \geq 0$ for every effective $C$.  Using the intersection numbers given in appendix~\ref{app:ell-int}.
Necessary conditions for $\omega$ to be in the K\"ahler cone are
\begin{eqnarray}
    \omega \cdot (E \times p) & = & x_2 + x_3 \geq 0,
    \\
    \omega \cdot (p \times E) & = & x_1 + x_3 \geq 0,
    \\
    \omega \cdot \Delta & = & x_1 + x_2 \geq 0.
\end{eqnarray}
}
Now, in our case, $\omega$ is the difference of pullbacks,
\begin{eqnarray}
    \omega & = & \pi_1^* L \: - \: \pi_2^* L,
    \\
    & = & x \left( E \times p - p \times E \right),
\end{eqnarray}
(This lies outside the K\"ahler cone.)
Take the holomorphic bundle ${\cal E}$ to be the line bundle ${\cal O}(D)$ for
\begin{equation}
    D \: = \: y_1 (E \times p) + y_2 (p \times E) + y_3 \Delta,
\end{equation}
where $y_1, y_2, y_3 \in {\mathbb Z}$.

The conditions~(\ref{eq:coiso}) relating ${\cal E}$ and $\omega$ are then
\begin{equation}  \label{eq:exe:coiso}
    D \cdot \omega \: = \: 0, \: \: \:
    D \cdot D \: = \: - \omega \cdot \omega,
\end{equation}
which imply\footnote{
For the more general symplectic forms of a previous footnote, the conditions~(\ref{eq:exe:coiso}) imply
\begin{equation}
    y_1( x_2 + x_3) \: + \: y_2 (x_1 + x_3) \: + \: y_3 (x_1 + x_2) \: = \: 0,
\end{equation}
\begin{equation}
    y_1 (y_2 + y_3) \: + \: y_2 (y_1 + y_3) \: + \: y_3 (y_1 + y_2)
    \: = \: - \left( 
    x_1 (x_2 + x_3) \: + \: x_2 (x_1 + x_3) \: + \: x_3 (x_1 + x_2)
    \right).
\end{equation}
}
\begin{equation}
    - x y_1 + x y_2 \: = \: 0,
\end{equation}
\begin{equation}
    y_1 (y_2 + y_3) + y_2 (y_1 + y_3) + y_3 (y_1 + y_2) \: = \: +2 x^2.
\end{equation}
For $x \neq 0$, so that the symplectic form is nontrivial, this implies 
\begin{equation}
y_2 \: = \: y_1 \: \in \: {\mathbb Z}, \: \: \:
    y_3 \: = \:  \frac{ x^2 - y_1^2 }{ 2 y_1} \: \in \: {\mathbb Z}.
\end{equation}

A family of solutions of these equations is 
\begin{equation}
    y_1 \: = \: y_2, \: \: \:
    x = n y_1, \: \: \:
    y_3 \: = \: \left( \frac{n^2 - 1}{2} \right) y_1,
\end{equation}
for $n$ odd.  For example,
\begin{center}
    \begin{tabular}{c|ccc}
    $n$ & $x$ & $y_1 = y_2$ & $y_3$ \\ \hline
    $1$ & $1$ & $1$ & $0$ \\
    $3$ & $3$ & $1$ & $4$ \\
    $5$ & $5$ & $1$ & $12$
    \end{tabular}
\end{center}

\subsection{Semisimplicity}  \label{sect:semisimp}

One of the defining properties of a fusion category is semisimplicity:
every object can be written as a (finite) direct sum of simple objects.

Now, as noted earlier in section~\ref{sect:semisimple0},
characterizing `simple' objects in a derived category is not entirely simple,
as any three objects related by an extension satisfy an equivalence relation, at least
at the level of Grothendieck groups.

Briefly, 
it is natural to propose that simple objects be identified with objects that are stable with respect to a fixed
Bridgeland stability condition. (Doing so takes us outside the realm of topological field
theories, as the stability condition involves a choice of K\"ahler metric.)

Due to the existence of a Harder-Narasimhan filtration, every object in a derived category is canonically equivalent in K theory to a direct sum of finitely many stable objects.

Thus, for this notion of simple objects, derived categories can be said to be semisimple,
in the sense above.

More generally, on a Fano space, it is expected that there are finitely many objects (in e.g.~an exceptional collection) that generate the derived category.  On a Calabi-Yau, this is not quite true -- one expects that one needs infinitely many objects to generate the derived category -- however, any one object can be expressed as a sum of finitely many stable objects.

Now, in the present case, there is a subtlety.  The statements above implicitly assume that stability is defined with respect to an ample line bundle, as arising e.g.~from a K\"ahler form.  For defects, however, the relevant symplectic form on $X \times X$ is the difference between the pullbacks from either factor (see (\ref{eq: symplectic form for folding})), not the sum.  Hence, one would want a notion of stability for a non-definite metric.

Now, mathematically, one ordinarily only defines stability for ample line bundles (positive-define metrics); other cases suffer from a variety of pathologies.
(For example, there may exist polarizations for which every sheaf is stable.)
This suggests that, perhaps, the relevant stability condition for defects involves an ample line bundle -- equivalently, a symplectic form that is the sum, not the difference, of pullbacks of symplectic forms on either factor.

For this paper, for simplicity, we will make this assumption, and work with stability defined with respect to an ample line bundle.  We leave a more thorough examination of notions of stability for defects for future work.

Before going on, it should be noted that stability depends upon moduli -- as one wanders around a moduli space, stable objects do not have to remain stable, and often become semistable or unstable.  Simple examples arise in heterotic string compactifications, where bundles that are stable (and hence satisfy the Donaldson-Uhlenbeck-Yau equation) may only be stable in part of the
K\"ahler cone, not all.  For details, see for example the physics papers \cite{Sharpe:1998zu,Anderson:2009sw,Anderson:2009nt} and references there.

\section{Noninvertible symmetries on elliptic curves}  \label{sect:ellcurve}

In this section, we discuss noninvertible symmetry structures arising on general elliptic curves,
not necessarily described by rational CFTs.  We begin with a discussion of fusion algebra structures,
then turn to a construction of Tambara-Yamagami structures.

\subsection{Basics of fusion on elliptic curves}

\subsubsection{Fusion on $K_{num}$}

For simplicity, rather than work with D-branes in $D^b(E \times E)$, we will work with D-brane charges, characterized by K-theory, in $K_{num}(E \times E)$.

We first consider the convolution algebra 
induced on $K_{num}(E \times E)$.
\begin{lemma}\label{lem:euler} We have the following
\begin{enumerate}[leftmargin=*]
\item
$K_{num}(E \times E) = \langle \OO, \OO_{D_1} = {\cal O}_{p \times E}, \OO_{D_2} = {\cal O}_{E \times p}, \OO_{\Delta}, \OO_q \rangle$, for $p \in E$ and $q \in E \times E$.
\item
The Euler pairing with respect to the basis 
$({\cal O}, {\cal O}_{p \times E}, {\cal O}_{E \times p},
{\cal O}_{\Delta}, {\cal O}_q)$ is given by
\[\chi = \begin{pmatrix}
0 & 0 & 0 & 0 & 1 \\
0 & 0 & -1 & -1 & 0 \\
0 & -1 & 0 & -1 & 0 \\
0 &-1 & -1 & 0 & 0 \\
1 & 0 & 0 & 0 & 0
\end{pmatrix}
\]
\end{enumerate}
\end{lemma}
The result of Lemma~\ref{lem:euler}(2) follows directly from directly from intersection theory on $E \times E$ via Hirzebruch-Riemann-Roch. For instance,
\begin{align*}
\chi(\OO, \OO_p) &= \langle (1,0,0), (0,0,1) \rangle = (1,0,0) \cdot (0,0,1) = 1  \\
\chi(\OO_{p\times E}, \OO_{E \times p}) &= \langle (0, p \times E, 0), (0, E \times p, 0) \rangle = (-1)(p\times E) \cdot (E \times p) = -1
\end{align*}
where the first equality is Riemann-Roch, the second applies the derived dual on Chern character, and the last is the intersection of algebraic cycles on $E \times E$.

For later use, the class of the antidiagonal $\overline{\Delta} = \{ (x, -x) \in E \times E \, | \, x \in E \}$ is given by 
\begin{equation}
    \overline{\Delta} = \{ (x,-x)\} = 2 (p \times E) + 2 (E \times p) - \Delta.
\end{equation}

\begin{lemma}
$\chi(\mathcal{E}, p_{13*}(p_{12}^* \mathcal{F}_1 \otimes p_{23*}^* \mathcal{F}_2)) = \chi(p_{13}^* \mathcal{E}, p_{12}^* \mathcal{F}_1 \otimes p_{23}^* \mathcal{F}_2)$
\end{lemma}

Taking fusion products commutes with passing to K theory, so we can describe fusion products of kernels in $D^b(E \times E)$ by merely listing products of elements of K theory.  
\begin{lemma}\label{lem:conv}
$K_{num}(E \times E) = \mathbb{Z}\langle x_1, x_2, x_3 ,x_4 ,x_5 \rangle $ with the multiplicative structure $*$ (the composition of kernels)
\begin{equation}   \label{eq:ExE:product}
A_{ij} \coloneqq x_i * x_j =
\begin{pmatrix}   
0& x_1 & 0& x_1 & x_3 \\
0 & x_2 & 0 & x_2 & x_5 \\
x_1 & 0 & x_3 & x_3 & 0 \\
x_1 & x_2 & x_3 & x_4& x_5\\
x_2 & 0 & x_5 & x_5 & 0
\end{pmatrix}
\end{equation}
where $x_1 = {\cal O}$, $x_2 = {\cal O}_{p \times E}$, $x_3 = {\cal O}_{E \times p}$, $x_4 = {\cal O}_{\Delta}$, $x_5 = {\cal O}_q$.
\end{lemma}
The computation for the $(1,2)$ element in Equation~(\ref{eq:ExE:product}) proceeds as follows. We have
\begin{align*}
x_1 * x_2 &\coloneqq p_{13*} (( p_{12}^* x_1) \otimes (p_{23}^* x_2)), \\
&= p_{13*} (( p_{12}^* (1,0, 0) ) \otimes (p_{23}^* (0,p\times E,0))), \\
&= p_{13*} ( ( 1,0, 0, 0) \cdot ( 0, E \times p \times E,0,0)), \\
&= p_{13*} ( 0, E \times p \times E, 0, 0), \\
&= (E\times E, 0, 0) = x_1.
\end{align*}

In the case that $E$ admits a complex multiplication, there can be additional contributions.
For example,
consider the complex torus $E = \mathbb{C}/\mathbb{Z} \langle 1, i \rangle$, which admits a complex multiplication. The graph $\Gamma$ of the morphism 
\begin{align*}
E &\rightarrow E, \\
 (x,y) &\mapsto i (x,y),
\end{align*}
gives an additional numerical class. The Euler pairing with respect to the basis
\[
\mathcal{B} = \{\OO, \OO_{D_1}, \OO_{D_2}, \OO_{\Delta}, \OO_{\Gamma}, \OO_q \}
\]
is given by 
\begin{equation}
\chi = \begin{pmatrix}
0 & 0 & 0 & 0 & 0&1 \\
0 & 0 & -1 & -1 &-1 &0 \\
0 & -1 & 0 & -1 & -1&0 \\
0 &-1 & -1 & 0 & -2&0 \\
0 & -1 & -1 & - 2 & 0 & 0 \\
1 & 0 & 0 & 0 & 0 & 0
\end{pmatrix}.
\end{equation}
The multiplicative structure, the $*$ product of kernels, is given by 
\begin{equation}
\begin{pmatrix}
0& y_1 & 0& y_1 &y_1 &y_3 \\
0 & y_2 & 0 & y_2 & y_2&y_6 \\
y_1 & 0 & y_3 & y_3 & y_3&0 \\
y_1 & y_2 & y_3 & y_4& y_5&y_6\\
y_1 & y_2 & y_3 & y_5 & 2 y_2 + 2y_3 - y_4& y_6\\
y_2 & 0 & y_6 & y_6 & y_6 &0
\end{pmatrix},
\end{equation}
where
\begin{equation}
    y_1 = {\cal O}_{E \times E}, \: \: \:
    y_2 = {\cal O}_{p \times E}, \: \: \:
    y_3 = {\cal O}_{E \times p}, \: \: \:
    y_4 = {\cal O}_{\Delta}, \: \: \:
    y_5 = {\cal O}_{\Gamma}, \: \: \:
    y_6 = {\cal O}_q,
\end{equation}
Note the product $y_5 \cdot y_5$ is the class of the antidiagonal subvariety $\overline{\Delta} \subset E \times E$. 

\subsubsection{Simple objects}

We identify simple objects with those that are stable with respect to a fixed Bridgeland stability condition $\sigma$.

Let $\mathcal{F} \in D^b(E \times E)$. Then,
following the discussion of section~\ref{sect:semisimp},
we claim that $\mathcal{F}$ is $\sigma$-semistable only if the Bogomolov inequality holds,  namely
\begin{equation}  \label{eq:stable:e}
- \chi(\mathcal{F}, \mathcal{F}) = \mathrm{c}_1^2(\mathcal{F}) - 2 \mathrm{r} \mathrm{ch}_2(\mathcal{F}) \geq 0.
\end{equation}
where the first equality follows from Hirzebruch-Riemann-Roch.

Indeed, assume that $\mathcal{F}$ is $\sigma$-stable. Then $\dim \mathrm{Hom}(E,E) = 1$, and by Serre duality, we have
\[
-\chi(\mathcal{F}, \mathcal{F}) = \dim \mathrm{Hom}(E,E[1]) - 2
\]
Note that $\dim \mathrm{Hom}(E,E[1])$ must be even, as Serre duality induces a non-degenerate symplectic form on the vector space $\mathrm{Hom}(E,E[1])$. By~\cite[Lemma 15.1]{bridgeland2006stabilityconditionsk3surfaces}, as $E \times E$ is an abelian surface, we must have that $\dim \mathrm{Hom}(E,E[1]) \neq 0$, and so the Bogomolov inequality must hold for any $\sigma$-stable object. The claim for $\sigma$-semistable objects follow by taking the Jordan-H\"older filtration of any $\sigma$-stable object.

In the example of $E \times E$, for a K-theory class of the form
\begin{equation}
\eta \: = \:
a_1 x_1 + a_2 x_2 + a_3 x_3 + a_4 x_4 + a_5 x_5
\end{equation}
the stability condition~(\ref{eq:stable:e}) implies that
\begin{equation}  \label{eq:stable:e2}
2 a_2 a_3 + 2 a_2 a_4 + 2 a_3 a_4 - 2 a_1 a_5 \: \geq \: 0.
\end{equation}

\subsubsection{Adjoints}

In this section, we will explicitly compute the adjoint of an object, physically speaking, the orientation reversal of $D^b(E \times E)$ represented in K-theory.

As discussed in section~\ref{sect:adj}, for any $K \in D^b(X \times X)$,
the adjoint $K^{\dag} = (\sigma^* K^{\vee})[\dim X]$ for $\sigma$ the transposition map.
So, we first need to compute some derived duals.

First, consider ${\cal O}_Y$ for $Y$ 
a divisor in $E \times E$.  This has the resolution
\begin{equation}
    0 \: \longrightarrow \: {\cal O}_{E \times E}(-Y) \: \longrightarrow \:
    {\cal O}_{E \times E},
\end{equation}
so the derived dual ${\cal O}_Y^{\vee}$ is given by the dual sequence
\begin{equation}
{\cal O}_{E \times E}(+Y) \: \longrightarrow \: {\cal O}_{E \times E},
\end{equation}
which (keeping track of where the cohomology is nonzero) we can identify with
${\cal O}_Y(+Y)[-1]$.  Furthermore, ${\cal O}(Y)|_Y$ is the normal bundle 
$N_{Y/ E \times E}$ of $Y$ in $E \times E$, hence 
\begin{equation}
    {\cal O}_Y^{\vee} \: = \: j_* N_{Y/E \times E} [-1],
\end{equation}
for $j: Y \hookrightarrow E \times E$ the embedding.
Thus, for example,
\begin{eqnarray}
    {\cal O}_{p \times E}^{\vee} \: = \: {\cal O}_{p \times E}[-1],
    \: \: \:
    {\cal O}_{E \times p}^{\vee} \: = \: {\cal O}_{E \times p}[-1],
\end{eqnarray}
hence
\begin{eqnarray}
    {\cal O}_{p \times E}^{\dag} \: = \: (\sigma^* {\cal O}_{p \times E}[-1])[1] \: = \:
    {\cal O}_{E \times p}, 
    \: \: \:
    {\cal O}_{E \times p}^{\dag} \: = \: (\sigma^* {\cal O}_{E \times p}[-1])[1] \: = \:
    {\cal O}_{p \times E}.
\end{eqnarray}

In the special case of the diagonal $\Delta \hookrightarrow E \times E$, 
the normal bundle $N_{\Delta/E \times E} = {\cal O}_E$, hence
${\cal O}_{\Delta}^{\vee} = {\cal O}_{\Delta}[-1]$ and
${\cal O}_{\Delta}^{\dag} = (\sigma^* {\cal O}_{\Delta}[-1])[1] = {\cal O}_{\Delta}$, correctly matching the specialization of equation~(\ref{eq:diagonal:derdual}).

It remains to compute the derived dual of ${\cal O}_q$.
Write $q = D_1 \cap D_2$, for $D_{1,2}$ a pair of divisors on $E \times E$.
Replacing ${\cal O}_q$ by its Koszul resolution
\begin{equation}
    0 \: \longrightarrow \: {\cal O}_{E \times E}(-D_1 - D_2) \: \longrightarrow \:
    {\cal O}_{E \times E}(-D_1) \oplus {\cal O}_{E \times E}(-D_2) \: \longrightarrow \: {\cal O}_{E \times E},
\end{equation}
it dualizes to
\begin{equation}
    {\cal O}_{E \times E} \: \longrightarrow \:
    {\cal O}_{E \times E}(+D_1) \oplus {\cal O}_{E \times E}(+D_2) 
    \: \longrightarrow \: {\cal O}_{E \times E}(+D_1+D_2),
\end{equation}
which we recognize as ${\cal O}_q(+D_1 + D_2)[-2]$.  Thus,
\begin{equation}
    {\cal O}_q^{\vee} \: = \: {\cal O}_q(+D_1 + D_2)[-2] \: = \: {\cal O}_q[-2].
\end{equation}
Similarly, $\sigma^* {\cal O}_q \cong {\cal O}_{q'}$ for some $q' \in E \times E$,
hence
\begin{equation}
    {\cal O}_q^{\dag} \: = \: (\sigma^* {\cal O}_q[-2])[1] \: = \: {\cal O}_{q'}[-1],
\end{equation}
whose K theory class matches that of $- {\cal O}_q$.

Putting this together, and expressing in terms of K theory classes, we see that for
\begin{equation}
    \eta \: = \: a_1 {\cal O} + a_2 {\cal O}_{p \times E} + a_3 {\cal O}_{E \times p} + a_4 {\cal O}_{\Delta} +
    a_5 {\cal O}_q,
\end{equation}
one has
\begin{equation}
    \sigma^* \eta^{\vee}  \: = \: a_1 {\cal O} - a_3 {\cal O}_{p \times E} - a_2 {\cal O}_{E \times p} - a_4 {\cal O}_{\Delta} +
    a_5 {\cal O}_q,
\end{equation}
and
\begin{equation}
\eta[1] \: = \: - \eta \: = \: - \left[ a_1 {\cal O} + a_2 {\cal O}_{p \times E} + a_3 {\cal O}_{E \times p} + a_4 {\cal O}_{\Delta} +
    a_5 {\cal O}_q \right].
\end{equation}
So, if we write
\begin{equation}
    D \: = \: b_1 {\cal O} + b_2 {\cal O}_{p \times E} + b_3 {\cal O}_{E \times p} + b_4 {\cal O}_{\Delta} +
    b_5 {\cal O}_q,
\end{equation}
then 
\begin{equation}
    D^{\dag} \: = \: - b_1 {\cal O} + b_3 {\cal O}_{p \times E} + b_2 {\cal O}_{E \times p} + b_4 {\cal O}_{\Delta} -
    b_5 {\cal O}_q,
\end{equation}

As a consistency check, it is straightforward to verify numerically that
$( \eta_1 * \eta_2)^{\dag} = \eta_2^{\dag} * \eta_1^{\dag}$, with the adjoint
given above and the convolution product ($*$) given in~(\ref{eq:ExE:product}).

In passing, if we do not include the transposition map $\sigma$ in the definition
of adjoints $K^{\dag}$, the expression above for $D^{\dag}$ would be changed only by
exchanging $b_2$ and $b_3$; however, it is straightfoward to show in this case
that $( \eta_1 * \eta_2)^{\dag} \neq \eta_2^{\dag} * \eta_1^{\dag}$.  This underlines the importance of the role of $\sigma$ in defining the adjoint.

\subsection{Tambara-Yamagami categories}

The paper \cite{Cordova:2023qei} observed a ${\mathbb Z}_2 \times {\mathbb Z}_2$ Tambara-Yamagami category structure
on elliptic curves, obtained by deforming away from the Gepner point to irrational SCFTs.  Now, in most dimensions, Gepner points are abstract SCFTs, related to nonlinear sigma models and geometry only by a K\"ahler deformation, but in the special case of K3s and elliptic curves, the Gepner model defines the same SCFT as a nonlinear sigma model.
As a result, the ${\mathbb Z}_2 \times {\mathbb Z}_2$ Tambara-Yamagami category structure observed by \cite{Cordova:2023qei} should also be visible in a nonlinear sigma model.

In this section, we will discuss how Tambara-Yamagami structures can be seen geometrically, directly in the nonlinear sigma model, for arbitrary elliptic curves -- not just in a neighborhood of the Gepner point.

Before walking through the details, let us first quickly review Tambara-Yamagami
structures.  Let $G$ be a finite abelian group.  We will construct a set of (invertible) lines
associated to group elements and realizing $G$, plus one additional noninvertible
line $D$, obeying the following rules 
\cite{Tambara:1998vmj}:
\begin{eqnarray}
    D * g & = & D \: = \: g * D \mbox{ for all }g \in G,
    \\
    D * D & = & \sum_{g \in G} g,
    \\
    D^{\dag} & = & D.
\end{eqnarray}

As a result, to find a Tambara-Yamagami structure for a group $G$, we must first construct a set of line operators / defects realizing $G$.  We will do that next,
for cyclic groups.

\subsubsection{${\mathbb Z}_n$ lines}

In this section we will construct some examples of ${\mathbb Z}_n$ lines, meaning,
defects $\eta$ such that $\eta^n = \eta * \eta * \cdots * \eta = {\cal O}_{\Delta}$.
We will work solely in K theory, where we can reduce these computations to simple linear algebra. However, to be clear, because we are working in K theory, we will only check that products of K theory classes match the image of ${\cal O}_{\Delta}$ in K theory, but we will not try to produce representatives in $D^b(E \times E)$, much less compute the products of such representatives in $D^b(E \times E)$, only in K theory.

We also note that because of the identity $\eta_1^{\dag} * \eta_2^{\dag} = (\eta_2 * \eta_1)^{\dag}$, the adjoint of a ${\mathbb Z}_n$ line will always be a ${\mathbb Z}_n$  line (though not necessarily the same line).  We will see this explicitly in examples, where it will serve as a self-consistency check.

First, let us construct ${\mathbb Z}_2$ lines.  Using the table of fusion products on $E \times E$ (for $E$ an elliptic curve) given in~(\ref{eq:ExE:product}), if we expand the $K$ theory class of a given line $\eta$ as
\begin{equation}
    \eta \: = \: a_1 {\cal O} + a_2 {\cal O}_{p \times E} + a_3 {\cal O}_{E \times p} + a_4 {\cal O}_{\Delta} +
    a_5 {\cal O}_q,
\end{equation}
for $p \in E$ and $q \in E \times E$, then requiring $\eta * \eta = 1 = {\cal O}_{\Delta}$ implies, at the level of $K$ theory classes,
\begin{eqnarray}
    a_1 a_2 + 2 a_1 a_4 + a_3 a_1 & = & 0,
    \\
    a_2^2 + 2 a_2 a_4 + a_1 a_5 & = & 0,
    \\
    a_1 a_5 + a_3^2 + 2 a_3 a_4 & = & 0,
    \\
    a_4^2 & = & 1,
    \\
    a_5 (a_2 + 2 a_4 + a_3) & = & 0.
\end{eqnarray}
Some integer solutions are listed in table~\ref{table:eta-solns}. As a consistency check, it is straightforward to compute that for every line $\eta$ listed in table~\ref{table:eta-solns}, the line $\eta^{\dag}$ also appears in the same table.

\begin{table}[h]
\begin{center}
\begin{tabular}{ccccc}
$a_1$ & $a_2$ & $a_3$ & $a_4$ & $a_5$ \\ \hline
0 & 0 & 0 & $\pm 1$ & 0 \\
0 & $\mp 2$ & $\mp 2$ & $\pm 1$ & 0 \\
0 & 0 & $\mp 2$ & $\pm 1$ & $n$ \\
0 & $\mp 2$ & 0 & $\pm 1$ & $n$ \\
$+n$ & $kn$ & $\mp 2 - kn$ & $\pm 1$ & $\mp 2k - nk^2$ \\
$-n$ & $\mp 2 - k n$ & $k n$ & $\pm 1$ & $\pm 2 k + n k^2$ \\
$-2n$ & $\mp 2 + 2kn$ & $-2kn$ & $\pm 1$ & $\mp 2k + 2nk^2$ \\
$+2n$ & $-2 k n$ & $\mp 2 + 2 k n$ & $\pm 1$ & $\pm 2k - 2 n k^2$
%
\end{tabular}
\caption{List of integer coefficients giving possible order-two $\eta \in D^b(E \times E)$.  In the table above, $k$ and
$n$ are arbitrary integers. 
The first two rows are invariant under the adjoint operation.  The remaining rows are interchanged by the adjoint.
\label{table:eta-solns} }
\end{center}
\end{table}

To be clear, table~\ref{table:eta-solns} only lists $K$-theory classes of ${\mathbb Z}_n$ lines.  To give a defect in the B model, one also needs to find representatives of a given $K$ theory class as an object of $D^b(E \times E)$.
In that spirit,
we point out two particularly noteworthy solutions listed in table~\ref{table:eta-solns}:
\begin{itemize}
    \item The trivial case of $\eta = {\cal O}_{\Delta}$ is listed, as the first entry.
    \item The antidiagonal $\overline{\Delta} = \{ (x,-x)\}$ has $K$ theory class  $2 (p \times E) + 2 (E \times p) - \Delta$, which also defines one of the solutions.
\end{itemize}

Similarly, requiring $\eta^3 = \eta * \eta * \eta = {\cal O}_{\Delta}$ implies, at the level of $K$ theory classes,
\begin{eqnarray}
    a_1 a_2^2 + a_1 a_2 a_3 + a_1 a_3^2 + 3 a_1 a_2 a_4 + 3 a_1 a_3 a_4 + 3 a_1 a_4^2 + a_1^2 a_5
    & = & 0,
    \\
    a_2^3 + 3 a_2^2 a_4 + 3 a_2 a_4^2 + 2 a_1 a_2 a_5 + a_1 a_3 a_5 + 3 a_1 a_4 a_5 & = & 0,
    \\
    a_3^3 + a a_3^2 a_4 + 3 a_3 a_4^2 + a_1 a_2 a_5 + 2 a_1 a_3 a_5 + 3 a_1 a_4 a_5 & = & 0,
    \\
    a_4^3 & = & 1,
    \\
    a_2^2 a_5 + a_2 a_3 a_5 + a_3^2 a_5 + 3 a_2 a_4 a_5 + 3 a_3 a_4 a_5 + 3 a_4^2 a_5 + a_1 a_5^2 & =  & 0.
\end{eqnarray}
We have listed some integer solutions in table~\ref{table:eta-3-solns}.

\begin{table}[h]
\begin{center}
\begin{tabular}{ccccc}
$a_1$ & $a_2$ & $a_3$ & $a_4$ & $a_5$ \\ \hline
0 & 0 & 0 & $1$ & 0 \\
$\pm 1$ & $-3 \mp k$ & $\pm k$ & $1$ & $\mp 3 - 3k \mp k^2$ \\
$\mp 1$ & $\pm k$ & $-3 \mp k$ & $1$ & $\pm 3 + 3k \pm k^2$ \\
$\pm 3$ & $-3 \mp 3k$ & $\pm 3k$ & $1$ & $\mp 1 - 3k \mp 3k^2$ \\
$\mp 3$ & $\pm 3k$ & $-3 \mp 3k$ & $1$ & $\pm 1 + 3k \pm 3k^2$
\end{tabular}
\caption{List of some integer coefficients giving possible order-three $\eta \in D^b(E \times E)$.  In the table above, $k$ is an arbitrary integer. 
The first row is invariant under the adjoint operation.  The remaining rows are interchanged by the adjoint.
\label{table:eta-3-solns} }
\end{center}
\end{table}

It is straightforward to also compute ${\mathbb Z}_4$ lines $\eta$, meaning lines such that
$\eta^4 = {\cal O}_{\Delta}$.  We summarize some integer solutions in table~\ref{table:eta-4-solns}.

\begin{table}[h]
\begin{center}
\begin{tabular}{ccccc}
$a_1$ & $a_2$ & $a_3$ & $a_4$ & $a_5$ \\ \hline
0 & 0 & 0 & $\pm 1$ & 0 \\
0 & $\mp 2$ & $\mp 2$ & $\pm 1$ & 0 \\
0 & 0 & $\mp 2$ & $\pm 1$ & $n$ \\
0 & $\mp 2$ & 0 & $\pm 1$ & $n$ \\
$+n$ & $kn$ & $\mp 2 - kn$ & $\pm 1$ & $\mp 2k - nk^2$ \\
$-n$ & $\mp 2 - k n$ & $k n$ & $\pm 1$ & $\pm 2 k + n k^2$ \\
$-2n$ & $\mp 2 + 2kn$ & $-2kn$ & $\pm 1$ & $\mp 2k + 2nk^2$ \\
$+2n$ & $-2 k n$ & $\mp 2 + 2 k n$ & $\pm 1$ & $\pm 2k - 2 n k^2$ \\ \hline
$+n  \: \: \: (|n| = 1)$ & $k$ & $\mp 2 - k$ & $\pm 1$ & $(-2 \mp 2k - k^2) / n$ \\
$-n \: \: \: (|n|=1)$ & $\mp 2 - k$ & $k$ & $\pm 1$ & $(2 \pm 2k + k^2)/n$  \\
$+n \: \: \: (|n| = 2)$ & $2k$ & $\mp 2 - 2k$ & $\pm 1$ & $(-2 \mp 4k - 4k^2)/n$ \\
$-n \: \: \: (|n| = 2)$ & $\mp 2 - 2k$ & $2k$ & $\pm 1$ & $(2 \pm 4k + 4k^2)/n$ 
\end{tabular}
\caption{List of integer coefficients giving some possible order-four $\eta \in D^b(E \times E)$.  In the table above, $k$, $n$, $a_1$, and
$a_2$ are arbitrary integers.  The first eight rows, up to the divider, are also ${\mathbb Z}_2$ lines; the four rows after that
are only ${\mathbb Z}_4$ lines. 
The first two lines are invariant under the adjoint operation.  Remaining lines are interchanged by the adjoint.  This table is not intended to be comprehensive, but rather lists a variety of sample K theory classes of potential ${\mathbb Z}_4$ lines.
\label{table:eta-4-solns} }
\end{center}
\end{table}

Now, not all of these lines correspond to simple (stable) objects.
From equation~(\ref{eq:stable:e2}), one can show that of the ${\mathbb Z}_2$ lines listed in table~\ref{table:eta-solns},
the only simple lines (stable sheaves) are
\begin{equation}
    (a_1, a_2, a_3, a_4, a_5) \: = \: 
    \pm (0, 2, 2, -1, 0), \: \: \:
    \pm (0, 0, 0, 1, 0),
\end{equation}
which are the ($K$-theory classes of the) structure sheaves of the diagonal $\Delta$, the antidiagonal $\overline{\Delta}$, and their shifts by 1.
By contrast, all of the ${\mathbb Z}_3$ lines listed in table~\ref{table:eta-3-solns} saturate the inequality~(\ref{eq:stable:e2}), and so are (semi-)stable.
Similarly, all of the strictly order-four elements listed in table~\ref{table:eta-4-solns}) also saturate the inequality~(\ref{eq:stable:e2}), and so are (semi-)stable.

\subsubsection{${\mathbb Z}_2 \times {\mathbb Z}_2$ Tambara-Yamagami structures}

In this section we will discuss ${\mathbb Z}_2 \times {\mathbb Z}_2$ Tambara-Yamagami structures.  These are defined by 
two commuting ${\mathbb Z}_2$ lines $\eta_1$, $\eta_2$, plus a noninvertible
line $D$ with fusion rules
\begin{equation}\label{eqn:z2ty}
    D * \eta_i \: = \: D \: = \: \eta_i * D,
    \: \: \:
    D \: = \: D^{\dag},
    \: \: \:
    D * D \: = \: 1 + \eta_1 + \eta_2 + \eta_1 * \eta_2.
\end{equation}

In K theory, solving these equations reduces to a straightforward linear algebra problem, and solutions seem to be relatively common.  We list here some simple examples of
${\mathbb Z}_2 \times {\mathbb Z}_2$ Tambara-Yamagami structures, at the level of K theory classes.  These examples have
\begin{equation}
    \eta_1 \: = \: {\cal O}_{\overline{\Delta}}[1] \: = \: -2 {\cal O}_{p \times E} - 2 {\cal O}_{E \times p} + {\cal O}_{\Delta},
\end{equation}
$\eta_2$ any line listed in table~\ref{table:z2z2tysolns}, and $D =\pm 2 {\cal O}_{p \times E} \pm 2 {\cal O}_{E \times p} \mp 2 {\cal O}_{\Delta}$.  The line $D$ is easily checked to be noninvertible, in the sense that there is no line $D^{-1}$ with
$D * D^{-1} = D^{-1} * D = {\cal O}_{\Delta}$.  Also, each line $\eta_2$ in table~\ref{table:z2z2tysolns} is easily checked to commutes with $\eta_1$ in K theory (meaning, $\eta_1 * \eta_2 = \eta_2 * \eta_1$), and
\begin{equation}
    \eta_1 * \eta_2 \: = \: {\cal O} - 2 {\cal O}_{E \times p} + {\cal O}_{\Delta},
\end{equation}
is easily shown to be another ${\mathbb Z}_2$ line, for each $\eta_2$ in
table~\ref{table:z2z2tysolns}.  These collections of lines all obey~(\ref{eqn:z2ty}).

\begin{table}
\begin{center}
    \begin{tabular}{ccccc}
    $b_1$ & $b_2$ & $b_3$ & $b_4$ & $b_5$  \\ \hline
    $\pm 1$ & $-2$ & $0$ & $1$ & $0$ \\
    $\pm 1$ & $-1$ & $-1$ & $+1$ & $\pm 1$  \\
    $\pm 1$ & $0$ & $-2$ & $+1$ & $0$  \\
    $\pm 1$ & $+1$ & $-3$ & $+1$ & $\mp 3$ \\
    $\pm 1$ & $+2$ & $-4$ & $+1$ & $\mp 8$ \\
    $\pm 1$ & $-3$ & $+1$ & $+1$ & $\mp 3$ \\
    $\pm 3$ & $+1$ & $-3$ & $+1$ & $\mp 1$ \\
    $0$ & $-2$ & $-2$ & $+1$ & $0$ \\
    $0$ & $-2$ & $0$ & $+1$ & $k$ \\
    \end{tabular}
    \caption{List of some sample ${\mathbb Z}_2$ lines $\eta_2$ that can be combined with the $\eta_1$, $D$ given in the text to form examples of ${\mathbb Z}_2 \times {\mathbb Z}_2$ Tambara-Yamagami structures on a generic elliptic curve.  Here, each line is
    $\eta_2 = b_1 {\cal O} + b_2 {\cal O}_{p \times E} + b_3 {\cal O}_{E \times p} + b_4 {\cal O}_{\Delta} + b_5 {\cal O}_q$, for values of $b_i$ listed above.  $k$ is an arbitrary integer.  We emphasize that this is merely a sample of lines, and is not intended to be in any sense complete.  
    \label{table:z2z2tysolns}
    }
    \end{center}
\end{table}

This all said, to be clear, the computation above has two limitations:
\begin{itemize}
    \item We have not discussed existence of representatives of these K-theory classes in $D^b(X \times X)$, 
    \item and we have not required any lines be stable, which as we will see is a strong restriction.  
\end{itemize}

Concerning stability specifically, the line $\eta_1$ is semistable from the criterion~(\ref{eq:stable:e2}), but none of the other lines in table~\ref{table:z2z2tysolns} is semistable, with the exception of one line whose K theory class is proportional to $\eta_1$.  This is a potential issue, as to be an interesting solution, one would want the lines defining the Tambara-Yamagami structure to be simple.  Now, table~\ref{table:z2z2tysolns}  is only a sample
list of solutions, a proof-of-concept, not a comprehensive list, so this does not imply that there are no `interesting' solutions, and furthermore, as discussed previously, our stability criterion itself may need to be reevaluated.  In any event, we mention this as a future left for future work.

Now, we should take a moment to compare to the results of
e.g.~\cite{Cordova:2023qei}.  There, it was argued that elliptic curves
admit ${\mathbb Z}_2 \times {\mathbb Z}_2$ Tambara-Yamagami structures along special loci in their moduli spaces.  This is certainly consistent with the proof-of-concept existence argument above.   It would be interesting to understand, in the present language, the precise Tambara-Yamagami structures of  \cite{Cordova:2023qei}, and how a stability criterion in our language arises that would only allow them to be simple along a special locus in moduli space, but that would require a better understanding of stability constraints in this context.  Furthermore, a more direct comparison is difficult, as \cite{Cordova:2023qei} described the defects with Gepner models, and not directly in geometry.  (Certainly the conformal field theory of the Gepner model of an elliptic curve is equivalent to that of a nonlinear sigma model for a corresponding elliptic curve at suitable moduli; however, any maps between branes only hold up to monodromy isomorphisms, making direct comparisons highly nontrivial at best.) We leave that for future work.

\subsubsection{${\mathbb Z}_3 \times {\mathbb Z}_3$ Tambara-Yamagami structures}

These structures are defined by a commuting pair of ${\mathbb Z}_3$ lines $\eta_a$, $\eta_b$, together with another noninvertible line $D$ satisfying
\begin{equation}
    \eta_{a,b} * D \: = \: D \: = \: D * \eta_{a,b},
    \: \: \:
    D \: = \: D^{\dag},
    \: \: \:
    D * D \: = \: \sum_{m,n = 0}^2 \eta_a^m * \eta_b^n.
\end{equation}

In K theory, solving these equations reduces to a straightforward linear algebra problem, and solutions seem to be relatively common.
We list here some simple examples of ${\mathbb Z}_3 \times {\mathbb Z}_3$ Tambara-Yamagami structures.  Take
\begin{equation}
    \eta_1 \: = \: {\cal O} - 3 {\cal O}_{p \times E} + {\cal O}_{\Delta} - 3 {\cal O}_q
\end{equation}
$\eta_2$ either
\begin{equation}
    \eta_2 \: = \: - {\cal O} -3 {\cal O}_{E \times p} + {\cal O}_{\Delta} + 3 {\cal O}_q
    \: \: \mbox{  or  }  \: \:
    \eta_2 \: = \: {\cal O} -3 {\cal O}_{p \times E} + {\cal O}_{\Delta} - 3 {\cal O}_q,
\end{equation}
(both of which are easily checked to be ${\mathbb Z}_3$ lines that commute with
$\eta_1$),
and $D = \mp 3 {\cal O}_{p \times E} \mp  3 {\cal O}_{E \times p} \pm 3 {\cal O}_{\Delta}$, which are easily checked to be noninvertible.
At least at the level of K theory classes,
the lines $(\eta_1,\eta_2,D)$ then define
a ${\mathbb Z}_3 \times {\mathbb Z}_3$ Tambara-Yamagami structure.

As before, we are only providing solutions in K theory.  To construct a physically-meaningful solution, one would also need to find representatives in $D^b(E \times E)$ that are semistable, so this should be interpreted merely as a proof-of-concept demonstration.

\subsection{Other fusion category structures}

In this section we will briefly discuss other noninvertible structures in defects on elliptic curves.  In particular, by reducing to $K$ theory, the fusion rules reduce to algebraic equations for the coefficients of an expansion in terms of $K$ theory, much as we have already seen for ${\mathbb Z}_n$ lines and Tambara-Yamagami structures.  As a result, it becomes very straightforward in this language to check for existence of given noninvertible symmetries, at least up to requiring stability or giving representative elements of the derived category.

\paragraph{Rep$(S_3)$.}
The Rep$(S_3)$ symmetry is defined by objects $x$, $y$, such that
\begin{equation}
    x * x \: = \: 1, \: \: \:
    x * y \: = \: y \: = \: y * x, \: \: \:
    y * y \: = \: 1 + x + y.
\end{equation}
(The ${\mathbb Z}_2$ line $x$ is identified with the nontrivial one-dimensional irreducible representation of $S_3$, and $y$ is identified with the irreducible two-dimensional representation of $S_3$.  The fusion products above correspond to tensor products of the corresponding representations.) 
If we expand
\begin{eqnarray}
    x & = & a_1 {\cal O} + a_2 {\cal O}_{p \times E} + a_3 {\cal O}_{E \times p} + a_4 {\cal O}_{\Delta} + a_5 {\cal O}_q,
    \\
    y & = & b_1 {\cal O} + b_2 {\cal O}_{p \times E} + b_3 {\cal O}_{E \times p} + b_4 {\cal O}_{\Delta} + b_5 {\cal O}_q
\end{eqnarray}
in K theory,
then some solutions for the coefficients are listed in table~\ref{table:ellcurve:rep-s3}.
(Of course, in principle one would also need to require that solutions be stable, and also find representatives in $D^b(E \times E)$, to be interesting; this is merely a proof-of-concept.)

\begin{table}
    \begin{center}
        \begin{tabular}{ccccc|ccccc}
        $a_1$ & $a_2$ & $a_3$ & $a_4$ & $a_5$ &
        $b_1$ & $b_2$ & $b_3$ & $b_4$ & $b_5$ \\ \hline
        0 & 0 & 0 & 1 & 0 & 0 & 0 & 0 & 2 & 0 \\
        0 & 0 & 0 & 1 & 0 & 0 & -3 & -3 & 2 & 0 \\
        0 & -2 & -2 & 1 & 0 & 0 & -2 & -2 & 2 & 0 \\
        0 & 2 & 2 & -1 & 0 & 0 & 2 & 2 & 0 & 0 \\
        0 & 0 & -2 & 1 & 0 & 0 & 0 & -2 & 2 & 0\\
        0 & 0 & -2 & 1 & 0 & 0 & -3 & -2 & 2 & 0 \\
        0 & 0 & 2 & -1 & 0 & 0 & 0 & 2 & 0 & 0 \\
        0 & -2 & 0 & 1 & 0 & 0 & -2 & 0 & 2 & 0 \\
        0 & -2 & 0 & 1 & 0 &  0 & -2 & -3 & 2 & 0 \\
        0 & 2 & 0 & -1 & 0 & 0 & 2 & 0 & 0 & 0 \\ 
        $-2n$ & $-2 + 2kn$ & $-2kn$ & $1$ & $-2k + 2nk^2$ &
        $n$ & $1-kn$ & $kn$ & $-1$ & $k-nk^2$ 
        \end{tabular}
        \caption{A few solutions for $K$-theory classes of lines implementing Rep$(S_3)$. In this list, $k$ and $n$ are arbitrary integers. This list is merely a set of sample solutions; it is not intended to be complete.  \label{table:ellcurve:rep-s3}}
    \end{center}
\end{table}

\paragraph{Fibonacci.}The Fibonacci fusion rule is defined by defects $W$ such that
\begin{equation}
    W * W \: = \: 1 + W.
\end{equation}
In terms of K theory classes, this relation defines a simple algebraic relation,
which one can solve.  Unfortunately, on an elliptic curve, in all solutions, at least one coefficient  is not an integer.  This is true both for generic elliptic curves as well as for elliptic curves with complex multiplication.
We conclude that elliptic curves do not admit Fibonacci structures.

\paragraph{Haagerup.}Haagerup structures are similar.
These are discussed in e.g.~\cite{h1,h2,h3,h4,h5,Huang:2021zvu,Huang:2021ytb,Huang:2021nvb,Lin:2022dhv}.
We will focus on the fusion ring for the categories ${\cal H}_2$ and ${\cal H}_3$, both of which have
six simple objects
\begin{equation}
    1, \: \: \: \alpha, \: \: \: \alpha^2, \: \: \: \rho, \: \: \:
    \alpha * \rho, \: \: \: \alpha^2 * \rho.
\end{equation}
obeying the fusion rules (see e.g., \cite[table 1]{h5} and \cite[section 3.1]{Huang:2021ytb})
\begin{equation}
    \alpha^3 \: = \: 1, \: \: \: \alpha * \rho \: = \: \rho * \alpha^2, \: \: \:
    \rho^2 \: = \: 1 + \rho + \alpha * \rho + \alpha^2 * \rho.
\end{equation}
As before, in terms of the coefficients in an expansion in K theory, these define simple algebraic relations, which can be solved exactly.
Also as before, all solutions have the property that at least one of the coefficients is not an integer, both for generic elliptic curves as well as elliptic curves with complex multiplication.
We conclude that elliptic curves do not admit the structure of the
${\cal H}_2$ or ${\cal H}_3$ Haagerup fusion category.

\paragraph{Non-unitary.}Further examples exist which involve non-unitary F-symbols, and so might be relevant for the B model TFT, which after all is a nonunitary theory.
One example is the Yang-Lee fusion category, which has the same fusion rules as Fibonacci, but different F symbols, see e.g.~\cite{Evans:2015zga}.  As no examples obeying the Fibonacci fusion rules exist on elliptic curves, Yang-Lee also cannot be realized.  Other non-unitary examples include
e.g.~categories of representations of Taft algebras, see for example \cite{bbknz}.

We conclude this section by reiterating that the noninvertible symmetries described by the B model are more general than fusion categories.  One can certainly discuss realizations of examples of fusion categories, but the B model itself is more general.

\subsection{Interface operators}

Next, we will discuss the action of interface operators on the $\mathbb{Z}_2$ defects that we have constructed. We note that our discussion complements~\cite{Bachas:2012bj}, which studied conformal and topological defects for string theory compactified on a $d$-dimensional torus. We focus on interfaces implementing the operation of T-duality
(see for example \cite[section 7]{Kapustin:2010zc}),
and so it can be naturally described in the language of this paper.
Up to factors of ${\mathbb Z}_2$, the T-duality group on an elliptic curve is 
$SL(2,{\mathbb Z}) \times SL(2,{\mathbb Z})$, one for K\"ahler moduli, the other for complex moduli.  We will focus on an $SL(2,{\mathbb Z})$ acting on complex moduli.
\subsubsection{T-duality}
The $SL(2,{\mathbb Z})$ generators 
can be described by line operators with the following kernels:
\begin{enumerate}
\item 
$T = \OO_{\Delta}(1) =   {\cal O}_{\Delta} + {\cal O}_q$ in K theory,
\item 
$S =  {\cal O}_{E \times E} + {\cal O}_{p \times E} + {\cal O}_{E \times p} - {\cal O}_{\Delta} - {\cal O}_q$ in K theory.
\end{enumerate}
Also, in the case that the elliptic curve $E$ admits a complex multiplication, the
automorphism from $\Gamma$ is described by the kernel $y_5 = {\cal O}_{\Gamma}$, the graph of the automorphism.

As a check on the $T$ transformation, note that
there exists an isomorphism of functors
\[
p_{2*}(p_1^*( -) \otimes \OO_{\Delta}(1)) \simeq ((-) \otimes \OO_E(1))
\]
and so $\Delta_*\mathcal{O}_E(1)$ is the Fourier-Mukai kernel for the functor $\otimes \mathcal{O}_E(1)$. This induces the following maps on a basis of numerical K-theory
\begin{align*}
[\mathcal{O}_E] &\mapsto [\mathcal{O}_E(1)] = [\mathcal{O}_E] + [k(p)], \\
[k(p)] &\mapsto [k(p)].
\end{align*}
With respect to this basis $\langle \mathcal{O}_E, k(p)\rangle$ for $K_{num}(E)$, this corresponds to the matrix
$\begin{pmatrix}
1 & 0 \\
1 & 1
\end{pmatrix}$
which is familiar from the usual form of T-duality. We refer to~\cite[Section 9.3]{huy} for a detailed discussion, and we note that $S$ and $\Gamma$ are of order $4$. 

Next, we consider how the T-duality line operators act on the line operators for the
${\mathbb Z}_2 \times {\mathbb Z}_2$ Tambara-Yamagami structure described in the previous section.  Briefly, the $S$ and $T$ transforms act on a given line $\eta$ by commuting past $\eta$.   For example, the image of a line $\eta$ under a $T$ transformation is the
line $\eta_T$ where
\begin{equation} \label{eq:ttrans}
    T * \eta \: = \: \eta_T * T,
\end{equation}
and the image of $\eta$ under a $S$ transformation is the line $\eta_S$ where
\begin{equation}  \label{eq:strans}
    S * \eta \: = \: \eta_S * S.
\end{equation}

We make this more explicit below.

First, expand a given line $\eta$ in terms of K theory classes as
\begin{equation}
    \eta \: = \: a_1 {\cal O} + a_2 {\cal O}_{p \times E} + a_3 {\cal O}_{E \times p} + a_4 {\cal O}_{\Delta} +
    a_5 {\cal O}_q,
\end{equation}
as before.  Using the matrix of fusion products
in~(\ref{eq:ExE:product}),
it is straightforward to compute
\begin{eqnarray}
    T * \eta & = &
    a_1 {\cal O} + (a_1 + a_2) {\cal O}_{p \times E} + a_3 {\cal O}_{e \times p} + a_4 {\cal O}_{\Delta} + ( a_3 + a_4 + a_5) {\cal O}_q,
    \\
    \eta' * T & = &
    a'_1 {\cal O} + a'_2 {\cal O}_{p \times E} + (a'_1 + a'_3) {\cal O}_{E \times p} + a'_4 {\cal O}_{\Delta} + (a'_2 + a'_4 + a'_5) {\cal O}_q,
    \\
    S * \eta & = &
    (a_2 + a_4) {\cal O} + (a_4 - a_1) {\cal O}_{p \times E} + (a_4 + a_5) {\cal O}_{E \times p} + (-a_4) {\cal O}_{\Delta} + (-a_3 - a_4) {\cal O}_q,
    \\
    \eta' * S & = & 
    (a'_3 + a'_4) {\cal O} + (a'_4 + a'_5) {\cal O}_{p \times E} + (a'_4 - a'_1) {\cal O}_{E \times p} + (-a'_4) {\cal O}_{\Delta} + (-a'_2 - a'_4) {\cal O}_q.
\end{eqnarray}
From the relations above, and a bit of algebra,
we find that the $T$ transform of the line $\eta$, defined by~(\ref{eq:ttrans}), is
\begin{equation}
    \eta_T \: = \: a_1 {\cal O} + (a_1 + a_2) {\cal O}_{p \times E} + (a_3 - a_1) {\cal O}_{E \times p} + a_4 {\cal O}_{\Delta} + (a_3 + a_5 - a_1 - a_2) {\cal O}_q.
\end{equation}
and the $S$ transform of the line $\eta$, defined by~(\ref{eq:strans}), is
\begin{equation}
    \eta_S \: = \: (-a_5) {\cal O} + a_3 {\cal O}_{p \times E} + a_2 {\cal O}_{E \times p} + a_4 {\cal O}_{\Delta} + (-a_1) {\cal O}_q.
\end{equation}

As a consistency check, note that the identity ${\cal O}_{\Delta}$ is invariant under both $T$ and $S$ transforms.
Also note from the definitions above that
\begin{equation}
    T * \eta^n \: = \: (\eta')^n * T, \: \: \:
    S * \eta^n \: = \: (\eta')^n * S.
\end{equation}
As a result, cyclic group relations are preserved.  This means that the $S$ and $T$ transforms always map ${\mathbb Z}_n$ lines to other ${\mathbb Z}_n$ lines.
It is straightforward to check this explicitly in the examples given.

\subsubsection{Generalities for abelian varieties}    \label{sect:noninv-tdual}

We recall that for toroidal sigma models on $T^d$, \cite{Bachas:2012bj} describes a rational extension of the duality group 
\[
O(d,d;\mathbb{Z}) \xhookrightarrow{} O(d,d;\mathbb{Q})
\]
to include non-invertible interfaces. The latter group acts on an integral charge vector in the usual way, but maps it to zero if the image is not integral. The extension of the symmetry group on the worldsheet theory holds to all orders in $\alpha'$, but are broken by non-zero $g_s$ effects. The mathematical results of~\cite{Polishchuk:2014} parallel this extension, and describe an extension of the auto-equivalence group of any abelian variety $A$:
\[
\mathrm{Aut}(A) \xhookrightarrow{} \mathrm{End}(A)
\]
to a subgroup of the group of endo-functors which induce an isomorphism on numerical K-theory. In particular, fusions of non-invertible interfaces implementing such endo-functors, $D$, with their parity conjugates, decompose into invertible interfaces $T_i$
\[
D \circ D^{\dagger} = \sum_{i} T_i,
\]
reminiscient of the corresponding property for general non-invertible defects in quantum field theories.

We briefly highlight the result for the target space of a product of $n$ isomorphic, general elliptic curves, $X = E^{\times n}$. There are three classes of generators for the symmetry group given by the symplectic group, $Sp(n, \mathbb{R})$, with the matrices 
\[
D(n) = \left\{  \left. \begin{pmatrix}
A & 0 \\
0 & (A^T)^{-1}
\end{pmatrix} \: \right\vert \: A \in \mathrm{GL(n, \mathbb{R})} \right\}, \quad N(n) = \left\{ \left. \begin{pmatrix}
I_n & 0 \\
B & I_n
\end{pmatrix} \: \right\vert \: B \in \mathrm{Sym(n, \mathbb{R})} \right\}, \quad \Omega = \begin{pmatrix}
0 & I_n \\
-I_n & 0
\end{pmatrix} 
\]
Roughly speaking, elements in $D(n)$ correspond to isogenies of $X$, and $N(n)$ correspond to the endo-functor $\otimes V_{ \varphi}$, where $V_{\varphi}$ is the semi-homogeneous bundle on $X$ of class $\varphi \in NS(X) \otimes \mathbb{Q}$. Here, $\Omega$ corresponds to the Fourier-Mukai functor $\varphi_L^* \circ \Phi_{\mathcal{P}}$, where 
\[
\varphi_L \colon X \rightarrow \hat{X}
\]
is the morphism induced by some polarization $L$ on the abelian variety $X$, and $\hat{X}$ denotes the dual abelian variety. We take $\mathcal{P}$ to be usual Poincar\`e bundle, and $\Phi_\mathcal{P}$ the corresponding Fourier-Mukai functor.

The rational extension of the duality group then corresponds to the inclusion
\[
\mathrm{Sp(n,\mathbb{Z})} \xhookrightarrow{} \mathrm{Sp(n,\mathbb{Q})}.
\]

\section{Noninvertible symmetries on K3s}  \label{sect:k3}

In this section we describe some basic examples of noninvertible algebra structures
on a K3 surface, at the level of K theory, closely following our analysis of elliptic curves.

Let $X \xhookrightarrow{} \mathbb{P}^3$ be a smooth quartic hypersurface. Then the restriction of the hyperplane class gives a curve $C$ with self-intersection $C^2 = 4$. We record the topological facts:
\[
\mathrm{td}(X) = (1,0,2), \quad \mathrm{td}(C) = (1,-2).
\]
We consider the numerical K-group of rank $10$, which is spanned by the class
\[
K_{num}(X \times X) = \langle X \times X,\  X \times C,\  C \times X,\  X \times p,\  p \times X,\  C \times C,\  \Delta,\ C \times p,\  p \times C,\ q \rangle.
\]
where $\Delta: X \hookrightarrow X \times X$ is the diagonal, $p$ is a point on $X$, and $q$ is a point on $X \times X$.

This gives the following convolution table
\[
\begin{pmatrix}
0 & 0 & 0 & 0 & x_1 & 0 & x_1 & 0 & x_2 & x_4 \\
0 & 0 & 4x_1 & 0 & 0 & 4x_2 & x_2 & 4x_4 & 0 & 0 \\
0 & 0 & 0 & 0 & x_3 & 0 & x_3 & 0 &  x_6 & x_8 \\
x_1 & x_2 & 0 & x_4 & 0 & 0 & x_4 & 0 & 0 & 0\\
0 & 0 & 0 & 0 & x_5 & 0 & x_5 & 0 & x_9 & x_{10} \\
0 & 0 & 4x_3 & 0 & 0 & 4 x_6 & x_6 & 4 x_8 & 0 & 0\\
x_1 & x_2 & x_3 & x_4 & x_5 & x_6 & x_7 & x_8 & x_9 & x_{10}\\
x_3 & x_6 & 0 & x_8 & 0 & 0 & x_8 & 0 & 0 & 0  \\
0 & 0 & 4x_5 & 0 & 0 & 4x_9 & x_9 & 4x_{10} & 0 & 0 \\
x_5 & x_9 & 0 & x_{10} & 0 & 0 & x_{10} & 0 & 0 & 0
\end{pmatrix}
\]

It is straightforward to find solutions for ${\mathbb Z}_2$ lines in K theory.
If we expand
\begin{equation}
    \eta \: = \: a_1 {\cal O}_{X \times X} + a_2 {\cal O}_{X \times C} + a_3 {\cal O}_{C \times X} + a_4 {\cal O}_{X \times p} + a_5  {\cal O}_{p \times X} + a_6 {\cal O}_{C \times C} + a_7 {\cal O}_{\Delta} + a_8 {\cal O}_{C \times p} + a_9 {\cal O}_{p \times C} + a_{10} {\cal O}_q,
    \nonumber
\end{equation}
then some solutions for $\eta * \eta = {\cal O}_{\Delta}$ in K theory are
listed in the table below:
\begin{center}
    \begin{tabular}{cccccccccc}
    $a_1$ & $a_2$ & $a_3$ & $a_4$ & $a_5$ & $a_6$ & $a_7$ & $a_8$ & $a_9$ & $a_{10}$ \\ \hline
    0 & 0 & 0 & 0 & 0 & 0 & $\pm 1$ & 0 & 0 & 0 \\
    $k$ & $\ell$ & 0 & $\mp 2$ & 0 & 0 & $\pm 1$ & 0 & 0 & 0 \\
    $k$ & $0$ & $-\ell$ & $0$ & $\mp 2$ & $0$ & $\pm 1$ & $0$ & $0$ & $0$ \\ 
    0 & 0 & 0 & $\mp 2$ & 0 & 0 & $\pm 1$ & $k$ & 0 & $\ell$ \\
    0 & 0 & 0 & 0 & $\mp 2$ & 0 & $\pm 1$ & 0 & $-k$ & $\ell$ \\ 
    0 & $k$ & 0 & $\mp 2$ & 0 & 0 & $\pm 1$ & 0 & $k \ell$ & $\mp 2 \ell$ \\
    0 & 0 & $-k$ & 0 & $\mp 2$ & 0 & $\pm 1$ &  $- k \ell$ & 0 & $\mp 2 \ell$ \\ 
    0 & 0 & 0 & 0 & $\mp 2$ & 0 & $\pm 1$ & 0 & $k$ & $\ell$ \\
    0 & 0 & 0 & $\mp 2$ & 0 & 0 & $\pm 1$ & $-k$ & 0 & $\ell$ \\ 
    0 & 0 & $k$ & $\mp 2$ & $\mp 2$ & 0 & $\pm 1$ & $\ell$ & 0 & 0 \\
    0 & $-k$ & 0 & $\mp 2$ & $\mp 2$ & 0 & $\pm 1$ & 0 & $-\ell$ & 0 \\ 
    0 & 0 & 0 & $\mp 2$ & $\mp 2$ & 0 & $\pm 1$ & $k$ & $\ell$ & $\pm 2 k \ell$ \\
%
    \end{tabular}
\end{center}
where $k, \ell$ are arbitrary integers.

We emphasize that this is not intended as a complete list, but rather a partial list of a few solutions, as a demonstration of existence.

Also, although a ${\mathbb Z}_2$ group action on $X$ would define a
${\mathbb Z}_2$ defect in $X \times X$ (as its graph), and hence an element of
K theory above, it is not clear to us whether the converse is true,
whether every ${\mathbb Z}_2$ line in K theory necessarily corresponds to a group action.  One issue is that not every element of K theory may have a representative in
$D^b(X \times X)$.  Another issue is that for those which can be represented by
elements of $D^b(X \times X)$, it is not clear to us whether they necessarily
correspond to the graph of a group action.  We leave this question for future work.

Next, we turn to ${\mathbb Z}_2 \times {\mathbb Z}_2$ Tambara-Yamagami structures.
The dual is given in K theory by
\begin{equation}
    \eta^{\dag} \: = \: 
    a_1 {\cal O}_{X \times X} - a_3 {\cal O}_{X \times C} - a_2 {\cal O}_{C \times X} + a_5 {\cal O}_{X \times p} + a_4  {\cal O}_{p \times X} + a_6 {\cal O}_{C \times C} + a_7 {\cal O}_{\Delta} - a_9 {\cal O}_{C \times p} - a_8 {\cal O}_{p \times C} + a_{10} {\cal O}_q.
    \nonumber
\end{equation}

As noted earlier, a ${\mathbb Z}_2 \times {\mathbb Z}_2$ Tambara-Yamagami structure is defined by two commuting ${\mathbb Z}_2$ lines $\eta_{1,2}$ together with another noninvertible line $D$ such that
\begin{equation}
    D * \eta_i \: = \: D \: = \: \eta_i * D, \: \: \:
    D \: = \: D^{\dag}, \: \: \:
    D * D \: = \: 1 + \eta_1 + \eta_2 + \eta_1 * \eta_2.
\end{equation}
For example, a set of K-theory representatives of lines obeying the fusion rules above is
\begin{equation}
    \eta_1 \: = \:  -2 {\cal O}_{X \times p} + {\cal O}_{\Delta},
    \: \: \:
    \eta_2 \: = \:  2 {\cal O}_{X \times p} + 2 {\cal O}_{p \times X} - {\cal O}_{\Delta},
    \: \: \:
    D \: = \: \pm 2 {\cal O}_{p \times X}.
\end{equation}
It is straightforward to check that $\eta_1 * \eta_2 = \eta_2 * \eta_1$, and that there does not exist a line $D^{-1}$ such that $D* D^{-1} = D^{-1} * D = {\cal O}_{\Delta}$.
We leave questions of both stability and realizations in $D^b(X \times X)$ for future work.

It is straightforward (using mathematica) to check that there are no integer solutions in K theory for Fibonacci lines on this K3.

\section{Examples of noninvertible symmetries on general Calabi-Yau's}
\label{sect:cy}

In this section we will discuss some examples of noninvertible symmetries that exist for Calabi-Yau's in any dimension,
namely a noninvertible $B$ field transformation, and also quantum symmetries.  Both have previously appeared in the literature; however, we believe our construction of quantum symmetry defects may be novel to the physics literature, and in any event, we include them both to emphasize that noninvertible symmetries exist in Calabi-Yau's in much greater generality than just RCFTs.

\subsection{Noninvertible $B$ field gauge transformations}

The two-form tensor potential $B$ is well-known to have a gauge transformation which locally can be written
\begin{equation}
    B \mapsto B + d \Lambda,
\end{equation}
where $\Lambda$ is a connection on a principal $U(1)$ bundle.
In fact, the $B$ field also has a noninvertible gauge transformation generalizing the
one above, as discussed in e.g.~\cite{waldorf07}, and utilized recently in for example
\cite{Perez-Lona:2024yih}.  We will see a close analogue in describing noninvertible quantum symmetries, so for completeness, we briefly review these defects here.

First, recall the defect for an ordinary $B$ field gauge transformation, of the form $B \mapsto B + 2 \pi c_1(L)$ for $L \rightarrow X$ some line bundle.
The interface describing this transformation is simply $\Delta_* L$, where $\Delta: X \rightarrow X \times X$ is the diagonal.

The condition for such an interface to be topological is~(\ref{eq:cpx-lg-condition}),
namely
\begin{equation}
    \pi_1^* B_1 \: + \: F(L) \: - \: \pi_2^* B_2 \: = \: 0,
\end{equation}
where $F(L)$ is the curvature\footnote{We gloss over technical subtleties in describing open strings with Chan-Paton factors of rank $>1$ in the expression above, as is common.} of $L$ (so that $[F(L)] = c_1(L)$).
In the special case that the $B$'s vanish, this is just the condition that the bundle be flat.  In the present case, we get a topological defect by taking $B_2 = B_1 + F$.  In this fashion, the defect realizes the gauge transformation of the $B$ field.

Noninvertible gauge transformations of $U(1)$ gerbes defined by vector bundles of rank greater than one have been previously discussed in \cite[definition 2]{waldorf07}, \cite{bss16}, and realized as defects in \cite{Fuchs:2007fw,Runkel:2008gr,Bunk:2018xwb}.
Following \cite[definition 2]{waldorf07},
the noninvertible analogue of a gauge transformation replaces the line bundle $L$ with a vector bundle $V$ (plus other data we will mostly suppress).  The topological condition is given in \cite[axiom (1M1)]{waldorf07}, which says
\begin{equation}
    \pi_1^* B_1 \: + \: \frac{1}{n} {\rm Tr}\, F(V) \: - \: \pi_2^* B_2 \: = \: 0
\end{equation}
where $F(V)$ is the curvature of $V$.
This is certainly the naive extrapolation of the topological condition to the case of
nonabelian Chan-Paton factors.  More formally, this is the statement that $B \mapsto B + 2 \pi c_1(V)/n$.

More can be said.  Considering D-branes in the presence of topologically
nontrivial $B$ fields leads to a notion of twisted bundles, see for example
\cite[section 13.4]{huy},
\cite{Bouwknegt:2000qt,andreithesis,Caldararu:2003kt}, and the discussion in section~\ref{sect:othertft}.
There are analogues here.  For example, \cite[section 3.2]{waldorf07}, on ``bundle gerbe modules,'' discusses twisted bundles, twisted by noninvertible symmetry group actions.
(See also \cite{Bouwknegt:2001vu}.)
Briefly, the idea is that given a (noninvertible) $B$ field transformation defined by a vector bundle $V \rightarrow X$, the corresponding action on D-branes is ${\cal E} \mapsto {\cal E} \otimes V$.

In passing, this fact also appears in discussions of noninvertible extensions of T-duality, as reviewed in section~\ref{sect:noninv-tdual}, in which $O(d,d;{\mathbb Z})$ (on a $d$-dimensional torus) is extended to $O(d,d;{\mathbb Q})$.
One piece of that T-duality is an action on $B$ fields, essentially, $B \mapsto B + 2 \pi$ and multiples thereof.  In the ${\mathbb Q}$-valued extension, this is replaced by a noninvertible analogue.  Here, we discuss noninvertible $B$ field gauge transformations on general spaces $X$, not necessarily tori.

\subsection{Quantum symmetries in orbifolds}

Consider the orbifold ${\mathfrak X} = [X/G]$, where $G$ is a finite group acting on a space $X$.  
We will construct a defect, an object in
$D( [X/G] \times [X/G] )$, that geometrically\footnote{
To be clear, defects giving quantum symmetries have been described in other sources, see for example
\cite[section 3.2]{Brunner:2013ota}, but to our knowledge previous presentations have been more algebraic in nature, not geometric.
}
realizes the quantum symmetry, for $G$ both abelian and nonabelian.

First, for any representation $\rho$ of $G$, we can construct a vector bundle $V_{\rho} \rightarrow [X/G]$, as the vector bundle associated to the principal $G$ bundle $E = (X \rightarrow [X/G])$ via the representation $\rho$, as  
\begin{equation}
    V_{\rho} \: = \: E \otimes_{\rho} {\mathbb C}^n,
\end{equation}
for $n$ the dimension of the representation $\rho$.
(For spaces $X/G$, $X \rightarrow X/G$ is only a bundle if $G$ acts freely, but for stacks, $X \rightarrow [X/G]$ is always a principal $G$ bundle, regardless of whether $G$ acts freely.)

We claim that the defect implementing the quantum symmetry is the object
\begin{equation}
    \Delta_* V,
\end{equation}
for $\Delta: {\mathfrak X} \rightarrow {\mathfrak X} \times {\mathfrak X}$ the diagonal embedding.

Now, to relate to the quantum symmetry, we must discuss the induced action on the closed-string states.  To that end, first recall that
the chiral primaries of the orbifold, the various twisted sector states, correspond to cohomology of the ``inertia stack'' $I_{\cal X}$:
\begin{equation}
    H^{\bullet}\left( I_{\mathfrak X} \right).
\end{equation}
Intuitively the inertia stack is the zero-momentum part of the loop space.

As discussed in section~\ref{sect:bulk}, for any bundle $V \rightarrow {\mathfrak X}$, the action induced by a defect $\Delta_* V$ on Hochschild cohomology is to map
\begin{equation}
\omega \: \mapsto \: \omega \otimes {\rm ch}^{\rm rep}(V),
\end{equation}
for any $\omega \in H^{\bullet}(I_{\mathfrak X})$.  
We can understand ch$^{\rm rep}$ in the expression above as follows.
Let $\pi: I_{\mathfrak X} \rightarrow {\mathfrak X}$ be the projection, then on each component of the
inertia stack $I_{\mathfrak X}$ with stabilizer $g$, the bundle
$\pi^* V$ decomposes into eigenbundles
\begin{equation}
    \pi^* V |_g \: = \: \bigoplus_{\chi} E_{g,\chi}.
\end{equation}
Then, ch$^{\rm rep}$ is defined on that component by
\begin{equation}
    {\rm ch}^{\rm rep}(V) |_g \: = \: \sum_{\chi} {\rm ch}(E_{g,\chi}) \otimes \chi(g).
\end{equation}
In particular, ch$^{\rm rep}$ has components with complex phases.

In the present case, there is no contribution to ch$^{\rm rep}(V)$ in degree greater than zero, essentially because $V$ is trivial on the atlas.  As a result, multiplication by ch$^{\rm rep}(V)$
merely multiplies by a complex number, of the form of a sum of ranks and phases.

If $G$ is abelian, so that the irreducible representations are one-dimensional, then the effect of multiplying by ch$^{\rm rep}(V)$ is to multiply twisted-sector states by characters, corresponding to the twisted sector in question.  This is precisely the same as an ordinary old-fashioned
quantum symmetry in an abelian orbifold, as discussed in e.g.~\cite[section 8.5]{Ginsparg:1988ui}.
For nonabelian $G$, the effect is closely analogous.  For these reasons, we identify this defect with the generator of the quantum symmetry.

Next, let us comment on products of these defects.  Recall from equation~(\ref{eq:fusion-diagonals}), 
given any two vector bundles $V_1$, $V_2$, the fusion product
\begin{equation}
    \left( \Delta_* V_1 \right) * \left( \Delta_* V_2 \right) \: = \: \Delta_* (V_1 \otimes V_2 ).
\end{equation}
In the present case, where each $V$ is associated to a representation, the product respects the product of the
representations:
\begin{equation}
    V_{\rho_1} \otimes V_{\rho_2} \: = \: V_{\rho_1 \otimes \rho_2}.
\end{equation}

In the case that $G$ is abelian, the irreducible representations $\rho$ are one-dimensional,
the associated bundles $V_{\rho}$ are line bundles, and so the multiplication above is clearly invertible.

In the case that $G$ is not abelian, the irreducible representations $\rho$ need not be one-dimensional,
and the multiplication need not be invertible.

In passing, we should mention that these defects have appeared in other
contexts recently.  For example, the papers \cite{Gutperle:2024vyp,Knighton:2024noc} constructs noninvertible symmetry defects in symmetric group orbifolds, as applicable to AdS/CFT (see for example \cite{Lerche:2023wkj}), but their constructions are, so far as we can tell, algebraic in nature, whereas the construction here in geometric.  Similarly, there is an algebraic construction of defects related to discrete torsion in \cite{Brunner:2014lua}; we leave a geometric construction for the future.

\section{Decomposition and condensation defects}   \label{sect:decomp}

We can also understand decomposition \cite{Hellerman:2006zs,Sharpe:2022ene} in this language.
Briefly and schematically, decomposition says that a $d$-dimensional theory with a global $(d-1)$-form symmetry is equivalent to a disjoint union of other $d$-dimensional theories.  We can see examples of this in sigma models whose targets are gerbes,
as outlined in section~\ref{sect:othertft}.  The derived category of a gerbe is equivalent to a disjoint union of twisted derived categories of spaces.  In this section, we will describe realizations of projectors onto universes and more general portals between universes in the language of derived categories.

For simplicity, let ${\mathfrak X}$ be a banded $G$ gerbe over a space $B$, of characteristic class $\omega \in H^2(B,G)$, for $G$ abelian.
(More general versions exist, but we will restrict to this special case for readability.)
In this case,
\begin{equation}  \label{eq:dercat:decomp}
    D^b({\mathfrak X}) \: = \: \coprod_{\chi} D^b({\mathfrak X})_{\chi},
\end{equation}
where the product is over all irreducible representations (here, characters) $\chi$ of $G$,
and
\begin{equation}
    D^b({\mathfrak X})_{\chi} \:\cong \: D^b(B, \chi_{\omega}),
\end{equation}
for $\chi_{\omega} \in H^2(B,U(1))$ given by the image of the characteristic class $\omega$ under the map defined by $\chi$.

Let us take a moment to clarify the meaning of the decomposition~(\ref{eq:dercat:decomp}).
This can be described as example of an orthogonal decomposition.
A triangulated category ${\cal T}$ has an orthogonal decomposition into full triangulated subcategories ${\cal A}$, ${\cal B}$ if
\begin{enumerate}
    \item ${\rm Hom}({\cal A}, {\cal B}) = 0 = {\rm Hom}({\cal B}, {\cal A})$, and
    \item for all $T \in {\cal T}$, $T = A \oplus B$ 
    where $A \in {\cal A}$, $B \in {\cal B}$.
\end{enumerate}
An orthogonal decomposition is a special case of a semiorthogonal decomposition, which arises more frequently.
A triangulated category ${\cal T}$ is said to have a semiorthogonal decomposition $\langle {\cal A}, {\cal B} \rangle$ if
\begin{enumerate}
    \item ${\rm Hom}({\cal A}, {\cal B}) = 0 $, and
    \item for all $T \in {\cal T}$, $T$ fits into a distinguished triangle
    \begin{equation}
        A \: \longrightarrow \: T \: \longrightarrow \: B \: \longrightarrow \: A[1],
    \end{equation}
    where $A \in {\cal A}$, $B \in {\cal B}$.
\end{enumerate}
In the case of an orthogonal decomposition, one adds the constraint Hom$({\cal B}, {\cal A}) = 0$, and the distinguished triangle is split,
because any map $B \rightarrow A[1]$ vanishes.

In effect, in a semiorthogonal decomposition, morphisms are allowed in one direction between different components, whereas in an orthogonal decomposition, morphisms are only allowed within a single component, and no morphisms exist, in any direction, between different components.  Phrased another way, the orthogonal decomposition above is a special case and more extreme form of a semiorthogonal decomposition.

Returning to our gerbe ${\mathfrak X}$, there exists a projector $\pi_{\chi}: D^b({\mathfrak X}) \rightarrow D^b({\mathfrak X})$ whose essential
image lies in $D^b({\mathfrak X})_{\chi}$.  It can be described as an integral transform, with kernel in $D^b({\mathfrak X} \times {\mathfrak X})$ we construct as follows.

First, ${\mathfrak X} \times {\mathfrak X}$ has a substack $S$ which is the pullback of the diagram
\begin{equation}
    \xymatrix{
    S \:\ar@{^{(}->}[r] \ar[d] & {\mathfrak X} \times {\mathfrak X} \ar[d] 
    \\
    B\: \ar@{^{(}->}[r]^-{\Delta} & B \times B.
    }
\end{equation}
In particular, $S$ is a $G \times G$ gerbe on $B$.

Now, the kernel we want, the object of $D^b({\mathfrak X} \times {\mathfrak X})$, can be described as a sheaf on $B$ twisted by a character of $G \times G$, evaluated on $\pi_1^* \omega \otimes \pi_2^* \omega$.  The desired kernel is twisted by the character $\pi_1^* \chi^{-1} \otimes \pi_2^* \chi$, which evaluates to the identity on $\pi_1^* \omega \otimes \pi_2^* \omega$, hence the twisting class will be trivial.

Finally, the object we want in $D^b(S) \subset D^b({\mathfrak X} \times {\mathfrak X})$ is just the
structure sheaf of $B$, ${\cal O}_B \in D^b(B) \subset D^b(S)$.  This object, viewed as an element of the $\pi_1^* \chi^{-1} \otimes \pi_2^* \chi$ universe, defines the kernel of an integral transform which acts as the projector $\pi_{\chi}: D^b({\mathfrak X}) \rightarrow D^b({\mathfrak X})$ projecting into universe $\chi$ in the decomposition of the gerbe ${\mathfrak X}$.

To be more concrete, let us consider a specific example.

Suppose ${\mathfrak X} = BG$, for $G$ abelian.
In this case, $D^b({\mathfrak X})$ is the derived category of representations of $G$.
The decomposition is indexed by irreducible representations (here, characters) of $G$,
where $D^b({\mathfrak X})_{\chi}$ is the piece of $D^b({\rm Rep}(G))$ which are sums of one-dimensional representations corresponding to $\chi$.  Phrased more compactly,
\begin{equation}
    D^b({\mathfrak X})_{\chi} \: = \: D^b({\rm Vec}) \otimes V_{\chi},
\end{equation}
where $V_{\chi}$ is the one-dimensional representation corresponding to the character $\chi$, and $D^b({\rm Vec})$ denotes the derived category of vector spaces -- complexes of vector spaces.

Now, any functor $D^b({\mathfrak X}) \rightarrow D^b({\mathfrak X})$ in this case is defined by a $G \times G$ bimodule $M$ (meaning, a module with both a left $G$ action and a right $G$ action).
Given $V \in {\rm Rep}(G)$, one acts by taking the invariants of $M \otimes V$ under the left $G$ action.  The result is a right $G$ module.

In this language, the projector $\pi_{\chi}$ is defined by the bimodule
$V_{\chi^{-1}} \otimes V_{\chi}$, for $V_{\chi}$ the one-dimensional representation
corresponding to the character $\chi$.  Taking left $G$ invariants under the tensor product with $V_{\chi^{-1}}$ projects out isotypic components not associated with character $\chi$, hence accomplishing the projection.

So far, we have discussed how projectors onto universes $\pi_{\chi}: D^b({\mathfrak X}) \rightarrow D^b({\mathfrak X})_{\chi}$ are constructed in derived categories.  Similar ideas can also be used to construct `portals' linking different universes, in the same way that nondynamical Wilson lines can link different universes in the decomposition of a gauge theory.
Specifically, returning to the description of banded abelian gerbes, if we take the kernel of an integral transform to be defined by an element of $D^b(S) \subset D^b({\mathfrak X} \times {\mathfrak X})$ which is not associated with the diagonal $\pi_1^* \chi^{-1} \otimes \pi_2^* \chi$, but rather with a more general element $\pi_1^* \chi_1^{-1} \otimes \pi_2^* \chi_2$, the resulting integral transform will map $D^b({\mathfrak X})_{\chi_1} \rightarrow D^b({\mathfrak X})_{\chi_2}$.

Finally, we will describe how derived categories can be used to give some simple examples of condensation defects.

First, recall that a condensation defect is obtained by gauging along a subvariety \cite{Roumpedakis:2022aik,Choi:2022zal,Lin:2022xod}.  More precisely, consider a $d$-dimensional quantum field theory with a global $k$-form symmetry, and restrict to a codimension $p$ submanifold $S$.  A condensation defect can be obtained by ``$p$-gauging'' the global $k$-form symmetry along $S$, where it appears to be a gauged $(k-p)$-form symmetry, a `condensation' of the $k$-form symmetry defects along the submanifold.

Following the spirit of \cite{Lin:2022xod}, suppose that the bulk worldsheet theory is a sigma model with target a $K$-gerbe ${\mathfrak X}$.  As such, the theory has a global 1-form symmetry.  Along a codimension-one submanifold, we can gauge that 1-form symmetry, where it appears to be a zero-form symmetry in one dimension.  Gauging that symmetry along the defect undoes the decomposition, and projects to a single universe, call it $Y$, as in \cite{Lin:2022xod}.  So, the bulk has target ${\mathfrak X}$, but the defect is an object of $D(Y \times Y)$.

Fusion products of such condensation defects are realized in the same fashion as before (composition of integral transforms), but those compositions are computed on $Y$, not ${\mathfrak X}$.

For completeness, we add that \cite[section 4.2]{Bah:2023ymy} also characterizes condensation defects in terms of tachyon condensation, albeit from a different perspective.

\section{Conclusions}

In this paper we have described how derived categories in the B model can be used to give a completely concrete understanding of defects and noninvertible symmetries in Calabi-Yau sigma models.  We have discussed both the topological B model as well as B-type defects in physical theories, and applied thee methods to study noninvertible symmetries in elliptic curves and a family of K3 surfaces, where we have provided K-theoretic evidence for e.g.~Tambara-Yamagami structures.  We have also argued that
some noninvertible symmetries (such as the Fibonacci structure) can not be realized at all.

\section{Acknowledgements}

We would like to thank P.~Aspinwall, D.~Ben-Zvi, R.~Bryant, R.~Donagi, P.~Horja, J.~J.~Heckman, E.~Heng, S.~Katz, J.~Knapp, I.~Melnikov, A.~Perez-Lona, D.~Robbins, and H.~Y.~Zhang for useful conversations.  
A.C.~was partially supported by the National Science Foundation through grant numbers DMS-2152088 and DMS-2202365.
T.P.~was supported in part by NSF/BSF grant DMS-2200914 and NSF FRG grant DMS 2244978.
E.S.~was partially supported
by NSF grant PHY-2310588. X.Y.~was supported
by NSF grant PHY-2310588.

\appendix

\section{Miscellaneous technical results}   \label{sect:misc}

In this appendix we collect a few technical results that are frequently utilized in computations in the text.

One result that is frequently used is the projection formula, which says, briefly,
\begin{equation}
    f_* \left( (f^* \alpha) \cdot \beta \right) \: = \: \alpha \cdot f_* \beta.
\end{equation}

Another identity that is frequently used relates derived duals of pushforwards to pushforwards of derived duals.  For ${\cal E} \in D^b(X \times Y)$,
\begin{equation}  \label{eq:relduality}
    \left( \pi_{X *} {\cal E} \right)^{\vee} \: = \:
    \pi_X^* \left( {\cal E}^{\vee} \otimes \pi_Y^* \omega_Y [\dim Y] \right),
\end{equation}
where $\omega_Y$ denotes the canonical bundle of $Y$.

More generally, for $f: X \rightarrow Y$ (a proper map of smooth projective varieties)
(see e.g.~\cite[section 3.1]{Ando:2010nm})
\begin{equation}
    f_*\left( {\cal E}^D \right) \: \cong \: \left( f_* {\cal E} \right)^D,
\end{equation}
for ${\cal E}^D = {\cal E}^{\vee} \otimes \omega_X[\dim X]$, where $\omega_X$ is the canonical bundle of $X$.
In the special case that $f = \pi: X \times Y \rightarrow X$,
\begin{eqnarray}
    \pi_*\left( {\cal E}^D \right) & = & \pi_* \left( {\cal E}^{\vee} \otimes \omega_{X \times Y}[\dim X + \dim Y]\right),
    \\
    \left( \pi_* {\cal E} \right)^D & = & \left( \pi_* {\cal E} \right)^{\vee} \otimes \omega_X[\dim X],
\end{eqnarray}
and equation~(\ref{eq:relduality}) follows.

Another frequently-used identity is Serre duality, which in the present context
is the statement that for any ${\cal E}, {\cal F} \in D^b(X)$,
\begin{equation}
    {\rm Hom}_X\left( {\cal E}, {\cal F} \right) \: \cong \:
    {\rm Hom}\left( {\cal F}, {\cal E} \otimes \omega_X[\dim X] \right),
\end{equation}
where $\omega_X$ is the canonical bundle of $X$.

Let us also recall the statement of Grothendieck-Riemann-Roch.  For a morphism $f: X \rightarrow Y$,
\begin{equation}
    {\rm ch}(f_* {\cal F}) \, {\rm td}(TY) \: = \: f_* \left( {\rm ch}({\cal F}) \, {\rm td}(TX) \right),
\end{equation}
where in ch$(f_* {\cal F})$, $f_*$ is interpreted as the alternating sum of cohomologies of the derived pushfforward, or equivalently
\begin{equation}
    {\rm ch}(f_* {\cal F})
    \: = \: f_* \left( {\rm ch}({\cal F}) \, {\rm td}(TX) f^* {\rm td}(TY)^{-1} \right).
\end{equation}

\section{Intersections on product of elliptic curves}
\label{app:ell-int}

Let $E$ be an elliptic curve.
From e.g.~\cite{rs}, for $E$ an elliptic curve,
\begin{itemize}
    \item for generic $E$, the Neron-Severi group of $E \times E)$ has dimension 3,
    generated by ${\rm point} \times E$, $E \times {\rm point}$, and $\Delta$,
    \item for $E$ an elliptic curve with complex multiplication, the Neron-Severi group of $E \times E$ has dimension 4, generated by the divisors above plus $\Gamma$, the graph of an endomorphism of $E$ arising due to the complex multiplication.
\end{itemize}

For reference, the intersections of those four divisors are as follows\footnote{
ES would like to thank R.~Donagi for a detailed explanation.
}:
\begin{equation}
     \begin{array}{c|cccc}
     & E \times p & p \times E & \Delta & \Gamma \\ \hline
     E \times p & 0 & 1 & 1 & {\rm deg} \, f \\
     p \times E & 1 & 0 & 1 & 1 \\
     \Delta & 1 & 1 & 0 & {\rm deg} \, (f - 1) \\
     \Gamma & {\rm deg}\, f & 1 & {\rm deg}\, (f-1) & 0
     \end{array}
\end{equation}
where $\Gamma$ is the graph of $f: E \rightarrow E$, and $p$ is a point of $E$.
For a generic elliptic curve, without complex multiplication, merely omit the row and column
for $\Gamma$.

To compute the matrix above, we record here a few notes:
\begin{itemize}
    \item In the intersection of $E \times p$ with itself, since we can change either $p$ and
    so move the curve, we can deform to not intersect.
    \item In the intersection of $E \times p$ with $p \times E$ and with the diagonal
    $\Delta$, there will be one shared point in common, namely $(p,p)$.
    \item In the intersection of $E \times p$ with $\Gamma = \{(x, f(x))$, there will be as many
    shared points as $f^{-1}(p)$, namely ${\rm deg}\, f$.
    \item In the intersection of $\Delta$ with $\Gamma$, there are as many shared points as
    solutions of $(p,p) = (p,f(p))$, meaning solutions of $f(p) - p = 0$, hence the
    degree of $f-1$.
    \item In the intersection of $\Delta$ with itself, we can deform either $\Delta$ to describe
    solutions $(x,y)$ of $y - x = \epsilon \in E$ for $\epsilon \neq 0$, which is disjoint.
    \item In the intersection of $\Gamma$ with itself, we can deform either $\Gamma$ to describe solutions of $y - f(x) = \epsilon \in E$ for $\epsilon \neq 0$, which is disjoint.
\end{itemize}

\end{document}